\documentclass[twocolumn]{aastex701}
\usepackage{verbatim}
\usepackage{graphicx}

\newcommand{\magphys}{\hbox{\sc magphys}}
\newcommand{\av}{\hbox{$A_\mathrm{V}$}}
\newcommand{\um}{\hbox{$\mu$m}}

\newcommand{\sigmstar}{\hbox{$\Sigma_{M_\ast}$}}

\newcommand{\sigsfr}{\hbox{$\Sigma_\mathrm{SFR}$}}

\newcommand{\ldust}{\hbox{$L_\mathrm{dust}$}}

\newcommand{\mtw}{\hbox{$M_{20}$}}

\newcommand{\Msun}{\mathrm{M}_{\odot}}

\begin{document}





\title{ALESS--JWST: Dust-driven Morphologies and Hidden Stellar Mass in $z\sim3$ Sub-millimeter Galaxies} 

\author[orcid=0000-0002-8184-5229,sname=Li,gname=Juno]{J. Li}
\email[show]{juno.li.research@gmail.com}
\affiliation{International Center for Radio Astronomy Research (ICRAR), The University of Western Australia, 35 Stirling Highway, Crawley WA 6009, Australia}
\affiliation{International Space Center (ISC), The University of Western Australia, 35 Stirling Highway, Crawley WA 6009, Australia}
\correspondingauthor{Juno Li}

\author[orcid=0000-0001-9759-4797,sname=da Cunha,gname=Elisabete]{E. da Cunha}
\email{elisabete.dacunha@uwa.edu.au}
\affiliation{International Center for Radio Astronomy Research (ICRAR), The University of Western Australia, 35 Stirling Highway, Crawley WA 6009, Australia}

\author[orcid=0000-0001-6586-8845]{J. A. Hodge}\email{hodge@strw.leidenuniv.nl}
\affiliation{Leiden Observatory, Leiden University, P.O. Box 9513, 2300 RA Leiden, The Netherlands}

\author[orcid=0000-0003-3037-257X]{I. Smail}
\email{ian.smail@durham.ac.uk}
\affiliation{Centre for Extragalactic Astronomy, Department of Physics, Durham University, South Road, Durham DH1 3LE, UK}

\author[orcid=0000-0002-7612-0469]{S. Kendrew}
\email{skendrew@stsci.edu}
\affiliation{European Space Agency (ESA), ESA Office, Space Telescope Science Institute, 3700 San Martin Drive, Baltimore, MD 21218, USA}

\author[orcid=0000-0003-4569-2285]{A. Battisti}
\email{andrew.battisti@uwa.edu.au}
\affiliation{International Center for Radio Astronomy Research (ICRAR), The University of Western Australia, 35 Stirling Highway, Crawley WA 6009, Australia}
\affiliation{Research School of Astronomy and Astrophysics, Australian National University, Cotter Road, Weston Creek, ACT 2611, Australia}

\author[orcid=0000-0002-7698-3002]{M. Cracraft}
\email{cracraft@stsci.edu}
\affiliation{Space Telescope Science Institute, 3700 San Martin Drive, Baltimore, MD 21218, USA}


\author[orcid=0000-0002-3952-8588]{L. A. Boogaard}
\email{boogaard@strw.leidenuniv.nl}
\affiliation{Leiden Observatory, Leiden University, P.O. Box 9513, 2300 RA Leiden, The Netherlands}

\author[orcid=0000-0002-0167-2453]{W. N. Brandt}
\email{wnbrandt@gmail.com}
\affiliation{Department of Astronomy and Astrophysics, 525 Davey Lab, The Pennsylvania State University, University Park, PA 16802, USA}
\affiliation{Institute for Gravitation and the Cosmos, The Pennsylvania State University, University Park, PA 16802, USA}
\affiliation{Department of Physics, 104 Davey Laboratory, The Pennsylvania State University, University Park, PA 16802, USA}

\author[orcid=0000-0002-3805-0789]{C.-C. Chen}
\email{ccchen@asiaa.sinica.edu.tw}
\affiliation{Academia Sinica Institute of Astronomy and Astrophysics (ASIAA), No. 1, Sec. 4, Roosevelt Road, Taipei 106216, Taiwan}

\author[orcid=0000-0003-2027-8221]{P. Cox}
\email{cox@iap.fr}
\affiliation{Sorbonne Université, UPMC Universit\'e Paris 6 and CNRS, UMR 7095, Institut d’Astrophysique de Paris, 98bis boulevard Arago, 75014 Paris, France}

\author[orcid=0000-0002-7821-8873]{K. K. Knudsen}
\email{kraiberg@chalmers.se}
\affiliation{Department of Space, Earth and Environment, Chalmers University of Technology, SE-412 96 Gothenburg, Sweden}

\author[orcid=0000-0002-5247-6639]{C.-L. Liao}
\email{liao@strw.leidenuniv.nl}
\affiliation{Leiden Observatory, Leiden University, P.O. Box 9513, 2300 RA Leiden, The Netherlands}

\author[orcid=0000-0003-0085-6346]{G. Calistro Rivera}
\email{gcalistrorivera@gmail.com}
\affiliation{German Aerospace Center (DLR), Institute of Communications and Navigation, 82234 Wessling, Germany}

\author[orcid=0000-0002-1383-0746]{M. Rybak}
\email{matus.rybak@gmail.com}
\affiliation{Leiden Observatory, Leiden University, P.O. Box 9513, 2300 RA Leiden, The Netherlands}
\affiliation{Faculty of Electrical Engineering, Mathematics and Computer Science, Delft University of Technology, Mekelweg 4, 2628 CD Delft, The Netherlands}
\affiliation{SRON–Netherlands Institute for Space Research, Niels Bohrweg 4, 2333 CA Leiden, The Netherlands}

\author[orcid=0000-0003-1192-5837]{A. M. Swinbank}
\email{a.m.swinbank@durham.ac.uk}
\affiliation{Centre for Extragalactic Astronomy, Department of Physics, Durham University, South Road, Durham DH1 3LE, UK}

\author[orcid=0000-0001-5434-5942]{P. van der Werf}
\email{pvdwerf@strw.leidenuniv.nl}
\affiliation{Leiden Observatory, Leiden University, P.O. Box 9513, 2300 RA Leiden, The Netherlands}

\author[orcid=0000-0003-4793-7880]{F. Walter}
\email{walter@mpia.de}
\affiliation{Max-Planck-Institut f\"ur Astronomie, K\"onigstuhl 17, 69117 Heidelberg, Germany}

\author[orcid=0000-0003-4678-3939]{A. Weiss}
\email{aweiss@mpifr-bonn.mpg.de}
\affiliation{Max-Planck-Institut f\"ur Radioastronomie, Auf dem H\"ugel 69, 53121 Bonn, Germany}

\author[orcid=0009-0004-3732-6394]{B. A. Westoby}
\email{westoby@strw.leidenuniv.nl}
\affiliation{Leiden Observatory, Leiden University, P.O. Box 9513, 2300 RA Leiden, The Netherlands}





\begin{abstract}
We present JWST/NIRCam and MIRI observations of twelve $z\sim3$ sub-millimeter galaxies (SMGs) from the ALESS survey, combined with high-resolution ($0.08''-0.16''$) ALMA 870\um\ imaging, enabling spatially resolved spectral energy distribution fitting (SED) on $\sim$\,kiloparsec scales.
We find a resolved star-forming main sequence linking the surface densities of star formation rate and stellar mass, suggesting that star formation remains tightly coupled to the local mass distribution even in heavily obscured systems.
%
Our resolved SED analysis reveals a systematic stellar mass bias in integrated fits (median $\sim0.2$\,dex offset), even when including rest-frame $\sim2$\um\ MIRI imaging. Rather than classical `outshining', this effect is primarily driven by spatially varying dust attenuation, indicating a `dust-obscuration bias' that causes highly obscured stellar mass to be missed.
We show that SMG morphologies are strongly wavelength-dependent. At rest-frame optical wavelengths, strong central obscuration produces large stellar-dust spatial offsets and inflated sizes, while at longer wavelengths these effects diminish. The rest-frame $\sim1.5-3\,\um$ MIRI imaging is far less affected by dust than NIRCam and reveals compact stellar structures that closely match the 870\um\ dust continuum.
We find that centrally concentrated dust attenuation drives both the apparent spatial offsets and size variations, demonstrating that dust geometry is the primary driver of the observed structural diversity. Consequently, morphologies inferred from rest-frame wavelengths $\lesssim1.6\,\um$ can be significantly biased without longer-wavelength constraints.
The intrinsic stellar mass and dust continuum sizes are consistent ($R_\mathrm{e,870\mu m}/R_\mathrm{e,\ast}=1.0\pm0.4$), supporting a picture in which SMGs host compact, heavily obscured star formation that builds dense stellar cores, consistent with their evolution into massive quiescent galaxies.
We suggest that such obscured structures and associated biases may also be common among massive star-forming galaxies at $z\gtrsim1$, implying that these effects are likely to be of broad relevance.
%
\end{abstract}



\section{Introduction}
\label{sec:Introduction}

Sub-millimeter galaxies (SMGs; $S_{850\mu{\rm m}}\gtrsim1$\,mJy) are the most star-forming, dusty galaxies in the high-redshift Universe. First uncovered in the late 1990s by blank-field single-dish sub-millimeter surveys (e.g., \citealt{Smail_1997,Barger1998,Hughes1998,Eales1999}), these systems host prodigious star formation (up to a thousand solar masses per year; e.g., \citealt{cunha2015,Dudze2020}) and are thought to play a major role in the early assembly of massive galaxies (e.g., \citealt{Blain_2002,Casey_2014}). Yet, despite their extreme infrared luminosities, SMGs are frequently faint or even undetected at rest-frame UV/optical wavelengths due to heavy dust obscuration, making their stellar populations, morphologies, and evolutionary pathways difficult to constrain.

The advent of the Atacama Large Millimeter/submillimeter Array (ALMA) transformed the study of SMGs. ALMA’s sensitivity and resolution enabled precise localization of SMGs, robust identification of their counterparts, and measurement of their redshift distribution (see, e.g., review in \citealt{Hodge2020}). Although recent ALMA studies provided major advances, many still relied on relatively coarse resolution or integrated measurements, limiting analyses to global properties (e.g., \citealt{hodge2013,swinbank2014,cunha2015,simpson2015,simpson2020,Dudze2020,dacunha2021}). Subsequent higher-resolution ALMA observations revealed the internal structure of SMGs, uncovering compact star-forming regions, clumpy dust morphologies, and spatial offsets between dust and stellar emission (e.g. \citealt{chen2015,hodge2016,Chen2017,CalistroRivera_2018,gullberg2018,hodge2019,gullberg2019}). These studies established the sub-kpc dust and gas properties of SMGs, but their stellar components remained poorly constrained: heavy dust attenuation meant that the Hubble Space Telescope (HST) often failed to detect the most obscured systems, and when detected, only dust-biased rest-frame UV/optical light was available (e.g., \citealt{Swinbank2010}). As a result, their intrinsic stellar mass distributions, morphologies, and sizes remained uncertain.

The James Webb Space Telescope (JWST) has significantly improved this situation. With its unprecedented sensitivity, JWST is detecting the optical/near-IR counterparts of SMGs up to $z\sim6$ (e.g., \citealt{chen2022b,cheng2022, gillman2023,rujopakarn2023,wu2023,Gillman2024,McKinney2025,Ikeda2026,Umehata2026}). The inclusion of JWST bands enables more robust constraints on the stellar mass, especially for the subset of sources with no prior optical/near-IR observations (e.g., \citealt{mckinney2023}). Furthermore, with the angular resolution achieved in the rest-frame optical and near-infrared, long wavelength JWST/NIRCam provides more direct measurements of the stellar light in these systems at scales comparable to ALMA. This capability has already demonstrated that SMGs often appear dramatically different depending on wavelength, highlighting the strong influence of dust attenuation (e.g., \citealt{smail2023,gillman2023,Gillman2024,hodge2025}). With NIRCam alone, morphological studies find clumpy and compact structures of SMGs at $z\sim2$ \citep{gillman2023,Gillman2024,lebail2024}. Based on NIRCam colors, \citet{lebail2024} point toward bulge formation in pre-existing disks, with sub-regions (cores) being quiescent or residing in quiescent regions. However, when combining matched resolution observations from both JWST NIRCam and ALMA 870\um\ (Band 7), \citet{hodge2025} find strong correlation between dust emission and reddening of NIRCam colors on $\sim$kpc scales, suggesting dust obscuration as the primary source of their red rest-frame optical colors (see also \citealt{Gillman2024}). 

Despite this progress, JWST studies of SMGs have so far relied almost entirely on NIRCam (up to F444W). Observations with MIRI remain rare (e.g. \citealt{Gillman2024,Boogaard2024,Gomez2024}), yet are crucial: at the typical redshifts of SMGs ($z\sim2-4$), MIRI probes rest-frame near-IR wavelengths where dust attenuation should be dramatically reduced. MIRI would thus provide the cleanest view of the underlying stellar mass distribution currently achievable.

In this paper, we present new JWST/MIRI observations of the ALESS-JWST sample, a uniquely well-observed set of SMGs with matched-resolution ALMA Band 7 and JWST/NIRCam imaging. By combining these datasets and performing spatially resolved SED fitting with {\sc magphys} \citep{magphys2008,cunha2015}, we recover two-dimensional maps of stellar mass, SFR, dust attenuation, and other key physical quantities. This enables the first systematic, dust-corrected view of SMG structure across rest-frame optical through far-IR wavelengths, and allows us to reassess their apparent morphologies, stellar-dust alignment, size measurements, and evolutionary implications.

This paper is structured as follows. In Section~\ref{sec:data}, we present the ALMA and JWST observations of our sources. In Section~\ref{sec:grids}, we describe the methodology to perform spatially-resolved SED modelling. Our results are presented in Section~\ref{sec:results}, and discussed in Section~\ref{sec:discussion}. In Section~\ref{sec:conclusion} we summarize our key findings. Throughout this work, we assume cosmological parameters $\Omega_M=0.3$, $\Omega_\Lambda=0.7$, and $H_0=70\,\mathrm{\,km\,s}^{-1}\mathrm{Mpc}^{-1}$, and we adopt AB magnitudes \citep{Oke1983} and a \cite{Chabrier2003} initial mass function (IMF).

\section{Observations and Data}
\label{sec:data}

\subsection{Sample selection}

\begin{deluxetable*}{lccccccc}
\tablewidth{0pt}
\tablecaption{The sample of ALESS galaxies used in this work.\label{tab:aless-sample}}
\tablehead{
\colhead{ID} & \colhead{R.A. (J2000)} & \colhead{Dec. (J2000)} & \colhead{$S_{870\mathrm{\mu m}}(\mathrm{mJy})$} & \colhead{$870\um$ FWHM} & \colhead{$z$\textsuperscript{a}} & \colhead{$m_\mathrm{F444W}$ (AB)} & \colhead{$m_\mathrm{F770W}$ (AB)}
}
\startdata
ALESS001.1 & 03:33:14.46 & $-$27:56:14.5 & $6.7\pm0.5$ & 0.16$''$ & 4.674 & $23.9\pm0.03$ & $23.1\pm0.04$ \\
ALESS001.2 & 03:33:14.41 & $-$27:56:11.6 & $3.5\pm0.4$ & 0.16$''$ & 4.669 & $23.0\pm0.01$ & $22.3\pm0.03$ \\
ALESS001.3 & 03:33:14.16 & $-$27:56:12.5 & $1.9\pm0.4$ & 0.16$''$ & $2.86^{+0.46}_{-1.53}$ & $23.5\pm0.02$ & $22.8\pm0.05$ \\
ALESS003.1 & 03:33:21.51 & $-$27:55:20.5 & $8.3\pm0.4$ & 0.08$''$ & 3.375 & $21.8\pm0.01$ & $21.0\pm0.01$ \\
ALESS009.1 & 03:32:11.33 & $-$27:52:12.0 & $8.8\pm0.5$ & 0.08$''$ & 3.694 & $22.4\pm0.01$ & $21.3\pm0.01$ \\
ALESS010.1 & 03:32:19.05 & $-$27:52:14.8 & $5.2\pm0.5$ & 0.16$''$ & $3.34^{+0.02}_{-0.16}$ & $21.6\pm0.01$ & $21.0\pm0.02$ \\
ALESS015.1 & 03:33:33.37 & $-$27:59:29.7 & $9.0\pm0.4$ & 0.08$''$ & $2.86^{+0.10}_{-0.20}$ & $21.0\pm0.01$ & $20.5\pm0.01$ \\
ALESS029.1 & 03:33:36.90 & $-$27:58:09.3 & $5.9\pm0.4$ & 0.16$''$ & $3.69^{+0.42}_{-0.45}$ & $21.4\pm0.01$ & $20.8\pm0.01$ \\
ALESS045.1 & 03:32:25.26 & $-$27:52:30.6 & $6.0\pm0.5$ & 0.16$''$ & 2.678 & $20.9\pm0.01$ & $20.5\pm0.01$ \\
ALESS076.1 & 03:33:32.35 & $-$27:59:55.7 & $6.4\pm0.6$ & 0.08$''$ & 3.389 & $23.2\pm0.02$ & $22.3\pm0.03$ \\
ALESS112.1 & 03:32:48.86 & $-$27:31:13.2 & $7.6\pm0.5$ & 0.08$''$ & 2.314 & $20.3\pm0.01$ & $20.3\pm0.01$ \\
\hline
ALESS3.1-comp & 03:33:21.43 & $-$27:55:25.4 & $1.07\pm0.07$ & 0.08$''$ & 3.374 & $23.2\pm0.02$ & $22.8\pm0.06$ \\
\enddata
\tablecomments{
\textsuperscript{a}Redshifts with no errors are spectroscopic, most taken from \cite{danielson2017,birkin2021} and the Dawn JWST Archive (DJA); for ALESS~3.1-comp the spectroscopic redshift is from Westoby et al., submitted. Others are updated photometric redshifts obtained with \textsc{magphys} in this work.
}
\end{deluxetable*}

Our sample consists of 12 SMGs from the ALESS survey \citep{Karim2013,hodge2013} that were targeted in high resolution, high fidelity ALMA 870\um\ follow-up observations by \cite{hodge2016,hodge2019}. The parent sample consistent of 99 SMGs identified in the Extended Chandra Deep Field South (ECDF-S) using single-dish observations \citep{weiss2009} and later observed with ALMA's most compact configuration. 
The sample selection for the \citet{hodge2016,hodge2019} high resolution follow up is based entirely on the 870\um\ brightness and the availability of HST coverage \citep{chen2015}, hence no selection was made on galaxy properties. Nevertheless, these effectively randomly selected targets span a wide range of redshifts ($z\simeq2.3-4.7$; median $z\simeq3.5$) and star formation rates ($\mathrm{SFR}\simeq200-1000~M_{\odot}\mathrm{yr}^{-1}$).

We note that only one of our sources, ALESS~45.1, may be associated with an X-ray active galactic nucleus (AGN) \citep{wang2013}; however, even this identification is ambiguous (see \citealt{hodge2025}). Therefore, we do not explicitly account for AGN in this analysis.

The properties of our targets are summarized in Table~\ref{tab:aless-sample}.

\subsection{JWST Observations and Data reduction}\label{sec:data-jwst}

Our targets were observed as part of the JWST Cycle 1 GO Program (PID:2516, PIs: J. A. Hodge \& E. da Cunha) with NIRCam and MIRI in four wide filters (F200W, F356W, F444W, and F770W)\footnote{We exclude one of the targets, ALESS~17.1, from the present analysis, because of source blending and ambiguous redshifts (Westoby et al., subm.).}.

The NIRCam observations were obtained in four pointings and imaged using a four-point dither pattern, and reduced using the JWST calibration pipeline (version=1.11.3; pmap=1119; \citealt{jwst_pipeline_1.11.3}), following the recipe and modifications from the CEERS survey (\citealt{ceers_survey}; more details about the NIRCam data calibration can be found in \citealt{hodge2025}).

The MIRI observations were obtained in six pointings. For each set of observations, the individual dithered images were processed with build 10.1 (version=1.13.4) of the JWST calibration pipeline \citep{jwst_pipeline_1.13.4}. They were run with mostly default parameters through pipelines \texttt{calwebb\_detector1} and \texttt{calwebb\_image2}. The only custom parameter used was setting the rejection threshold for the jump step to 5.0. After \texttt{calwebb\_image2} was run, the individual \texttt{*cal.fits} images were stacked and a median sky image was created (per pixel), and then subtracted from each cal image to create a set of median sky (background) subtracted images. This subtraction removes the background as well as any remaining detector effects. These background-subtracted images were then combined in the next stage of the pipeline. In \texttt{calwebb\_image3}, the stage in the pipeline that resamples and combines the different dithers, 
the resampled pixel scale was set to an absolute pixel scale of 0.06$''$. The final output image is a resampled version of the combined dithered, background-subtracted images. For each filter, there is one output image, with sky levels approximately zero. 

We present the resulting NIRCam and MIRI images of our sources in Appendix~\ref{appendix_images}.

\subsection{ALMA Data}\label{sec:data-alma}

In this work, we use the ALMA continuum observations taken as part of programs 2012.1.00307.S and 2016.1.00048.S (PI: J. A. Hodge), which were previously presented in \cite{hodge2016,hodge2019}, respectively. In both programs, the targets were observed in Band 7 at 344 GHz (870\um, i.e., rest-frame $\sim$200\um\ at $z\sim3.5$) over the 8 GHz bandwidth. The naturally weighted continuum data achieved synthesized beam sizes of 0.17$\times$0.15$''$ \citep{hodge2016}, with a follow-up of half the sample by \cite{hodge2019} yielding beam sizes of 0.10$''$$\times$0.07$''$. These angular resolutions are comparable to NIRCam from 2.0--4.4\um\ (0.07$''$--0.14$''$), but sharper than MIRI resolution at 7.7\um\ (0.27$''$). The ALMA images of our sources are presented as flux contours and overlaid with the JWST images in Appendix~\ref{appendix_images}.

\subsection{Astrometric Alignment}
Accurate astrometric alignment is required for multi-wavelength morphological analysis and spatially resolved SED studies.
For the 870\,$\mu$m ALMA data, the nominal astrometric accuracy is expected to be a few to $\sim10$~mas for $0.08''-0.16''$ resolution images\footnote{ALMA Cycle 11 Technical Handbook, Chapter 10.5.2}, significantly smaller than the 60 mas pixel scale of the NIRCam and MIRI images used here.

For the JWST images, alignment is challenging due to the scarcity of point sources in our fields. For NIRCam, we used \texttt{tweakreg} to perform both relative and absolute alignment, adopting Gaia DR3 \citep{Gaia2023} as the absolute reference. To optimize the results for individual pointings, we removed Gaia sources lacking proper-motion information\footnote{See \url{https://github.com/spacetelescope/jwst/issues/8168} for a discussion of this issue} and excluded poorly centroided sources from the pipeline catalogs. In some cases, PSF photometry was used to improve centroid estimates, particularly for partially saturated sources.
We achieved an absolute accuracy of $0.02–0.065''$ ($\lesssim1$ pixel) relative to Gaia DR3 and a relative accuracy of $\lesssim0.015''$ among NIRCam filters (Table 3 of \citealt{hodge2025}).

This method could not be applied to MIRI, as too few Gaia sources are present in the MIRI images. Instead, we aligned the MIRI data to NIRCam by matching bright, compact sources in the F444W and F770W images. We reran the final calibration step (Section~\ref{sec:data-jwst}), using these sources as the astrometric reference in the \texttt{tweakreg} step. This anchors the MIRI astrometry to NIRCam rather than directly to Gaia, but the larger number of reference sources yields a more robust solution. We verified that this approach improves the astrometric alignment.

\subsection{Multi-wavelength photometry and redshifts}\label{sec:pz}

The redshifts for our targets (Table~\ref{tab:aless-sample}) are drawn from a mix of spectroscopic measurements and updated photometric estimates based on our JWST data. Eight of the galaxies have spectroscopic redshifts from optical/UV spectroscopy \citep{danielson2017} and/or CO spectroscopy (\citealt{birkin2021}; Westoby et al., subm.). For the remaining four sources, we update the photometric redshifts \citep{cunha2015} using the new JWST photometry and the \magphys\ photometric-redshift extension \citep{magphys2019}. As noted by \cite{hodge2025}, two of the targets, ALESS~10.1 and ALESS~29.1, have spectroscopic redshifts, but these are likely misidentifications: their reported spectroscopic values are inconsistent with the JWST/NIRCam and MIRI SED shapes, and we therefore adopt the revised photometric redshifts (listed in Table~\ref{tab:aless-sample}).

To construct the SEDs for the photometric-redshift fitting, we obtained integrated fluxes from the $B$ band to 1.4 GHz from the catalogs of \citet{simpson2014}, \citet{swinbank2014}, and \citet{dacunha2021}, and replaced the ground-based $K$-band and Spitzer/IRAC photometry with our new NIRCam and MIRI measurements (though we note that the IRAC photometry is broadly consistent with the NIRCam and MIRI photometry). In the rest-frame UV/optical, we also include deeper limits from the VIDEO survey \citep{Jarvis2013} from the ESO archive.

The only exception is ALESS~10.1, whose near-IR and optical photometry is likely contaminated by a nearby low-redshift interloper that is indistinguishable in low-resolution ground-based UV/optical imaging. For this source, we therefore use archival HST imaging in four CANDELS filters in GOODS–South \citep{Koekemoer2011}. We mask the contaminating north clump in all HST and NIRCam bands and run \magphys-photoz on the updated, contamination-free SED, combining the HST and JWST fluxes with the FIR–radio photometry (24$\mu$m to 1.4\,GHz). We note that ALESS~3.1-comp was not in the original ALESS catalog and therefore has only JWST and ALMA 870\um\ data.


\section{Spatially Resolved SED modeling}\label{sec:grids}

\subsection{PSF modeling and Homogenization}
We followed \citet{chen2022b} to create PSF models for all NIRCam filters. For each of the four pointings, we visually selected the best observed stars and modeled them with Gaussian profiles using {\sc imfit} \citep{erwin2015}, adopting simulated {\sc STPSF} (formerly {\sc WebbPSF}) models \citep{perrin2014} matched to the observation date, time, and detector position. For pointings 1, 3, and 4, we verified consistency between different reference stars. Pointing 2 lacked a suitable isolated star, so we adopted the average Gaussian profile derived from pointings 1, 3, and 4 to convolve the simulated PSF.

For MIRI, we directly used simulated {\sc STPSF} PSFs (v1.4.0), which include improved treatment of position-dependent detector effects (e.g., the cruciform artifact). PSFs were generated for each galaxy using the observation date, time, position angle, and detector location.

The ALMA primary beam was approximated as a 2D elliptical Gaussian, as all data were obtained in a single configuration.

Finally, following \citet{li2024}, we constructed matching kernels between each filter PSF and the target MIRI/F770W PSF using \texttt{create\_matching\_kernel} from {\sc photutils} \citep{photutils2022}, applying a low-pass window to avoid spectral leakage. All images were convolved to the MIRI/F770W resolution using \texttt{convolve} in {\sc Astropy} \citep{astropy2022}. The final angular resolution is $\sim0.27''$ (FWHM), which corresponds to $\sim2$~kpc at the median redshift of our sample.

\subsection{MAGPHYS modeling}

We model both the integrated and spatially-resolved multi-wavelength SEDs of our sources using the \magphys\ code (described in detail in \citealt{magphys2008,cunha2015}). Briefly, \magphys\ is a physically motivated, energy‐balance SED fitting code that links the attenuated stellar emission of galaxies to the re-emitted dust emission across the UV–to–FIR/radio. \magphys\ combines analytical star formation histories with stochastic bursts, a two-component dust attenuation model, and empirical dust emission templates to derive posterior probability distributions for key galaxy parameters such as stellar age, stellar mass ($M_\ast$), SFR, specific SFR (sSFR), dust attenuation (\av), and dust luminosity (\ldust). We use the high-redshift extension described in \citet{cunha2015,magphys2020}, which includes priors that are appropriate for intensely star-forming, dusty systems\footnote{Only ALESS~45.1 is identified as an X-ray source \citep{wang2013}, and its AGN nature is uncertain \citep{hodge2025}; we therefore do not include an AGN component in the SED modeling.}.

After aligning and PSF-matching all JWST and ALMA images to a common astrometric and resolution framework (see \citealt{hodge2025}), we perform spatially resolved SED modeling by extracting photometry in small, regularly spaced apertures across each galaxy and fitting each aperture independently with \magphys. Following \citet{li2024}, we apply a uniform square tessellation (hereafter ``aperture grid") within a visually defined circular region that encompasses the full extent of the galaxy in all filters. Each aperture has a size of 0.06$''$, corresponding to a single pixel in the JWST images; the ALMA images are resampled to the same pixel grid. At the redshifts of our targets, these apertures probe physical scales of $\lesssim500$\,pc.
 
For analyses requiring spatial independence, we also construct a coarser grid with 0.24$''$ apertures, comparable to the FWHM of the MIRI/F770W PSF. These $\sim2$\,kpc apertures reduce the covariance between adjacent bins as there is less flux bleeding beyond this scale; this binning is used where explicitly noted.

Because the galaxies suffer strong, spatially variable dust attenuation across a wide wavelength range (Fig.\,\ref{fig:montage}), no single filter can serve as a universal detection image. Shorter-wavelength data may miss deeply obscured regions that are bright in the mid-IR or sub-mm, and vice versa. We therefore model a given aperture only if it is detected at $>3\sigma$ in at least two filters, following \citet{smail2023} and \citet{li2024}. This is essential for sources such as ALESS~29.1, where some regions are visible only at F770W and 870\um\ due to extreme central obscuration. We visually confirm that this selection recovers all physically meaningful sightlines.

Most of our SMGs also have unresolved detections in at least one of  the following far-IR bands, including Spitzer/MIPS at 24\um, Herschel/SPIRE at 250, 350, and 500\um\ \citep{swinbank2014}, and ALMA Band~4 at 2\,mm \citep{dacunha2021}. While unresolved, these observations provide crucial constraints on the total dust luminosity, which helps break the age-dust degeneracy and yields more reliable star formation rates (SFRs). Hence, we incorporate them as ``augmented filters" in the resolved fitting by assuming the same spatial morphology as the ALMA 870\um\ emission. In practice, we scale the 870\um\ image by the ratio of integrated flux in each band to the integrated flux at 870\um, and assign the resulting map a (low) effective S/N of 1.5 per aperture. For non-detections, we also scaled the unresolved upper limits based on the same ratio as above, and incorporated them as resolved upper limits at $3\sigma$ level. This ensures consistency with the total far-IR flux, while still allowing for some spatial variations in dust temperature.

We present the integrated SED modelling results in Section~\ref{sec:iSED}, and the spatially resolved results in Section~\ref{sec:rSED}.

\section{Analysis \& Results}\label{sec:results}

\subsection{Integrated Properties}\label{sec:iSED}

\begin{figure}[t]
    \centering
    \includegraphics[width=\linewidth]{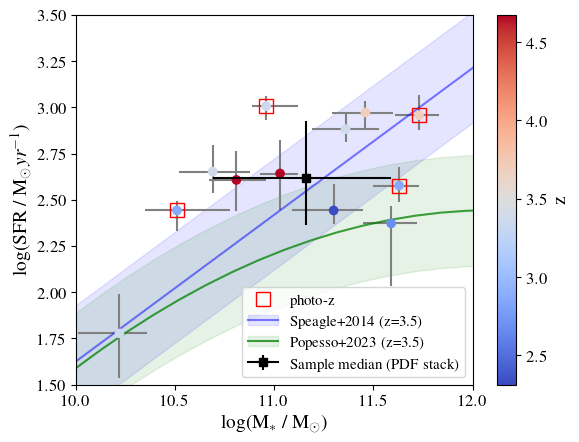}
    \caption{The stellar masses and SFRs of our targets, derived from \magphys\ SED modeling, color-coded by redshift (red squares indicate targets with photometric redshifts).
    The blue and green lines and filled regions show the star-forming main sequence at $z=3.5$ and its $\pm0.3$ dex region, from \citealt{Speagle2014}  and \citealt{Popesso2023}  respectively.}\label{fig:SFMS}
\end{figure}

\begin{deluxetable*}{lcccccc}
\tablecaption{Physical properties of ALESS galaxies from integrated SED fitting.}
\tablehead{
\colhead{ID} & \colhead{$\log(M_\ast/\Msun)$} & \colhead{$\log(\mathrm{SFR}/\Msun\,\mathrm{yr}^{-1})$} & \colhead{\av} & \colhead{$\log(L_\mathrm{dust}/\rm L_{\odot})$\textsuperscript{d}} & 
\colhead{$\log(M_\mathrm{dust}/\Msun)$} & \colhead{$\log(\mathrm{sSFR}/\mathrm{yr}^{-1})$}
}
\startdata
ALESS001.1 & 10.81$^{+0.15}_{-0.14}$ & 2.61$^{+0.15}_{-0.17}$ & 3.46$^{+0.35}_{-0.30}$ & 12.73$^{+0.16}_{-0.18}$ & 9.10$^{+0.11}_{-0.11}$ & $-8.18^{+0.20}_{-0.25}$ \\
ALESS001.2 & 11.03$^{+0.09}_{-0.10}$ & 2.64$^{+0.18}_{-0.19}$ & 2.74$^{+0.27}_{-0.35}$ & 12.75$^{+0.16}_{-0.20}$ & 8.64$^{+0.10}_{-0.08}$ & $-8.38^{+0.20}_{-0.25}$ \\
ALESS001.3 & 10.51$^{+0.27}_{-0.16}$ & 2.44$^{+0.05}_{-0.11}$ & 4.06$^{+0.25}_{-0.22}$ & 12.59$^{+0.07}_{-0.08}$ & 8.11$^{+0.04}_{-0.04}$ & $-8.07^{+0.15}_{-0.20}$ \\
ALESS003.1 & 11.36$^{+0.17}_{-0.17}$ & 2.88$^{+0.08}_{-0.07}$ & 3.76$^{+0.18}_{-0.17}$ & 12.95$^{+0.08}_{-0.05}$ & 9.20$^{+0.06}_{-0.05}$ & $-8.47^{+0.20}_{-0.20}$ \\
ALESS009.1 & 11.46$^{+0.14}_{-0.17}$ & 2.97$^{+0.06}_{-0.08}$ & 4.39$^{+0.20}_{-0.22}$ & 13.04$^{+0.06}_{-0.07}$ & 9.15$^{+0.07}_{-0.06}$ & $-8.47^{+0.20}_{-0.20}$ \\
ALESS010.1\textsuperscript{a} & 10.96$^{+0.16}_{-0.07}$ & 3.01$^{+0.05}_{-0.08}$ & 3.64$^{+0.15}_{-0.32}$ & 13.18$^{+0.03}_{-0.13}$ & 8.82$^{+0.10}_{-0.11}$ & $-7.92^{+0.00}_{-0.15}$ \\
ALESS015.1 & 11.63$^{+0.10}_{-0.13}$ & 2.58$^{+0.10}_{-0.09}$ & 3.39$^{+0.25}_{-0.20}$ & 12.71$^{+0.07}_{-0.03}$ & 9.25$^{+0.04}_{-0.04}$ & $-9.03^{+0.20}_{-0.20}$ \\
ALESS029.1 & 11.73$^{+0.10}_{-0.12}$ & 2.96$^{+0.11}_{-0.08}$ & 3.59$^{+0.10}_{-0.20}$ & 13.10$^{+0.09}_{-0.04}$ & 8.82$^{+0.05}_{-0.05}$ & $-8.78^{+0.20}_{-0.15}$ \\
ALESS045.1 & 11.59$^{+0.13}_{-0.14}$ & 2.37$^{+0.10}_{-0.34}$ & 3.14$^{+0.30}_{-0.17}$ & 12.54$^{+0.05}_{-0.05}$ & 9.08$^{+0.03}_{-0.09}$ & $-9.22^{+0.25}_{-0.40}$ \\
ALESS076.1 & 10.69$^{+0.19}_{-0.17}$ & 2.65$^{+0.15}_{-0.12}$ & 4.09$^{+0.30}_{-0.27}$ & 12.82$^{+0.15}_{-0.11}$ & 9.00$^{+0.11}_{-0.08}$ & $-7.97^{+0.05}_{-0.25}$ \\
ALESS112.1 & 11.30$^{+0.15}_{-0.21}$ & 2.44$^{+0.15}_{-0.08}$ & 2.01$^{+0.25}_{-0.25}$ & 12.52$^{+0.15}_{-0.05}$ & 9.34$^{+0.06}_{-0.08}$ & $-8.82^{+0.25}_{-0.25}$ \\
ALESS003.1-comp\textsuperscript{b} & 10.22$^{+0.14}_{-0.21}$ & 1.78$^{+0.22}_{-0.24}$ & 1.44$^{+0.43}_{-0.38}$ & 11.81$^{+0.25}_{-0.25}$ & 8.18$^{+0.37}_{-0.51}$ & $-8.41^{+0.35}_{-0.35}$ \\
\hline 
Whole sample\textsuperscript{c} & 11.08$^{+0.49}_{-0.56}$ & 2.58$^{+0.34}_{-0.3}$ & 3.40$^{+0.57}_{-1.25}$ & 12.70$^{+0.32}_{-0.25}$ & 8.96$^{+0.23}_{-0.49}$ & -8.51$^{+0.46}_{-0.52}$\\
\enddata
\tablecomments{The quotes values are the median, and the errors are the 16th- and 84th-percentiles of the likelihood distribution of each parameter. \\
\textsuperscript{a} Using HST filters F814W, F105W, F125W and F160W, together with NIRCam, MIRI and all FIR filters. The blue galaxy to the north of ALESS~10.1 that is not visible in MIRI F770W and ALMA Band 7 are masked from all HST and NIRCam filters. \\
\textsuperscript{b} Only has NIRCam, MIRI and ALMA Band 7 observations. \\
\textsuperscript{c} Median and 16th-84th percentile range obtained from stacking the likelihood distribution of each parameter for the whole sample. \\
\textsuperscript{d} The dust luminosity is integrated between 3 and 1000\um.}
\label{tab:properties}
\end{deluxetable*}

We first revisit the integrated physical properties of the galaxies in our sample (previously presented in \citealt{cunha2015}) using the new JWST photometry combined with existing multi-wavelength data. In the near-infrared, we replace the previously used ground-based $K$-band and Spitzer/IRAC measurements with JWST/NIRCam and MIRI fluxes, which provide substantially improved spatial resolution and sensitivity. At longer wavelengths, we adopt the mid- and far-infrared photometry from Spitzer/MIPS and Herschel/PACS and SPIRE, together with the ALMA 870\um\ and 2\,mm continuum measurements and the VLA 1.4\,GHz data compiled in previous studies of the ALESS sample \citep{hodge2013,miller2013,swinbank2014,dacunha2021}. The only exception is ALESS~3.1-comp, which was not in the original ALESS catalog and therefore has only JWST and ALMA 870\um\ data. We adopt the updated spectroscopic and photometric redshifts (Table~\ref{tab:aless-sample}). Updated values of stellar mass, SFR, \av, and \ldust\ are listed in Table~\ref{tab:properties}. The derived stellar masses span $\log(M_\ast/\Msun)\simeq10.7-11.3$, while SFRs range from $\simeq200$ to $\sim1000\,\Msun,\mathrm{yr}^{-1}$, consistent with the extreme star-forming nature of sub-millimeter galaxies at these redshifts.

As shown in Fig.\,\ref{fig:SFMS}, our sources lie on or above the star-forming main sequence at $z\sim3.5$, depending on the reference relation adopted. They are broadly consistent with the main sequence in the \citet{Speagle2014} parameterization, but lie systematically above the more recent \citet{Popesso2023} relation. This reflects current uncertainty in the high-mass end of the high-redshift main sequence. If we adopt the most recent calibration, we conclude that 
most of our sources lie significantly above the main sequence, confirming that they are undergoing intense star-formation activity, and suggesting that these systems represent short-lived phases of rapid growth in massive high-redshift galaxies.

Overall, the integrated properties derived with the updated JWST photometry remain broadly consistent with those previously reported for the ALESS sample (\citealt{cunha2015}; see also, \citealt{danielson2017}). The improved wavelength coverage and photometric accuracy provided by JWST lead to modest refinements in the stellar mass and SFR estimates, but the galaxies remain among the most massive and intensely star-forming systems at $z\sim2-4$. This confirms that the subset of ALESS SMGs studied here remains representative of the parent population.

The close agreement between our updated integrated properties and pre-JWST results partly reflects the consistency between MIRI and Spitzer/IRAC photometry. To test this, we compared F770W fluxes in our six MIRI fields with IRAC 8\um\ measurements from the MUSYC catalog in the ECDF-S \citep{Cardamone2010}. We find good overall correspondence, including for our SMGs, with increased scatter only at $m_{\rm F770W}\gtrsim24$, consistent with recent comparisons of MIRI and IRAC photometry in other surveys \citep[e.g.,][]{Boogaard2024}.

While the integrated properties place these galaxies above the star-forming main sequence, it remains unclear whether this reflects global or centrally concentrated star formation. We address this question using spatially resolved SED modeling.

\subsection{Resolved SEDs}\label{sec:rSED}
\subsubsection{Physical parameter maps}\label{sec:pmap}

\begin{figure*}
    \centering
    \includegraphics[width=0.85\linewidth]{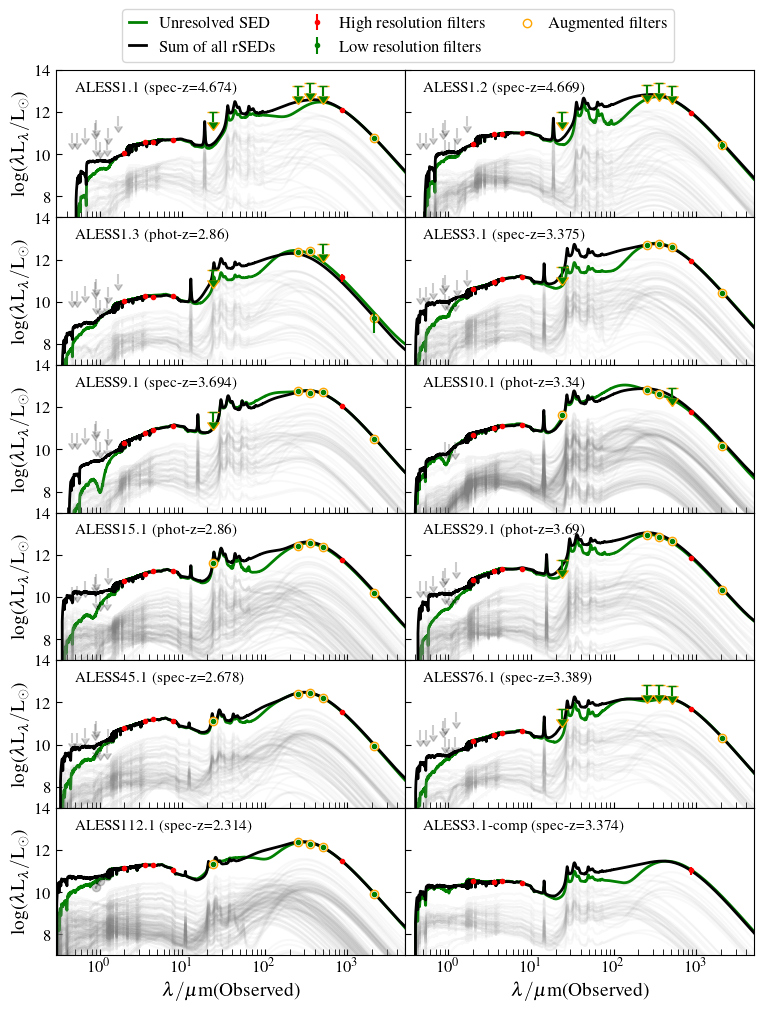}
    \caption{For each galaxy, we show the best-fit \magphys\ SEDs for all spatial bins (grey), their sum (black), and the best-fit integrated SED (green). Red and green points denote integrated JWST/ALMA and unresolved Spitzer, Herschel, and ALMA photometry, respectively. Grey arrows indicate $3\sigma$ upper limits from ground-based optical/near-IR data \citep{simpson2014}, supplemented by deeper limits from the VIDEO survey \citep{Jarvis2013} (ESO archive).}
    \label{fig:seds_all}
\end{figure*}

\begin{figure*}
    \centering
    
    \includegraphics[width=0.85\linewidth,trim=0 3mm 0 0,clip]{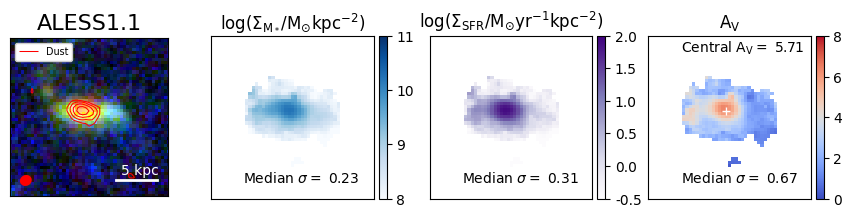}
    \includegraphics[width=0.85\linewidth,trim=0 3mm 0 0,clip]{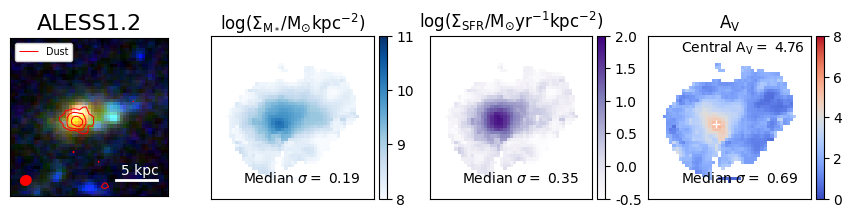}
    \includegraphics[width=0.85\linewidth,trim=0 3mm 0 0,clip]{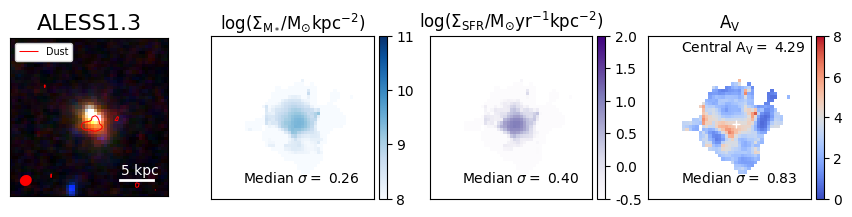}
    \includegraphics[width=0.85\linewidth,trim=0 3mm 0 0,clip]{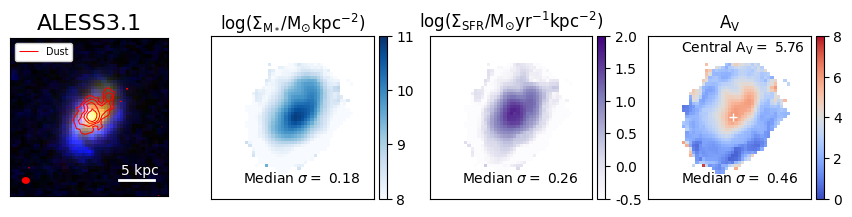}
    \includegraphics[width=0.85\linewidth,trim=0 3mm 0 0,clip]{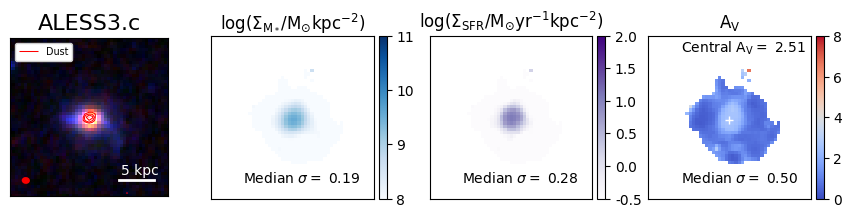}
    \includegraphics[width=0.85\linewidth,trim=0 3mm 0 0,clip]{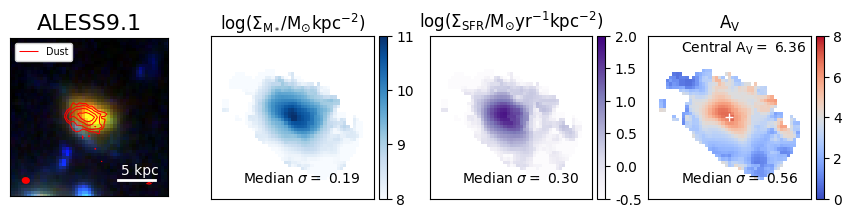}
    \caption{For each source, the left-hand panel shows an RGB composite using JWST F770W, F444W, and F200W, respectively, with superimposed ALMA 870\um\ contours ($2-24\sigma$). The middle and right-hand panels show maps of stellar mass, SFR, and dust attenuation (\av) from our spatially-resolved SED fitting. The 850\um\ flux peak location is marked with white cross in the right panels, where the median \av\ within one MIRI PSF-sized aperture is measured (i.e. the `central \av'). Median parameter uncertainties are indicated in each panel.}
    \label{fig:maps_all}
\end{figure*}

\addtocounter{figure}{-1}
\begin{figure*}
    \centering
    
    \includegraphics[width=0.85\linewidth,trim=0 3mm 0 0,clip]{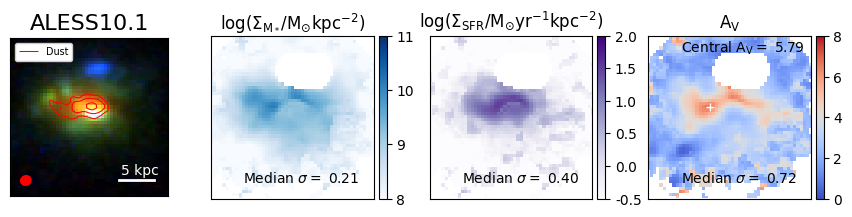}
    \includegraphics[width=0.85\linewidth,trim=0 3mm 0 0,clip]{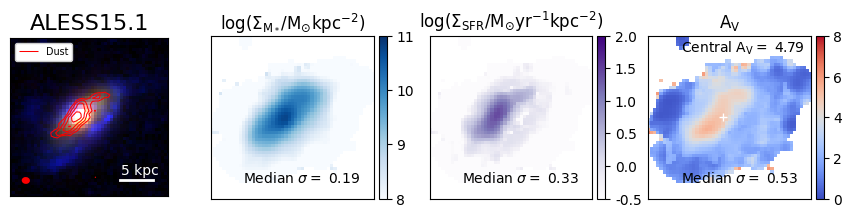}
    \includegraphics[width=0.85\linewidth,trim=0 3mm 0 0,clip]{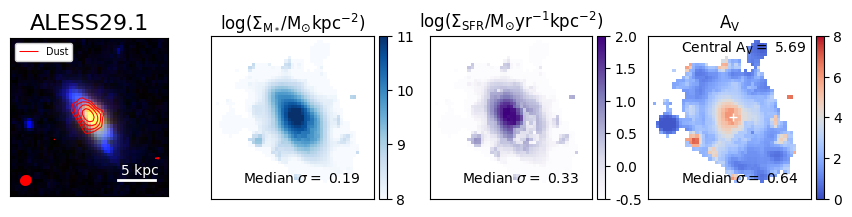}
    \includegraphics[width=0.85\linewidth,trim=0 3mm 0 0,clip]{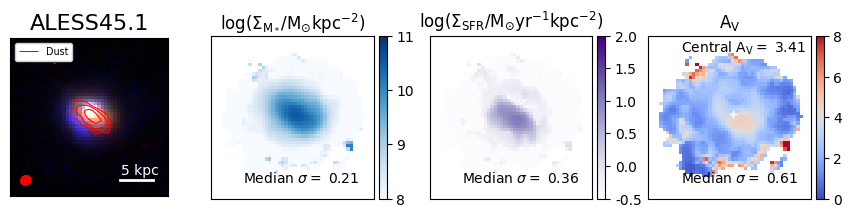}
    \includegraphics[width=0.85\linewidth,trim=0 3mm 0 0,clip]{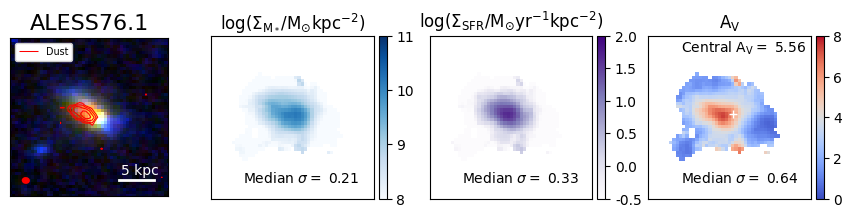}
    \includegraphics[width=0.85\linewidth,trim=0 3mm 0 0,clip]{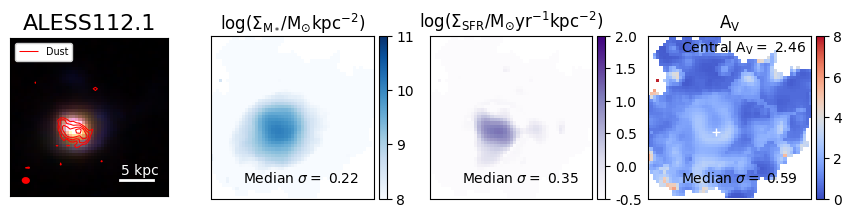}
    \caption{Continued.}
    \label{fig:maps_all}
\end{figure*}

To investigate the internal physical properties of the galaxies, we perform spatially resolved SED fitting using the JWST and ALMA imaging described in Section~\ref{sec:grids}. All images are convolved to a common resolution and resampled onto the same pixel grid before fitting. We then construct resolved SEDs for each spatial element using the available multi-wavelength photometry and model them with the same SED fitting framework used for the integrated analysis.

Figure~\ref{fig:seds_all} shows the resulting resolved SEDs across different regions of the galaxies. The photometry spans rest-frame optical to far-infrared wavelengths, allowing us to constrain both the stellar populations and the dust emission in each spatial element. The models reproduce the observed photometry well across most of the wavelength range, indicating that the adopted SED modeling approach provides a satisfactory description of the spatially resolved emission.

As a consistency check, we sum the best-fitting resolved SEDs across all spatial elements and compare the result with the integrated SED derived from the global photometry. In general, the summed resolved SEDs agree with the integrated SEDs over the wavelength range where resolved observations are available.
The main discrepancy occurs in the rest-frame ultraviolet and, where resolved observations are not available. In this regime, the summed resolved SEDs tend to slightly overpredict the integrated fluxes. This likely reflects the fact that the UV emission arises from a small number of relatively unobscured sightlines that dominate the integrated flux. The summed resolved SEDs also tend to overpredict the rest-frame $\sim5-50\um$ emission compared to the integrated SEDs, although no resolved constraints are available at those wavelengths. This could be due to the MAGPHYS dust optical depth and/or temperature priors.

Overall, the reasonable agreement between the summed resolved SEDs and the integrated SEDs provides justification that the spatially resolved modeling reliably captures the global energy balance of the galaxies. This enables us to use the resolved SED fits to derive maps of key physical quantities, including stellar mass, star formation rate, dust luminosity, and dust attenuation, which we analyze in the following sections.

Fig.\,\ref{fig:maps_all} shows examples of these parameter maps for stellar mass, star-formation rate, and average $V$-band dust attenuation, alongside the JWST+ALMA images for each source. Overall, the 7.7\um\ emission tends to peak where the 870\um\ flux peaks. The stellar mass and SFR (as well as the dust luminosity, not shown) also peak near the 870\um\ continuum peak and show similar extended features, demonstrating that both MIRI and ALMA trace the central, dust-enshrouded mass concentration. 


Dust attenuation peaks sharply at the ALMA continuum maximum (where optical/near-IR colors are red), reaching high central \av\ values for most sources. We quantify this by measuring the median \av\ within one MIRI PSF-sized aperture centered on the ALMA peak (`central \av'), indicated in the right-hand panels of Fig.\,\ref{fig:maps_all}. These strong central dust attenuations play a key role in the wavelength-dependent morphologies discussed in Section~\ref{sec:dust-morph}.

We note that even in the most dust-obscured centers, the effective attenuations inferred with \magphys\ are much lower than those inferred from the dust column densities along the line of sight assuming a foreground-screen extinction (e.g., \citealt{Simpson2017}), which can imply \av\ values of hundreds to thousands of magnitudes\footnote{For typical values of our sample, a dust mass $M_\mathrm{dust}=10^{8.5}\,M_\odot$ distributed within a radius $R=1$\,kpc, assuming a gas-to-dust ratio of $\delta_\mathrm{gdr}=90$, implies a gas column density $N_\mathrm{H}\simeq2.6\times10^{24}\,\mathrm{cm}^{-2}$, corresponding to a line-of-sight $\av\sim400$ under a simple, uniform-screen geometry (see \citealt{Simpson2017}).}. This apparent discrepancy arises because these two quantities probe different aspects of the dust distribution. Estimates based on the dust column density trace the total mass of dust along the line of sight, as inferred from the far-infrared emission. In contrast, the attenuation derived by \magphys\ represents an effective, luminosity-weighted attenuation experienced by the stellar populations that contribute to the observed rest-frame optical and near-infrared SED, and is therefore inherently biased toward the less obscured sightlines from which stellar light can escape. As a result, even in systems with extremely large dust columns, the observed stellar continuum is expected to be dominated by regions of lower effective optical depth and/or by less embedded stellar populations. A purely uniform, optically thick screen would obscure most of the stellar emission and lead to a severe underestimate of the stellar mass. The fact that substantial stellar light is still detected therefore points to a dust-star configuration that departs from a simple foreground screen, likely reflecting a strongly inhomogeneous distribution of dust relative to the stars (see also figure 9 and discussion in \citealt{hodge2025}).

\subsubsection{Resolved star-forming main sequence}\label{sec:rSFMS}
\begin{figure*}[ht!]
    \centering
    \includegraphics[width=\linewidth]{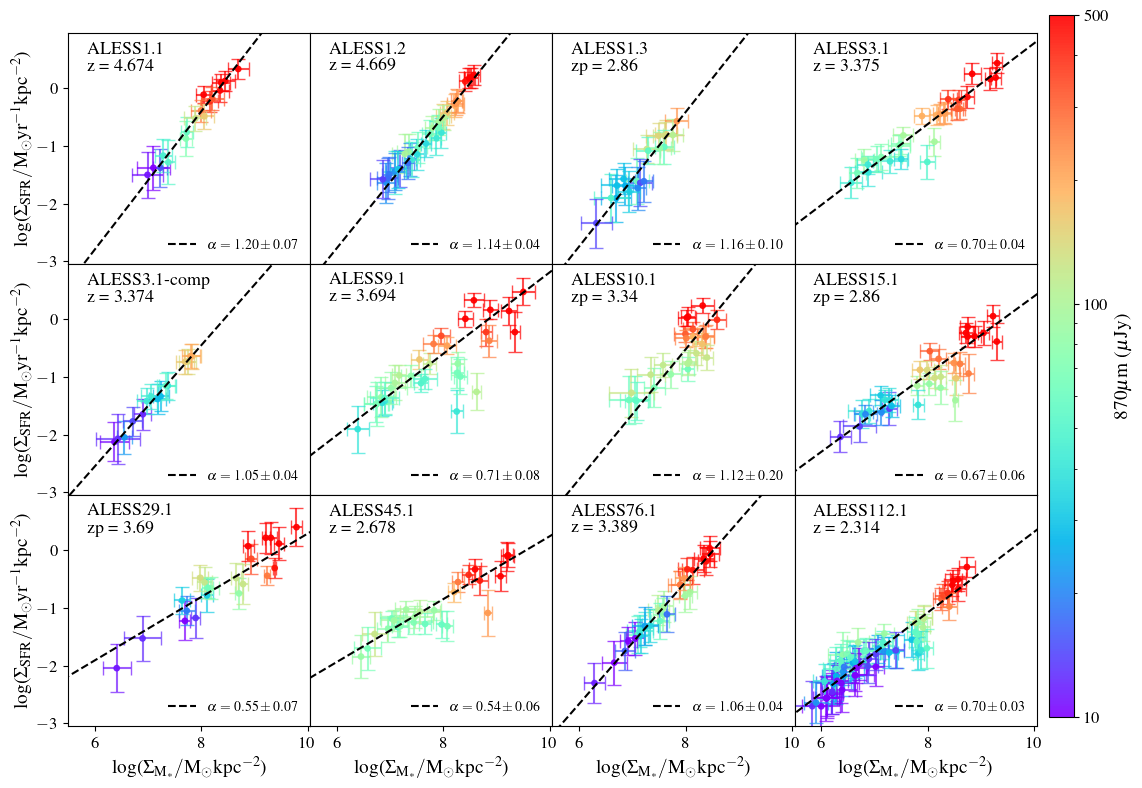}
    \caption{Resolved star-forming main sequence (rSFMS) for each target. Colored points show independent 0.24$''$ ($\sim2$\,kpc) bins, color-coded by their 870\um\ flux. The black dashed line shows the best-fit power-law relation for each galaxy, with the slope $\alpha$ given in the legend.
    }
    \label{fig:rSFMS-aless}
\end{figure*}

The resolved star-forming main sequence (rSFMS), i.e., the correlation between stellar mass surface density (\sigmstar) and SFR surface density (\sigsfr), has been established in nearby galaxies on kiloparsec scales and, increasingly, at high redshift (e.g., \citealt{Cano-Diaz2016,Abdurrouf2017,pessa2021,Sanchez2021,Mun2024,Koller2024}). Previous work has claimed that the rSFMS persists at least to $z\sim2$ \citep{Wuyts2013}, and recent JWST+ALMA studies have extended this relation to $z>4$ \citep{li2024}, with slopes comparable to those measured locally. The physical origin of the rSFMS remains debated: it may reflect spatially uniform specific SFRs across galaxy disks, or arise from the combination of the resolved Kennicutt–Schmidt relation and the correlation between stellar mass and molecular gas surface densities (e.g., \citealt{Baker2022}).

In heavily obscured systems such as SMGs, however, dust attenuation can strongly affect the observed spatial distribution of both stellar light and star formation. The resolved stellar mass maps derived from our SED modeling can provide a more reliable view of the intrinsic stellar structure of these galaxies.

We investigate the rSFMS for the ALESS SMGs by comparing the stellar mass surface density and star-formation rate surface density on resolved spatial scales.
Figure \ref{fig:rSFMS-aless} shows \sigsfr\ as a function of \sigmstar\ for all spatial bins in each our of ALESS galaxies (we show the 0.24$''$, i.e., $\sim2$\,kpc aperture scales, comparable to the MIRI/F770W PSF)to minimize covariance between neighboring bins). 
In most galaxies we find a clear positive correlation between these quantities.

For each galaxy we fit a power law, $\sigsfr\propto\sigmstar^\alpha$. The best-fit slopes span $\alpha\simeq0.7-1.2$, broadly consistent with previous measurements of the rSFMS in both nearby galaxies and high-redshift systems \citep{pessa2021,li2024}.

In most sources, the highest \sigmstar\ regions correspond to the heavily obscured central regions traced by the ALMA 870\um\ emission. These regions also exhibit the highest \sigsfr\ values, indicating that the intense star formation in SMGs is concentrated within compact, massive central regions.

These results indicate that the rSFMS may extend into the heavily obscured, extreme star-formation regime of SMGs. However, the derived slopes and scatter depend on the available far-infrared constraints, which influence the inferred dust temperatures and therefore the SFR estimates. In particular, galaxies with brighter Herschel detections tend to exhibit slightly shallower relations, likely reflecting improved constraints on the dust emission (hence, the shallower slopes are more reliable). This highlights the importance of resolving dust temperature when characterizing the rSFMS at high redshift. We discuss the implications and limitations of this analysis further in Section~\ref{sec:discussion}.



\subsubsection{Outshining or Dust-Obscuration Bias?}\label{sec:outshining}

\begin{figure*}
    \centering
    \includegraphics[width=0.49\linewidth]{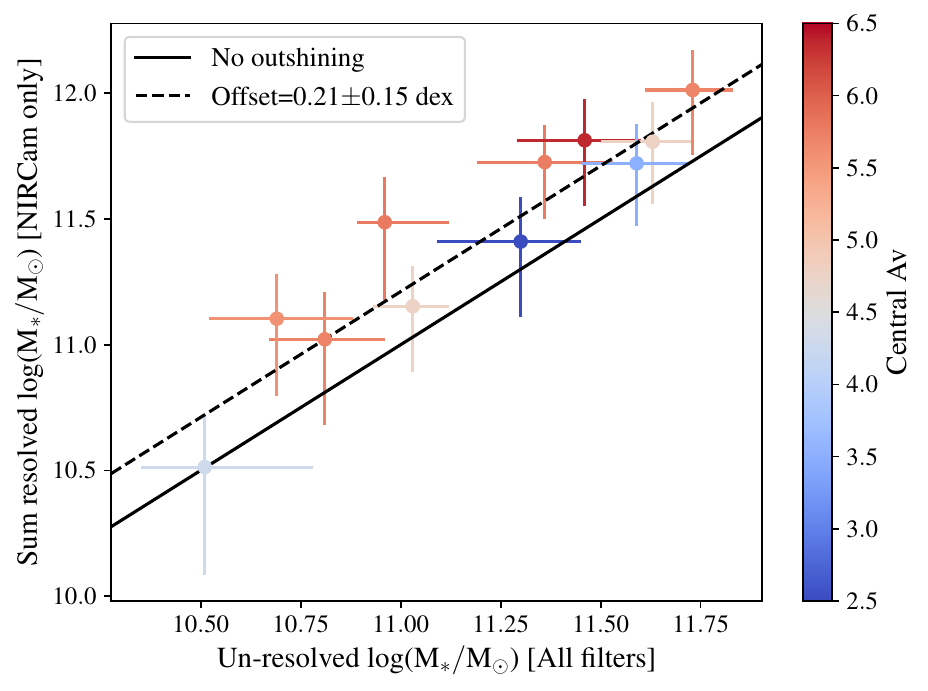}
    \includegraphics[width=0.49\linewidth]{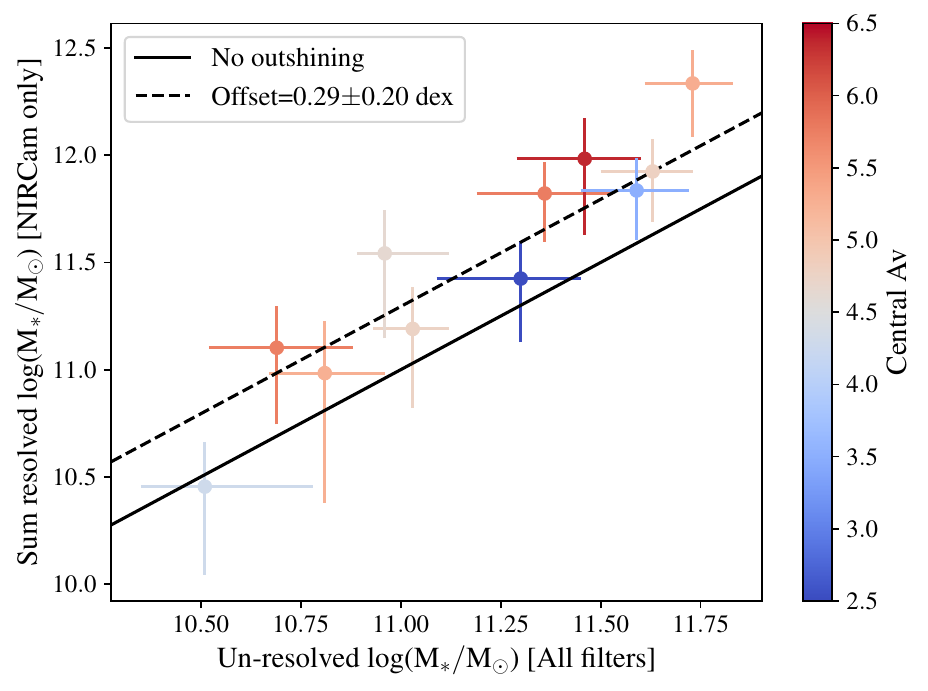}
    \caption{Comparisons between the total stellar masses of our galaxies obtained from the integrated SED fitting (x-axes) with resolved SED fitting (y-axes). {\em Left:} All filters included in the resolved SED fit. {\em Right:} Only NIRCam filters used in the resolved SED fits. In both cases, the sum of stellar masses from the resolved SED fits exceed the stellar mass obtained from fitting the integrated photometry.}
    \label{fig:outshining}
\end{figure*}

Previous spatially-resolved studies have shown that integrated SED fitting can underestimate stellar masses, an effect commonly attributed to outshining by young, luminous stellar populations (e.g., \citealt{Sorba2015,Sorba2018}; see also \citealt{gimenez-arteaga2024}). To test whether this effect is also present in SMGs, we compare stellar masses obtained by summing all spatially-resolved SED bins with those derived from fitting the total integrated photometry.

In the left-hand panel of Fig.\,\ref{fig:outshining}, we compare the stellar masses obtained from integrated fits to the full SEDs of our sources (Section~\ref{sec:iSED}) with the sum of the stellar masses of all resolution elements derived from our spatially-resolved SED fitting. We find that the resolved stellar masses exceed the integrated values for all galaxies, with a median offset of 0.21$\pm0.15$\,dex, corresponding to resolved stellar masses that are $\sim1.6^{+0.7}_{-0.5}$ times higher than those inferred from the integrated fits. This implies that integrated SED fitting recovers only $\sim60\%$ of the stellar mass. This behavior is consistent with the patchy dust geometry discussed in Section~\ref{sec:pmap}, where the effective attenuations inferred from the SEDs are much lower than those implied by the dust column densities. In such a geometry, a substantial fraction of the stellar mass may lie behind heavily obscured regions that contribute little to the observed optical/near-infrared SED.

If classical outshining by young stellar populations were the dominant mechanism, the offset would be expected to correlate with the specific star formation rate (sSFR; \citealt{Sorba2018}). However, we do not find a strong dependence on sSFR. Instead, the offset correlates more strongly with the total dust luminosity of our sources (Pearson $R=0.84$, $p=0.0013$) and with the central V-band dust attenuation ($R=0.7$, $p=0.016$). Galaxies with central $\av>5$ show a median offset of $\sim0.36\pm0.10$\,dex, whereas those with $\av<5$ are broadly consistent with no significant offset. This suggests that high and spatially-variable dust attenuation plays a major role in driving the mass deficit seen in integrated fits. Using coarser, independent 2-kpc apertures yields similar results, confirming that the effect is not driven by covariance among smaller spatial bins.

To assess the impact of wavelength coverage, we repeat the analysis using only NIRCam bands (Fig.\,\ref{fig:outshining}, right-hand panel), a regime typical of many JWST-based studies lacking mid-IR or far-IR constraints (e.g., \citealt{gimenez-arteaga2024,Lines2025,harvey2025}). In this case, the median offset increases to 0.29$\pm0.2$\,dex. The correlations with central \av\ and dust luminosity remain similar.

\begin{figure}
    \centering
    \vspace{0.5cm}
    \includegraphics[width=\linewidth]{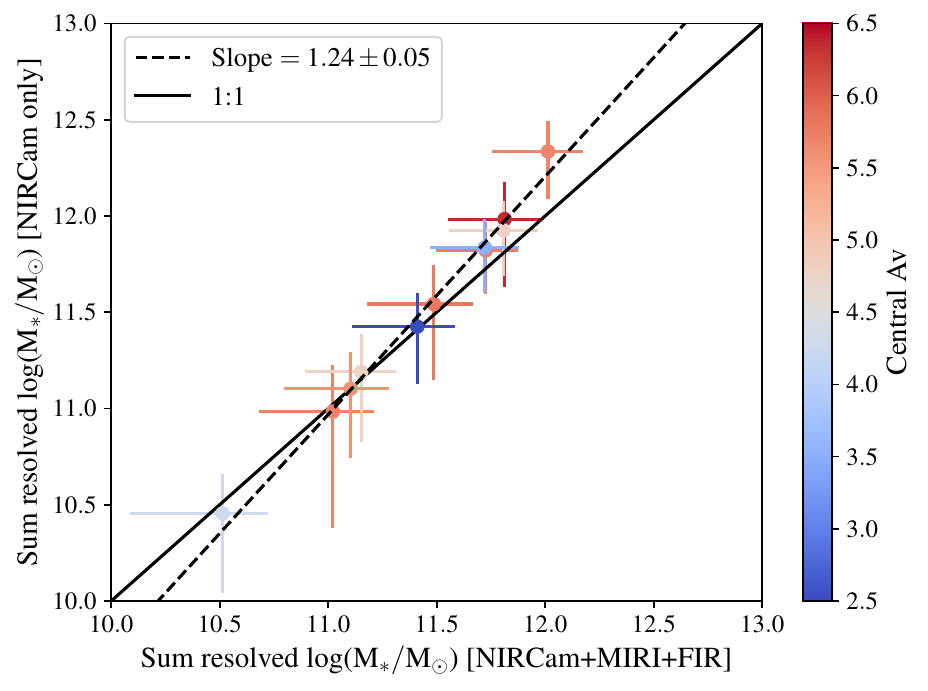}
    \caption{Comparison between the total stellar masses obtained by summing the spatially resolved SED fits using all available filters (NIRCam+MIRI+ALMA; x-axis) and using only NIRCam bands (y-axis). The solid line shows the one-to-one relation. The dashed line indicates the best-fitting linear relation to the data, with a slope of $1.24 \pm 0.05$, indicating a superlinear relation between the two estimates.}
    \label{fig:nircam_all}
\end{figure}

We also compare the summed resolved stellar masses obtained when fitting NIRCam-only SEDs with those derived when the full NIRCam+MIRI+ALMA dataset is used (Fig.\,\ref{fig:nircam_all}). While we find no strong systematic offset between the two, the relation is superlinear. This likely reflects biases in the NIRCam-only fits: these tend to underestimate dust attenuation in the lower-\av\ cases and overestimate it in the higher-\av\ cases (see also \citealt{li2024}).

Overall, our results show that integrated SED fitting systematically underestimates the stellar masses of SMGs, with typical offsets of $\sim40-50\%$, and that the magnitude of this bias depends on the available wavelength coverage. This bias systematically shifts their inferred location on the star-forming main sequence (Fig.\,\ref{fig:SFMS}). Unlike previous studies that primarily associate this effect with high sSFR, we find that the stellar mass deficit correlates more strongly with dust luminosity and central dust attenuation. This suggests that, in SMGs, the apparent outshining bias is largely driven by strong and spatially variable dust obscuration, which can hide a substantial fraction of the stellar mass from integrated SED fits. In this sense, the effect observed in our sample differs from the classical interpretation of outshining and is more accurately described as a `dust-obscuration bias'. These results highlight the importance of spatially resolved analyses and the need to include mid- and far-infrared constraints when estimating stellar masses in dusty high-redshift galaxies (e.g., \citealt{smail2023}).

\subsection{Wavelength-dependent Morphology}\label{sec:morphology}

We now examine the morphologies of the ALESS SMGs across the JWST and ALMA observations and investigate how they depend on wavelength and dust attenuation. We first quantify the spatial offsets between the stellar and dust emission, then compare the observed and intrinsic sizes of the galaxies, and finally examine their structural properties using non-parametric morphology indicators and dust-corrected stellar emission maps.

\subsubsection{Spatial offsets}\label{sec:offset}

\begin{figure*}
\centering
    \includegraphics[width=0.85\linewidth]{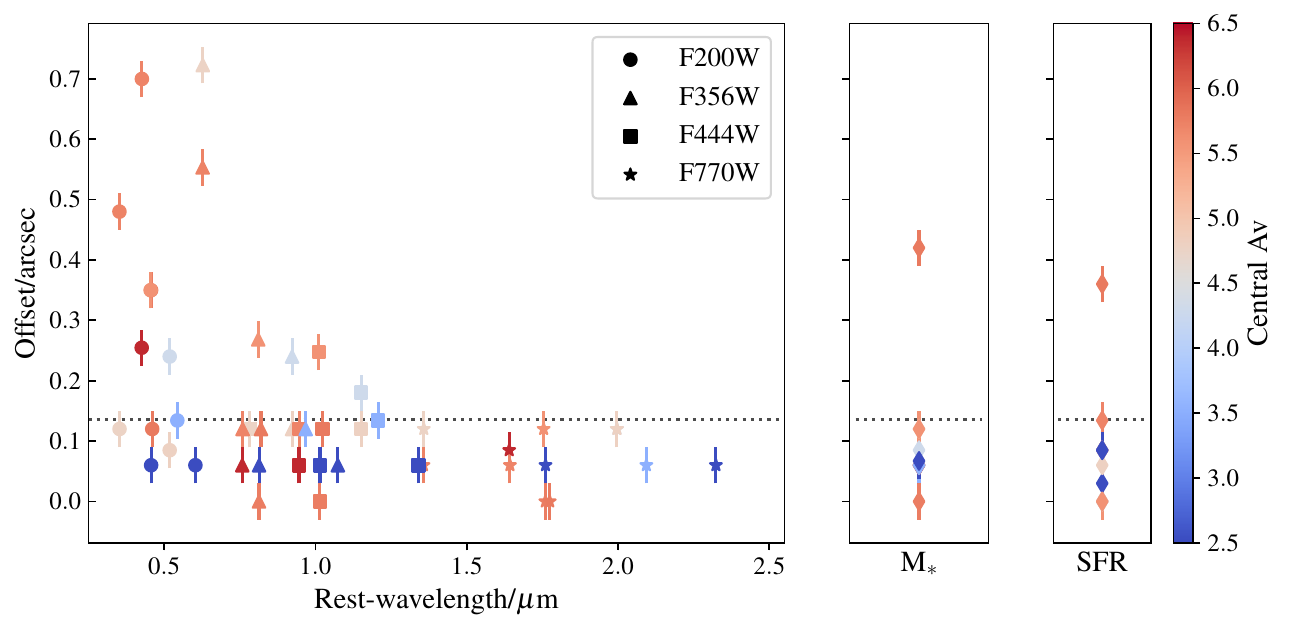}
    \caption{Spatial offsets relative to the ALMA 870\um\ peak. We show the peak positions in the four JWST bands (left-hand panel), as a function of rest-frame wavelength, and in the stellar mass (center) and SFR map (right-hand panel), measured relative to the 870\um\ peak. The dotted lines mark the half-width at half-maximum of the F770W PSF. In all panels, points are colored by central \av\ inferred from integrated SED fitting. Spatial offsets decrease toward longer rest-frame wavelengths and are largest in the most obscured galaxies, while the peaks of the stellar mass and SFR maps are generally consistent with the ALMA position.
    }
    \label{fig:offset}
\end{figure*}

Spatial offsets between rest-frame UV/optical emission and the submillimeter continuum have been widely reported in high-redshift galaxies (e.g., \citealt{hodge2012,chen2015,hodge2016,Buat2019,CalistroRivera_2018,Cheng_2020,Cochrane_2021,inami2022,Killi2024}). These offsets have been interpreted in terms of dust geometry, stellar feedback, and large-scale gas flows. While such mechanisms may operate in some systems, spatially structured dust obscuration can produce similar signatures, is more plausible, and should be considered before invoking more complex explanations. We therefore examine the wavelength dependence of the flux peak positions in our sample from rest-frame optical ($\sim0.5\um$) to near-IR ($\sim3\um$), where dust attenuation is expected to decline.

In Fig.\,\ref{fig:offset}, we show the offsets of the JWST NIRCam and MIRI emission peaks relative to the ALMA 870\um\ continuum peak as a function of rest-frame wavelength, as well as in our \sigmstar\ and \sigsfr\ maps. Large offsets are common in the shorter-wavelength NIRCam bands (F200W, F356W), particularly in the most dust-obscured galaxies, and decrease toward longer wavelengths. F770W is effectively coincident with the ALMA peak within the F770W PSF (FWHM $\sim0.27''$). Offsets exceeding the F770W PSF radius occur only at rest wavelengths $\lesssim1\,\um$ and primarily in systems with central $\av>4$. This behavior is consistent with heavy, centrally concentrated dust obscuration shifting the apparent location of the stellar emission at shorter wavelengths. Similar trends have been predicted by radiative transfer simulations (e.g., \citealt{cochrane2019,popping2022}), and the combination of MIRI and ALMA at comparable resolution now provides strong observational support for this interpretation.

The peak locations in the \sigmstar\ and \sigsfr\ maps are generally consistent with the ALMA peak, apart from only one exception (ALESS~10.1), where the morphology is complicated by a foreground companion.

\subsubsection{Observed sizes}\label{sec:cog}

\begin{figure*}
    \centering
    \includegraphics[width=0.5\linewidth]{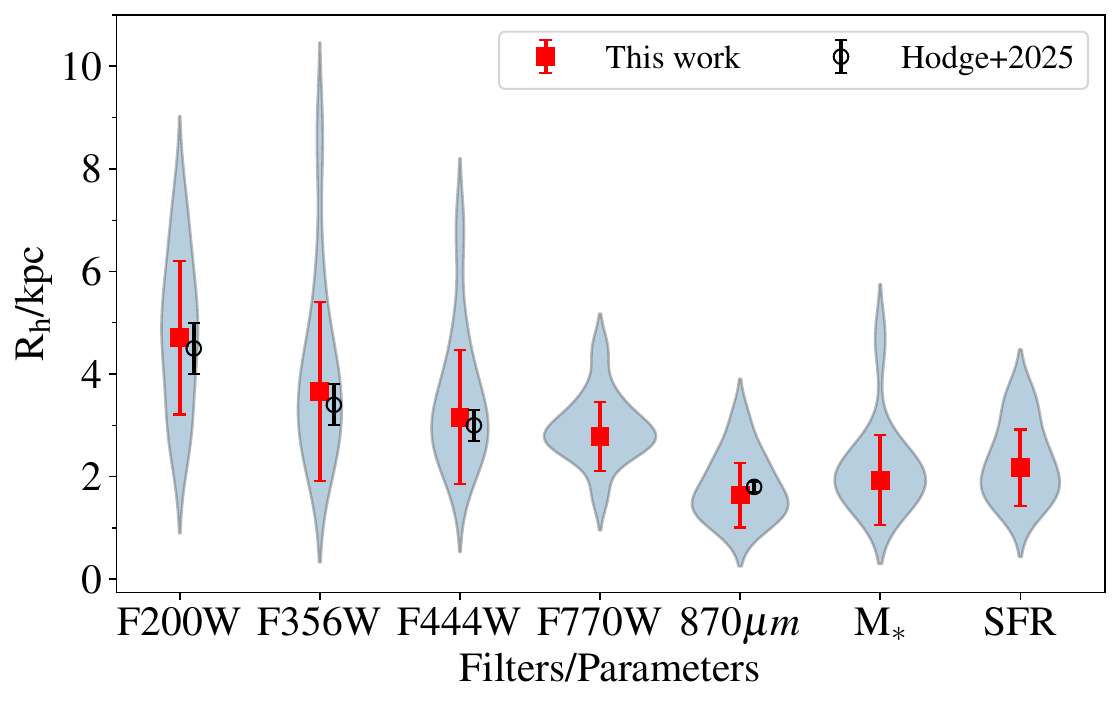}
    \includegraphics[width=0.48\linewidth]{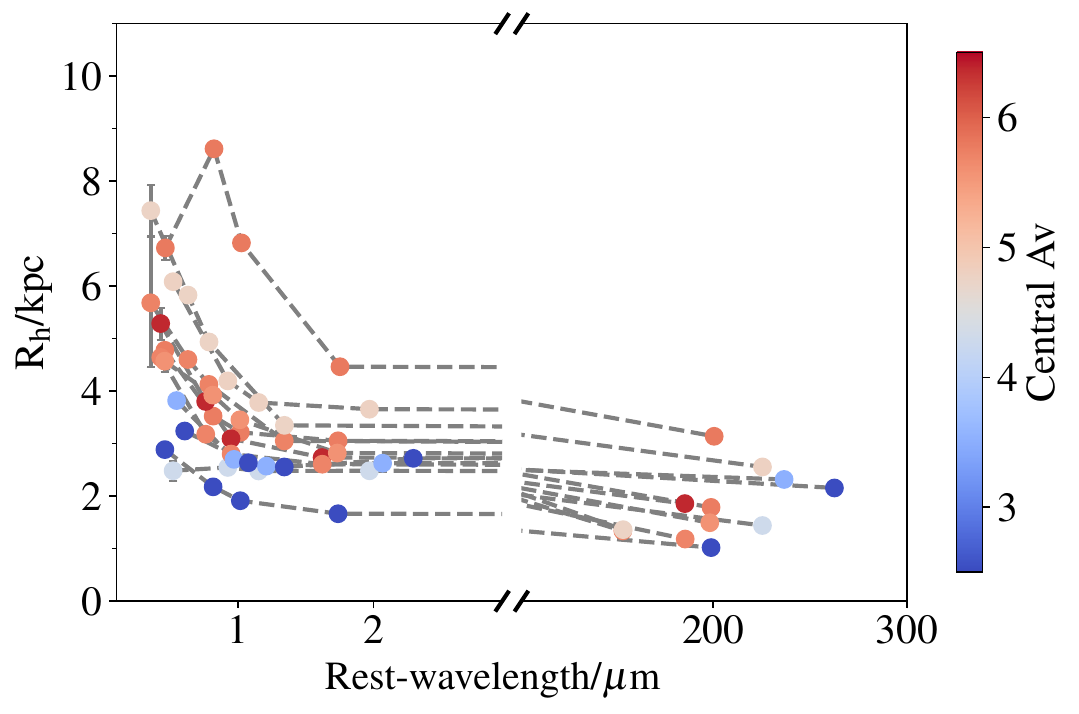}
    \caption{Half-light radii as a function of wavelength. {\em Left:} distributions of the half-light radii for all ALESS SMGs measured in each JWST band, the ALMA 870\um\ map, and the stellar mass and SFR maps. Black points show independent measurements from  \citet{hodge2025}. {\em Right:} half-light radius versus rest-frame wavelength for the same galaxies. Points connected by dashed lines correspond to individual galaxies and are colored by central \av.
    }\label{fig:cog}
\end{figure*}

To quantify the wavelength dependence of galaxy structure, we measured half-light radii from curves of growth in all JWST bands, the ALMA 870\um\ map, and the stellar mass and SFR maps. Source positions and elliptical apertures were defined on the F770W image using SEP \citep{Barbary2016}, the Python implementation of \texttt{Source Extractor} \citep{SExtractor1996}, as this band is minimally affected by dust and traces the stellar structure well. Fluxes were then extracted in concentric elliptical apertures using \textsc{photutils} \citep{photutils2022}. Total fluxes were measured within five times the F770W effective radius, and the same apertures were applied to all other images to ensure consistent measurements.

The left-hand panel of Figure \ref{fig:cog} shows the distribution of half-light radii ($R_\mathrm{h}$) in each band. We find a clear decrease in galaxy size with increasing wavelength. Our measurements are consistent with those of \citet{hodge2025} for the JWST/NIRCam and ALMA images. The median F444W radius is nearly twice that at 870\um, while the median F200W radius is $\sim60\%$ larger than at F770W. The scatter also increases toward shorter wavelengths, consistent with larger dust-driven peak offsets at bluer rest-frame wavelengths (Section \ref{sec:offset}). ALESS~10.1 is an outlier in all bands, likely due to deblending issues; excluding it reduces the median radii by $\lesssim0.1$\,kpc and the scatter by $0.1-0.4$\,kpc, depending on the filter.
The stellar mass and SFR maps yield similar radii, $1.9\pm0.9$\,kpc and $2.2\pm0.7$\,kpc, respectively, intermediate between the F770W and 870\um\ measurements.

The right-hand panel of Figure \ref{fig:cog} shows the half-light radius as a function of rest-frame wavelength. Galaxies with higher central \av\ exhibit the strongest wavelength dependence: for those with $\av>5$, the F200W radii are $50-100\%$ larger than the F770W radii ($R_\mathrm{h,F200W}/R_\mathrm{h,F770W}=1.70\pm0.15$). This is driven by severe central attenuation in the rest-frame optical, which suppresses the core and artificially broadens the light profile. At rest-frame $\sim2\,\um$, dust effects diminish, and the radii converge to more compact and less dispersed values ($R_\mathrm{h}\lesssim3$\,kpc). The only exceptions are ALESS~10.1 (large radii at all wavelengths) and the faint ALESS~3.1 companion (very compact).

Overall, the strong size gradient with wavelength, particularly in the highest-\av\ systems, indicates that dust attenuation plays a dominant role in shaping the apparent morphology of SMGs at rest-frame optical and near-infrared wavelengths, extending to longer wavelengths than is typically assumed.

\subsubsection{Stellar sizes}

\begin{figure*}
    \centering
    \includegraphics[width=0.9\linewidth]{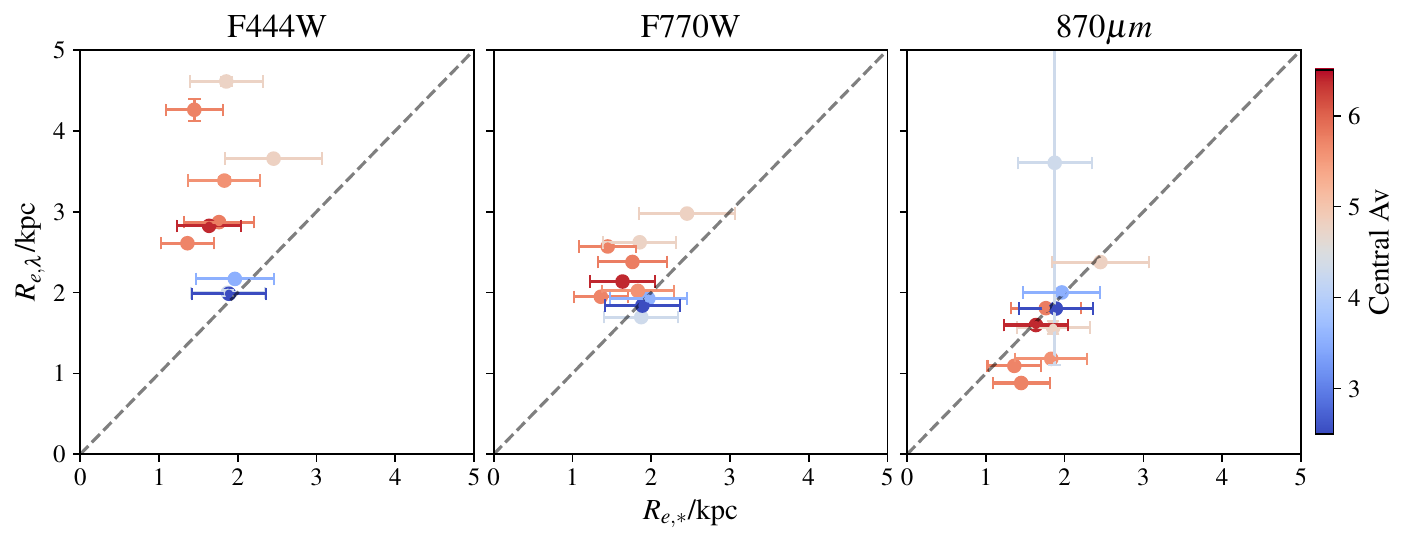}
    \caption{Comparison between the intrinsic stellar effective radii, $R_\mathrm{e,\ast}$, with those measured from F444W (left), F770W (middle), and 870\um\ (right) images. The x-axis in all panels shows the stellar-mass effective radii from S\'ersic fits to the \sigmstar\ maps using \texttt{IMFIT} at 240\,mas pixel scale (including a $25\%$ error floor). The y-axes show the effective radii from S\'ersic modelling of each observed image with \texttt{statmorph} at 60\,mas pixel scale. The dashed line indicates the 1:1 relation. Colored points (blue–red) are our measurements, color-coded by central \av\ (as in Figures~\ref{fig:offset} and \ref{fig:cog}).} \label{fig:sersic-sizes}
\end{figure*}

The observed half-light radii show a strong dependence on wavelength (Section~\ref{sec:cog}), suggesting these measurements may be biased by dust attenuation. We therefore estimate the intrinsic stellar sizes using S\'ersic modeling of the stellar mass maps.

We performed S\'ersic modelling with \texttt{statmorph} \citep{statmorph2019}, focusing on the longer-wavelength images (F444W, F770W, and ALMA 870\um) as well as the \sigmstar\ and \sigsfr\ maps.
For the imaging data, S\'ersic fits were carried out at the native pixel scales (60\,mas for JWST and 20\,mas for ALMA) using the appropriate PSFs and noise maps. Our measurements are consistent with those reported by \citet{hodge2025} at overlapping wavelengths, despite differences in pixel scale and fitting code.
For the stellar mass maps, we adopted the F770W PSF, since all images were convolved to this resolution before SED fitting. We generated 1000 Monte-Carlo realizations by perturbing each map with its uncertainty image and fitting each realization with \texttt{statmorph}. In each iteration, we used the square root of the stellar mass map as the weighting map (equivalent to setting gain=1). The recovered structural parameters are generally well behaved, although the S\'ersic indices can become artificially large when the intrinsic central profiles are sharper than the PSF.

As an independent check, we also fit S\'ersic models using \texttt{IMFIT} on a coarser $0.24''$ grid (matching the FWHM of the F770W PSF). In these fits, we restricted the S\'ersic index to the range $0.5–4$ to avoid unphysical solutions at the lower resolution.

These three size estimators -- S\'ersic radii from \texttt{statmorph} at 60\,mas ($R_\mathrm{e,\texttt{statmorph}}$), from \texttt{IMFIT} at 240\,mas ($R_\mathrm{e,\texttt{IMFIT}}$), and curve-of-growth radii from Section~\ref{sec:cog} ($R_\mathrm{e,\texttt{CoG}}$) -- correlate well with each other but show small systematic offsets. The median ratios are

\begin{itemize}
\item $R_\mathrm{e,\texttt{IMFIT}} / R_\mathrm{e,\texttt{statmorph}} \simeq 1.2 \pm 0.30$\,,
\item $R_\mathrm{e,\texttt{IMFIT}} / R_\mathrm{e,CoG} \simeq 1.0 \pm 0.30$\,.
\end{itemize}

The slightly larger CoG radii reflect the absence of PSF correction. Given that $R_\mathrm{e,\texttt{IMFIT}}$ lies between the two estimates and is PSF-corrected, we adopt it as our fiducial intrinsic stellar size $R_\mathrm{e,\ast}$, with a conservative $25\%$ systematic uncertainty. We exclude ALESS~3.1-companion and ALESS~10.1 due to unreliable fits.

Figure~\ref{fig:sersic-sizes} compares $R_\mathrm{e,\ast}$ with S\'ersic radii measured in F444W, F770W, and at 870\um. Consistent with radiative transfer simulations \citep{popping2022,Baes2024}, the observed rest-frame optical and near-infrared sizes are systematically larger than the intrinsic stellar sizes:

\begin{itemize}
\item $R_\mathrm{e,\ast} / R_\mathrm{e,F444W} \simeq 0.6 \pm 0.2$\,, 
\item $R_\mathrm{e,\ast} / R_\mathrm{e,F770W} \simeq 0.8 \pm 0.2$\,.
\end{itemize}

This difference reflects strong central dust attenuation, which inflates the observed sizes at shorter wavelengths. By contrast, the ALMA 870\um\ sizes closely track $R_\mathrm{e,\ast}$, indicating that the FIR continuum traces the intrinsic stellar mass distribution well. We verified that this is not an artifact of including 870\um\ flux in the SED fitting: stellar mass maps derived using JWST-only photometry yield half-mass radii consistent to within $\lesssim0.1$\,kpc.

\subsubsection{Non-parametric morphologies}\label{sec:statmorph}

\begin{figure*}
    \centering
    \includegraphics[width=\linewidth]{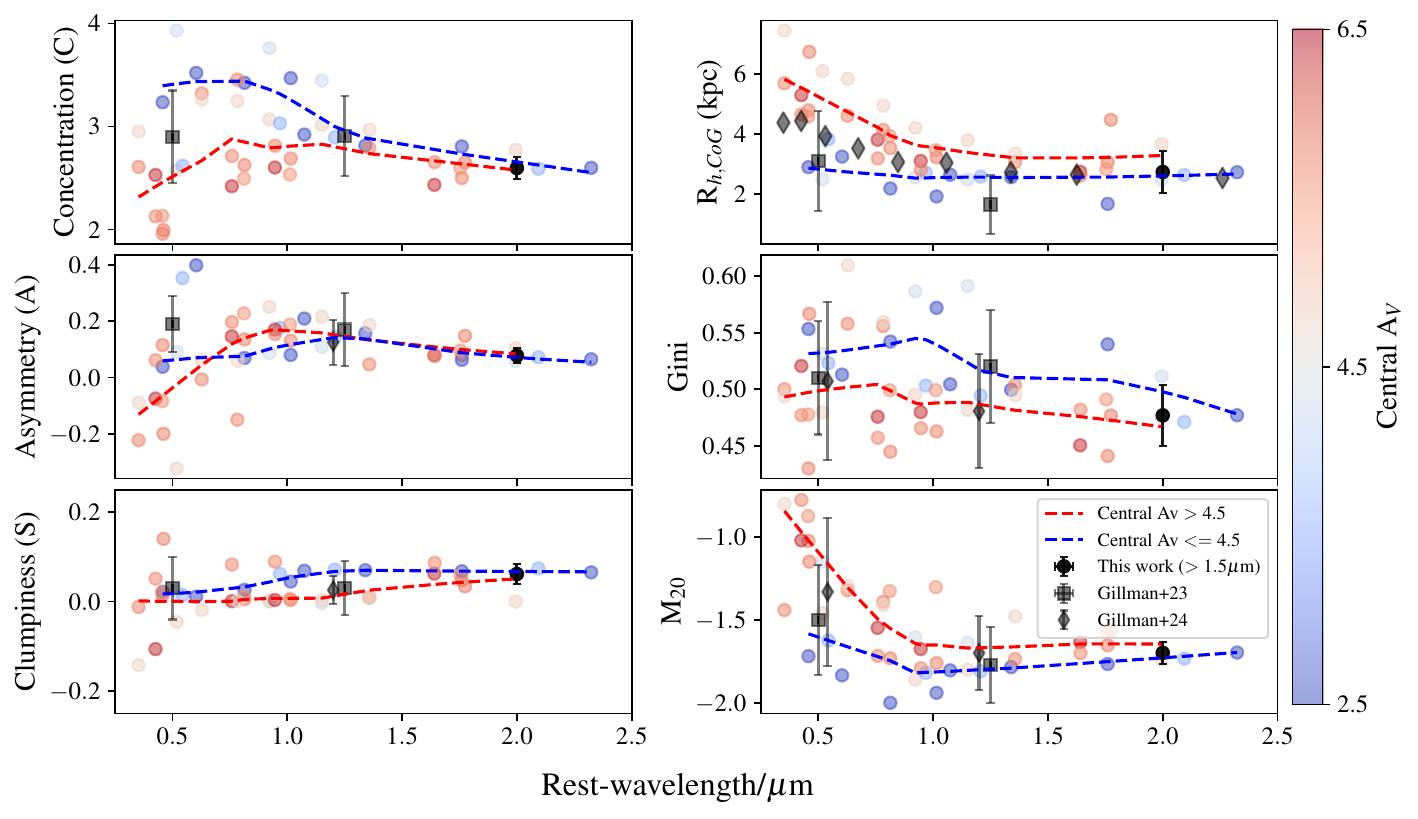}
    \caption{Distribution of half-light radius, CAS parameters, and Gini–\mtw\ values as a function of rest-frame wavelength (see Section~\ref{sec:statmorph}). Points are colored by central \av\ (red: $\av>4.5$; blue: $\av\leq4.5$), with locally-weighted smoothing curves for each group. Grey symbols show comparable measurements for $z\sim2$ SMGs from \citet{gillman2023,Gillman2024} at similar rest-frame wavelengths. Black points mark the median measurements for our sample at rest-frame $\sim2\,\um$.}\label{fig:morph_lambda}
\end{figure*}

We further quantified the wavelength-dependent morphologies of our galaxies using \texttt{statmorph} to compute standard non-parametric morphology diagnostics, including the CAS parameters (concentration, asymmetry, clumpiness; \citealt{Conselice2003,Lotz2004}) and the Gini–\mtw\ statistics \citep{Lotz2004}, which together characterize the central concentration, symmetry, and distribution of light within a galaxy.

For each source, we constructed segmentation maps and elliptical apertures using SEP \citep{Barbary2016}, ensuring consistent definitions across filters. To estimate uncertainties, we repeated the \texttt{statmorph} measurements 1000 times per filter, perturbing each image with Gaussian noise drawn from the local rms and adopting the median of the successful realizations. Concentration and half-light radii from \texttt{statmorph} agree well (within $5–10\%$) with the independent curve-of-growth analysis (Section~\ref{sec:cog}; $C_\mathrm{CoG}/C_\texttt{statmorph}=1.05\pm0.02$ and $R_\mathrm{h,\texttt{statmorph}}/R_\mathrm{h,CoG}=0.96\pm0.08$), confirming the robustness of the measurements.

Gini and \mtw\ were computed following the standard definitions in \citet{Lotz2004}. As
\mtw\ is sensitive to off-center bright structure while Gini quantifies the inequality of the full flux distribution, the two parameters provide complementary information regarding substructure and central concentration.

Figure~\ref{fig:morph_lambda} shows the morphological parameters as a function of rest-frame wavelength, with points color-coded by central \av. We also include comparable measurements for $z\sim2$ SMGs from \citet{gillman2023,Gillman2024}. A clear trend emerges: dustier galaxies exhibit more extended, less concentrated, and less symmetric morphologies at short wavelengths, while these differences diminish toward the near-infrared.

Galaxies with high central \av\ ($>4.5$) show systematically lower concentrations and larger half-light radii at rest-frame $\lambda\lesssim1\,\um$, reflecting strong central obscuration that suppresses the inner light and broadens the apparent profile. The dispersion in both $C$ and $R_\mathrm{h}$ is also largest at short wavelengths, consistent with the diverse dust distributions inferred from the SED-fitting maps. At longer wavelengths ($\gtrsim1.5\,\um$), concentrations increase and the scatter decreases for all sources, indicating that rest-frame near-IR observations are far less affected by dust structure.

Asymmetry ($A$) also decreases with central \av: heavily obscured systems appear smoother in the rest-frame optical, likely because dust both hides small-scale features and correlates spatially with asymmetric structures \citep[e.g.][]{hodge2025}. Clumpiness ($S$) remains low ($\lesssim0.1$) at all wavelengths, similar to \citet{gillman2023,Gillman2024}, suggesting that SMGs are generally smooth outside their central regions.

The Gini and \mtw\ parameters are consistent with the CAS results. Gini increases with concentration, while \mtw, which traces the distribution of the brightest regions, tends to be higher (less negative) in the dustier galaxies, particularly at short wavelengths. At rest-frame $\gtrsim1.5,\um$, both parameters converge to a narrow range regardless of \av, highlighting the stabilizing effect of rest-frame near-IR on the observed morphology.

Our measurements at overlapping wavelengths are broadly consistent with those of \citet{gillman2023}. However, their SMGs most closely resemble the less dusty subset of our sample (blue curves in Fig.\,\ref{fig:morph_lambda}), whereas our dustier, far-IR luminous systems show systematically lower concentrations, larger radii, and higher \mtw. This difference likely reflects sample selection: the \citet{gillman2023} sources are considerably fainter in the sub-millimeter (median deboosted $S_\mathrm{850\mu m}=1.62^{+0.81}_{-0.79}$\,mJy) than our targets (median $S\mathrm{870\mu m}=6.4\pm2.6$\,mJy), implying less extreme obscuration and lower central \av. By contrast, the brighter SMG sample studied by \citet{Gillman2024} ($S_\mathrm{850\mu m}=3.8\pm0.4$\,mJy) shows structural parameters more consistent with our higher-\av\ galaxies.


\subsubsection{Dust-corrected model images}\label{sec:intrinsic-sizes}

\begin{figure*}
    \centering
    \includegraphics[width=0.85\linewidth]{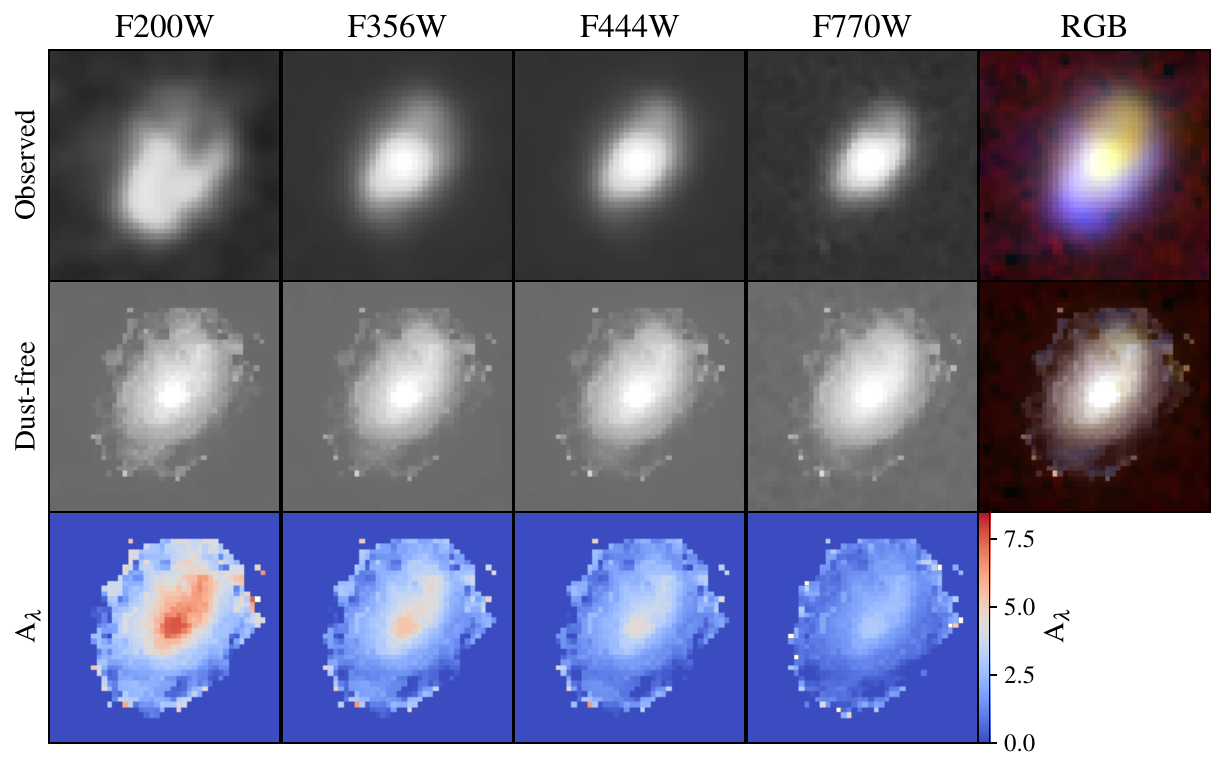}
    \caption{Observed, dust-corrected model images, and resolved $A_\lambda$ maps for ALESS~3.1. {\em Top row:} observed JWST images (each convolved to the MIRI/F770W resolution), plus an RGB composite using F770W, F444W, and F200W. {\em Middle row:} corresponding dust-corrected model images produced from the best-fit resolved SEDs, shown in identical format. All images are displayed with a logarithmic stretch. {\em Bottom row:} wavelength-dependent attenuation maps, $A_\lambda$, derived from the ratio of observed to dust-corrected images via $A_\lambda=-2.5\log(I_\mathrm{obs}/I_\mathrm{dust-free})$. The `dust-free' morphologies are similar, and dust can remain optically-thick even at 7.7\um.}
    \label{fig:dust-free}
\end{figure*}

To isolate the impact of dust attenuation on the observed morphologies of our SMGs, we constructed dust-corrected (`intrinsic') stellar emission maps using the resolved SED models. For each spatial bin, we extracted the unattenuated best-fit SED and generated model images in each JWST filter by integrating through the corresponding throughput curves. These model images, shown for ALESS~3.1 in Fig.\,\ref{fig:dust-free}, are compared directly to the observed images (all convolved to the MIRI/F770W resolution). Pixel-by-pixel ratios between the observed and model images then yield resolved, wavelength-dependent attenuation maps ($A_\lambda$).

After correcting for dust attenuation, the morphologies become strikingly similar across filters, indicating that intrinsic age and metallicity gradients are modest compared to the spatial structure of the dust.

Importantly, dust attenuation remains substantial even in MIRI/F770W. We find median $A_\mathrm{F770W}\simeq1.6$ and central values exceeding $2.4$\,mag. This contrasts with recent suggestions that rest-frame $1.6\,\um$ emission from dusty star-forming galaxies is largely unaffected by dust (e.g., \citealt{rujopakarn2023,lebail2024}).
Moreover, attenuation remains high ($A_\lambda\gtrsim1$) out to radii of $\sim3-4$\,kpc, consistent with previous studies (e.g., \citealt{hodge2016,smail2023,kokorev2023a}). This suggests that dust is not confined to compact nuclear regions but instead extends over kiloparsec scales, possibly tracing dusty disks or large-scale outflows in SMGs (see discussion in \citealt{smail2023} and references therein).

We also explored how dust affects parametric morphology measurements by fitting S\'ersic profiles to the dust-corrected F444W and F770W model images. These comparisons, presented in Appendix~\ref{app:sersic}, show that dust strongly biases S\'ersic parameters in the observed images (particularly ellipticity and S\'ersic index), whereas the dust-corrected images yield parameters more consistent with those measured from the 870\um\ continuum. Position angles, however, are generally robust to dust.


\section{Discussion}\label{sec:discussion}

The combined JWST (NIRCam+MIRI) and high-resolution ALMA observations allow us to trace the stellar, dust, and star-forming structure of SMGs on consistent physical scales. We find that their morphologies and structural parameters vary strongly with wavelength due to intense, centrally concentrated dust attenuation. Once this dust is accounted for through resolved SED modeling, the stellar distribution becomes markedly more compact and closely aligned with the dust continuum, implying that most of the stellar mass and star formation are co-located in heavily obscured central regions. This has key implications for interpreting JWST morphologies of dusty galaxies, and how they relate to their evolution.

\subsection{Dust-Driven Morphology}\label{sec:dust-morph}

We find that the morphologies of SMGs change strikingly with wavelength, as noted in previous work (e.g., \citealt{Gillman2024,hodge2025,Boogaard2026}). As shown in Fig.\,\ref{fig:montage}, the galaxies appear irregular and diverse at rest-frame optical wavelengths but become progressively smoother and more disk-like toward the near-IR, indicating that dust attenuation drives the strongly wavelength-dependent morphologies of SMGs.

MIRI/F770W provides a key view of this transition. Unlike NIRCam/F444W, the morphologies at 7.7\um\ closely resemble the 870\um\ dust continuum. Although attenuation still affects F770W in these dusty systems, the galaxy centers are no longer completely hidden, and then stellar-dust offsets commonly seen before JWST (e.g., \citealt{Hodge2015}) largely disappear. The near-IR flux peaks coincide with the far-IR dust peaks, supporting the picture that the bulk of star formation occurs in heavily obscured nuclear regions. Consistently, our curve-of-growth analysis shows that effective radii decrease smoothly with increasing wavelength, with the strongest wavelength dependence in galaxies with the highest central \av.

Our non-parametric \texttt{statmorph} analysis reinforces this interpretation. Galaxies with higher central \av\ show lower concentration, larger half-light radii, and lower Gini coefficients at rest-frame $\lesssim1\,\um$, consistent with dust suppressing central light while leaving outer regions relatively unaffected (see also \citealt{Gillman2024}). Consequently, rest-frame optical morphologies alone can misrepresent the intrinsic stellar structure of SMGs.

By contrast, rest-frame near-IR morphologies (especially F770W) are much more consistent with the dust continuum. F770W and 870\um\ sizes correlate strongly, both in curve-of-growth and S\'ersic measurements, and after applying our resolved SED dust corrections the structural parameters converge even further toward the ALMA values. F770W therefore provides the most reliable proxy for the stellar mass distribution, while F444W measurements (e.g., \citealt{cheng2022,cheng2023,gillman2023,sun2024,bodansky2025,chan2025,price2025}) may remain biased by dust attenuation, particularly in the most obscured systems. This close alignment between the F770W stellar distribution and the dust continuum also explains the tight spatial correlation we observe between \sigmstar\ and \sigsfr\ in the resolved star-forming main sequence (Section~\ref{sec:rSFMS}).

\subsection{Stellar and Dust Sizes}

Using our resolved SED-derived stellar mass maps, we measure compact intrinsic stellar sizes for the ALESS SMGs, with a median $R_\mathrm{e,\ast}=1.8\pm0.5$\,kpc. These are smaller than many previous estimates based on rest-frame optical colors (typically $\sim3$\,kpc at $z\simeq2$; \citealt{lang2019,Tadaki2020ApJ}) or some JWST color-gradient analyses reaching $\sim5$\,kpc \citep{wu2023}. Such optical-color methods are highly sensitive to central dust extinction and therefore tend to overestimate the true stellar extent.

Near-IR measurements from NIRCam/F444W generally yield $R_\mathrm{e}\sim1-5$\,kpc \citep{chen2022b,cheng2023,gillman2023,rujopakarn2023,Gillman2024,Boogaard2024,lebail2024,hodge2025,bodansky2025}, consistent with our observed F444W sizes (Fig.\,\ref{fig:sersic-sizes}), but still larger than the mass-weighted radii derived from our stellar-mass maps. As in recent resolved SED studies \citep{smail2023,sun2024}, we find the intrinsic stellar distribution to be more compact, typically $1-2$\,kpc.

The dust continuum emission is similarly compact. We measure $R_\mathrm{e,870\mu m}=1.7\pm0.7$\,kpc, consistent with \citet{hodge2025} and with previous ALMA measurements ($\sim1-2$\,kpc; \citealt{simpson2015,gullberg2019,lang2019,wu2023}). The close match between stellar and dust sizes in our sample ($R_\mathrm{e,870\mu m}/R_\mathrm{e,\ast}\simeq1.0\pm0.4$) reflects the structural similarity between the MIRI/F770W and ALMA dust emission.

Earlier studies often reported smaller dust-to-stellar size ratios ($R_\mathrm{e,dust}/R_\mathrm{e,\ast}\sim0.4$–$0.6$; \citealt{lang2019,Tadaki2020ApJ,wu2023,sun2021}), largely because stellar sizes were derived from rest-frame UV/optical imaging that remains strongly affected by dust. More recent studies using resolved SED modeling find ratios $\gtrsim1$ \citep{sun2024,smail2023,Gillman2024}, consistent with hydrodynamical simulations predicting dust emission as extended as, or more extended than, the stellar mass distribution \citep{cochrane2019,popping2022}.

Both stellar and dust size estimates may still be affected by extreme central obscuration \citep{Simpson2017}, which can flatten the observed profiles. We discuss these optical-depth effects further in Section~\ref{sec:caveat}.

\subsubsection{Implications for SMG evolution}

\begin{figure*}
    \centering
    \includegraphics[width=0.85\linewidth]{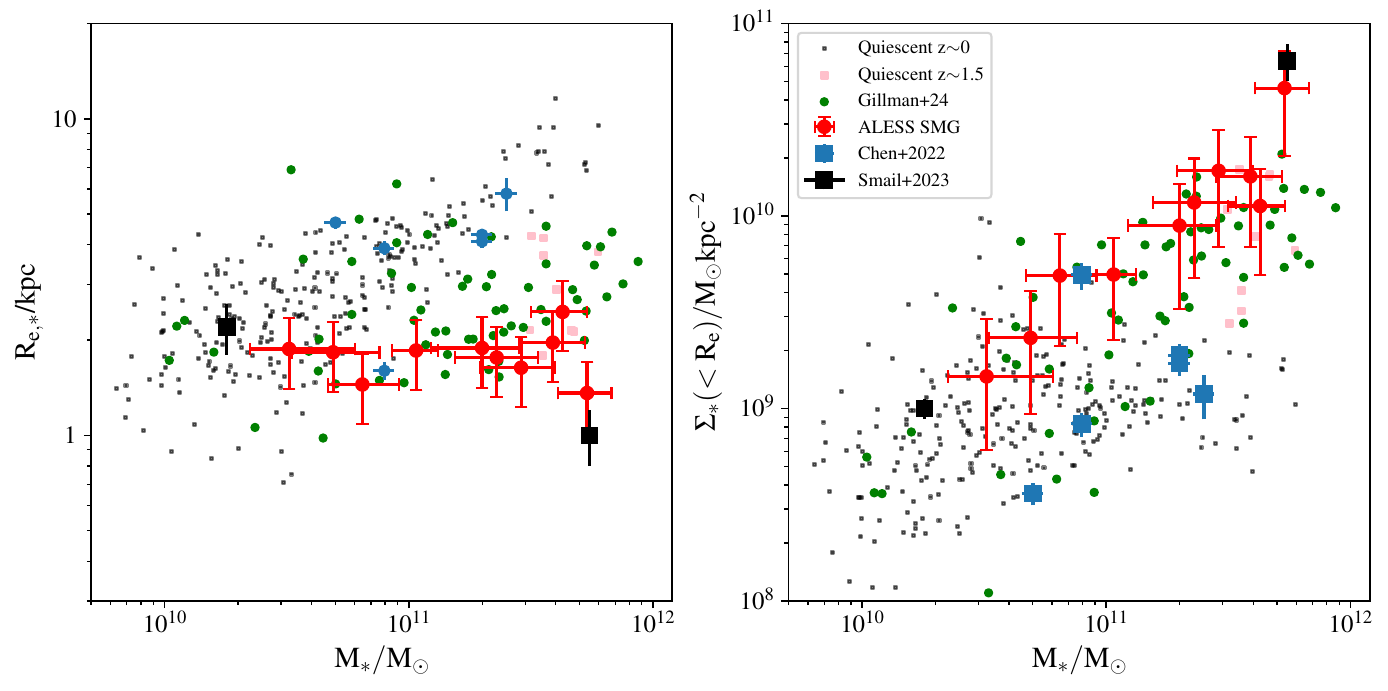}
    \caption{Comparison of stellar mass effective radii (left) and stellar mass surface densities within $R_\mathrm{e,\ast}$ (right). Red points show results from this work based on SED-derived stellar mass maps. Grey points are $z\sim0$ quiescent galaxies from ATLAS-3D \citep{Cappellari2011,Cappellari2013}, and pink points are $z\simeq1.5$ quiescent galaxies from \citet{Longhetti2007}. Blue squares and green circles indicate F444W-based SMG sizes at $z\sim1-4$ from \citet{chen2022b} and \citet{Gillman2024}, respectively, and black squares show stellar mass sizes of two $z=4.26$ SMGs \citep{smail2023}. SMGs occupy a region consistent with being progenitors of lower-redshift quiescent galaxies.
    }\label{fig:size-mass-density}
\end{figure*}

Galaxy size evolution provides a key diagnostic of massive galaxy formation. Our dust-corrected stellar mass radii ($R_\mathrm{e,\ast}=1.8\pm0.5$\,kpc) place the ALESS SMGs among the most compact massive systems at $z\sim1-3$, comparable to the densest quiescent galaxies at similar epochs \citep{Trujillo2006,Longhetti2007,Lustig2021,vdWel2024}. Such compact quiescent galaxies are $\sim3-4$ times smaller than their $z\sim0$ descendants \citep{Newman2012,vanderWel2014,Ito2024,Martorano2024}, suggesting that systems like our SMGs likely quench early and subsequently grow via minor gas-poor mergers \citep{Hopkins2009,Naab2009}.

As shown in Fig.\,\ref{fig:size-mass-density}, our SMGs occupy the $\lesssim2$\,kpc regime, smaller than the F444W-based sizes of SMGs at $z\sim1-3$ \citep{chen2022b,Gillman2024}. This difference is expected because F444W measurements can still be inflated by heavy central obscuration (we note that our curve-of-growth half light radii measured on the F444W images are comparable to those in the aforementioned studies). Stellar-mass-based sizes therefore provide a more reliable estimate of the intrinsic SMG structure.

Empirical relations for massive quiescent galaxies ($R_\mathrm{e}\propto(1+z)^{-1.4}$; \citealt{Trujillo2007,Ito2024,vdWel2024}) imply that our $z\sim3$ systems should grow by a factor of $\sim2$ by $z\sim1.5$ and $\sim7$ by $z=0$, naturally reaching the $\sim10$\,kpc sizes of present-day massive ellipticals (see \citealt{smail2023}). For lower-mass SMGs in our sample, size evolution may also involve stellar mass growth through major mergers (e.g., ALESS~1.1 and ALESS~1.2; \citealt{hodge2025}). Overall, their location in the size-mass plane is consistent with an evolutionary sequence linking compact dusty starbursts at $z\sim3$ to compact quiescent galaxies at $z\sim1-3$, followed by merger-driven size growth.

Our SMGs also reach stellar mass surface densities comparable to, and sometimes exceeding, those of compact quiescent galaxies at similar redshifts, and nearly an order of magnitude higher than local early-type galaxies\footnote{Literature measurements are not always derived using identical apertures or stellar-mass maps, which may introduce systematic offsets relative to our resolved SED-based estimates.}. Such dense stellar cores are predicted to precede rapid quenching (\citealt{vanDokkum2014,Barro2017}; see also \citealt{smail2023}).

Together, the compact stellar sizes, high central densities, and close spatial alignment between stellar mass and dust-obscured star formation support a scenario in which SMGs are rapidly building the dense stellar cores that later evolve into massive quiescent galaxies.

\subsection{Merger-driven starbursts?}

\begin{figure*}
    \centering
    \includegraphics[width=\linewidth]{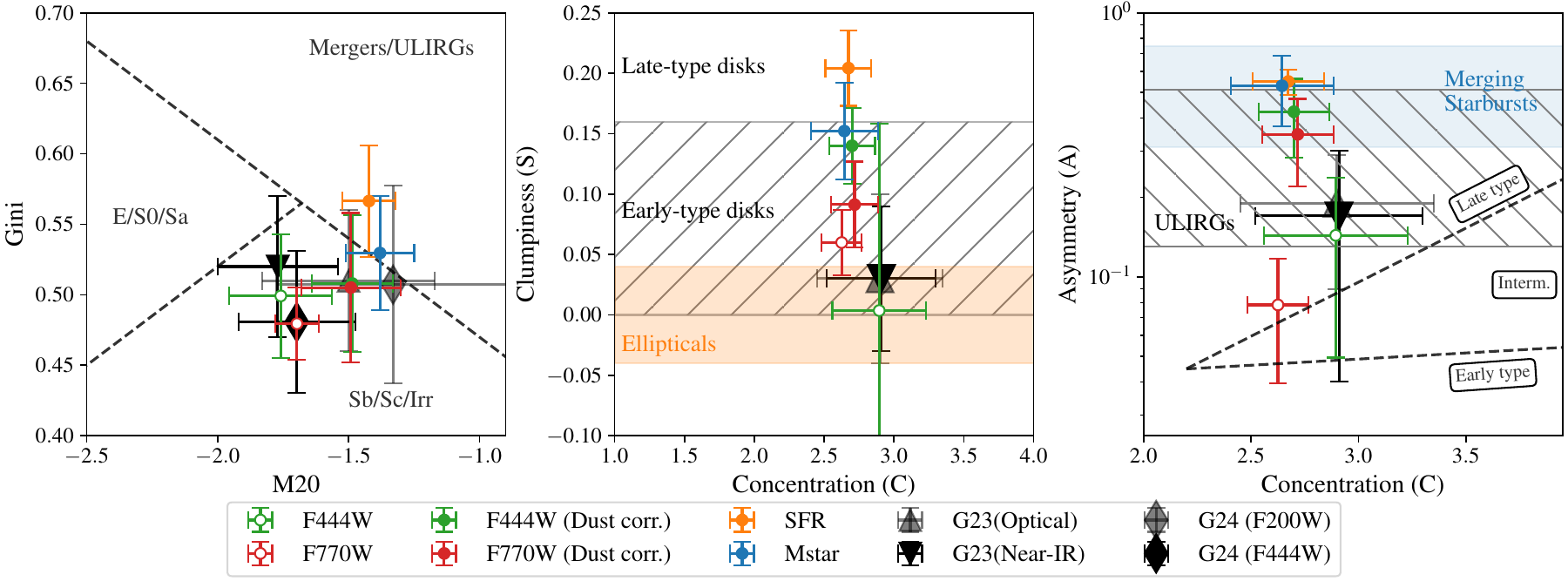}
    \caption{Non-parametric classifications of the ALESS SMGs. {\em Left:} Gini-\mtw\ plane with \citet{Lotz2008} boundaries. {\em Middle:} Clumpiness-Concentration (S-C) plane showing regions for disks and ellipticals \citep{Conselice2003,Conselice2014}. {\em Right:} Concentration-Asymmetry (C-A) plane with Hubble sequence regions and merger thresholds \citep{Bershady2000,Conselice2003}. Black/grey points show NIRCam-based SMG measurements from \citet{gillman2023,Gillman2024}. Dust-corrected bands, stellar mass, and SFR maps lie in or near the merger regime.}\label{fig:cas-gini_m20}
\end{figure*}

Having established that dust strongly affects the observed morphologies of SMGs, we now investigate whether their intrinsic structures show evidence for merger-driven starbursts. To do so, we examine their locations in the Gini-\mtw\ and CAS parameter spaces (Section~\ref{sec:statmorph}), using measurements from observed F444W and F770W images, their dust-corrected counterparts, and the stellar mass and SFR maps derived from our resolved SED fitting.

Across all three diagnostic planes (Fig.\,\ref{fig:cas-gini_m20}), the observed F444W and F770W morphologies fall broadly within the regions occupied by disk galaxies, consistent with previous JWST studies (e.g., \citealt{gillman2023,Gillman2024}). Differences with NIRCam-based measurements from those studies likely reflect the lower angular resolution of F770W, particularly in concentration.

After dust correction, the morphologies shift systematically: Gini and \mtw\ increase, and both clumpiness and asymmetry rise in the S-C and C-A planes. The dust-corrected F770W morphologies approach the regions associated with ULIRGs and merger-driven starbursts, indicating that a substantial fraction of the intrinsic asymmetric structure is hidden by dust at shorter wavelengths.

The strongest merger signatures are seen in the physical parameter maps. Both the \sigmstar\ and \sigsfr\ maps lie firmly within the merger/ULIRG regime in Gini-\mtw\ and C-A, and appear more clumpy in S-C than any single observed band. The \sigsfr\ morphologies are the most asymmetric, consistent with very recent ($\lesssim100$\,Myr) starburst activity. Although these diagnostics were not calibrated for physical parameter maps, the qualitative trends are robust: once dust effects are removed and all wavelengths are combined, SMGs appear more irregular, asymmetric, and centrally concentrated.

We note that increase in asymmetry and clumpiness in the dust-corrected and physical-parameter maps does not contradict the wavelength-dependent trends discussed in Section~\ref{sec:morphology}. Patchy/inhomogeneous dust attenuation can both introduce apparent asymmetries in the observed light and obscure intrinsically asymmetric structures associated with heavily obscured star-forming regions. Dust-corrected stellar mass maps remove the former while revealing the latter.

Our results are consistent with previous evidence for sub-galactic structure in SMGs. High-resolution ALMA imaging has revealed bars, rings, and spiral features in several ALESS sources \citep{hodge2019}, and \citet{hodge2025} showed that dust substructures coincide with residual features in NIRCam S\'ersic fits. Although F770W lacks the resolution to reveal similar structures, our analysis indicates that F444W remains significantly affected by dust, and that the dust-corrected and physical-parameter maps reveal disturbed, asymmetric stellar and star-forming components consistent with recent or ongoing mergers/interactions (see also Westoby et al., subm.).

\subsection{Extreme central obscuration}\label{sec:caveat}

Our resolved SED modelling shows that the central regions of SMGs remain optically thick even in the near-IR: typical attenuations are $A_\mathrm{F444W}>1$ at NIRCam/F444W and $A_\mathrm{F770W}\sim1$ for the emission detectable at MIRI/F770W. Independent constraints from sub-millimeter continuum show even more extreme dust columns. Under a uniform screen assumption, several high-redshift SMGs reach $\av\sim500$, and in the most obscured cases $\av\gtrsim1000$ \citep[e.g.][]{Simpson2017,Gomez-Guijarro_2018}. Such conditions challenge standard SED assumptions, including optically thin dust emission in the FIR (e.g., \citealt{Riechers2013,Boogaard2026}).

Optical and near-IR observations probe only sightlines where some stellar continuum escapes. As a result, we likely underestimate the stellar mass in the most obscured central regions, where the brightest 870\um\ emission often peaks. If this hidden mass is substantial, the intrinsic stellar distribution could be even more compact, implying smaller true radii, denser cores, and potentially larger dust-to-stellar size ratios than inferred here.

Similarly, if the dust is optically thick at far-IR wavelengths, the observed continuum will be biased. Optical depth suppresses emission on the Wien side of the SED, making the dust appear cooler and biasing temperature and mass estimates \citep{dacunha2021}. The true dust distribution may therefore be more compact than implied by the observed 870\um\ emission, effectively lowering the dust-to-stellar size ratios.

Resolving these effects will require improved SED modeling and future multi-band, high-resolution FIR imaging capable of probing the Wien side of the dust SED and breaking age-dust-temperature degeneracies. Such data will also provide key constraints for developing optically thick extensions of \textsc{magphys} and for testing the resolved rSFMS (Section~\ref{sec:rSFMS}).

\section{Summary and Conclusion} \label{sec:conclusion}

In this work, we have used spatially resolved SED fitting of JWST (NIRCam+MIRI) and high-resolution ALMA Band 7 continuum observations to investigate the internal structure of 12 ALESS SMGs ($2.0 \lesssim z \lesssim 4.5$) on $\sim$kiloparsec scales. Our analysis shows that their observed morphologies and inferred stellar masses are strongly shaped by a highly inhomogeneous (patchy) dust geometry, in which moderate effective attenuations inferred from the SEDs coexist with extremely large dust columns along some sightlines. This configuration produces strongly wavelength-dependent morphologies and can hide a substantial fraction of the stellar mass from integrated SED fits, highlighting the limitations of integrated analyses for heavily obscured systems. Our main results are summarized as follows:

\begin{itemize}

\item We find a strong correlation between \sigmstar\ and \sigsfr\ across most galaxies, establishing a resolved star-forming main sequence (rSFMS). The slopes range from $\sim0.7$ to $\gtrsim1$, with smaller values for those with better FIR constraints. This behavior is broadly consistent with other high-redshift systems, but highlights the sensitivity of resolved SFR estimates to dust temperature priors. High-resolution, multi-band FIR data will be required to constrain spatial variations in dust temperature and robustly test the rSFMS.

\item By comparing integrated stellar masses to the sum of resolved stellar masses, we find a systematic underestimation in integrated fits (median $\sim0.2$\,dex). While often attributed to classical `outshining', this effect correlates more strongly with dust luminosity and attenuation than with sSFR, indicating that stellar mass biases in SMGs are primarily driven by dust obscuration rather than classical outshining. Highly obscured stellar mass is therefore missed by integrated SED fits, reinforcing the limitations of integrated analyses in dusty starbursts. When only NIRCam bands are used, the discrepancy increases further, underscoring the importance of longer-wavelength constraints. Resolved SED fitting thus provides a more complete view of the stellar mass budget.

\item The structural and morphological appearance of SMGs changes strongly with wavelength. At rest-frame optical wavelengths, heavy central obscuration produces large apparent offsets between stellar and dust emission and inflates half-light radii. These offsets decrease rapidly toward longer wavelengths and disappear in MIRI/F770W, where attenuation becomes more modest. Non-parametric morphology indicators (Gini, \mtw, concentration, asymmetry) show similar behavior, driven primarily by dust rather than intrinsic structural differences. Recovering meaningful stellar morphologies therefore requires rest-frame $\gtrsim1.6\,\um$ data and SED modeling that accounts for dust.

\item Intrinsic stellar mass sizes derived from resolved SED fitting agree closely with ALMA 870\um\ dust continuum sizes, in contrast with earlier rest-frame optical-based studies that found larger stellar radii. However, extreme central obscuration may still hide stellar mass in some sightlines, implying that our stellar sizes could be modest underestimates. At the same time, if the dust is optically thick in the FIR, the true dust-emitting region may be more compact than inferred. Disentangling these effects will require high-resolution, multi-band FIR observations that probe the Wien side of the SED and enable optically thick dust modeling.

\item The strong central obscuration implies intense, centrally concentrated star formation, favoring rapid growth of dense stellar cores. This is consistent with theoretical predictions and supports the idea that SMGs at $z\simeq3$ are progenitors of compact quiescent galaxies, linking these systems to early bulge formation and potential inside-out quenching.

\end{itemize}

Together, these results demonstrate that resolving dust and stellar structure is essential for accurately interpreting the physical properties of heavily obscured galaxies in the early Universe. Given that most massive star-forming galaxies at $z\gtrsim1$ are significantly dust-obscured (e.g., \citealt{Whitaker2017,vdWel2025}), the conditions driving the biases identified here are not unique to sub-millimetre selected systems. Our results therefore have broad applicability to the general population of massive star-forming galaxies, implying that even `normal' galaxies are likely to be subject to similar dust-obscuration biases when analysed using unresolved data.

MIRI/F770W plays a key role in this analysis, providing the link between heavily obscured rest-frame optical emission and the dust continuum, and enabling physically consistent SED modeling across the full JWST wavelength range. However, ALMA Band 7 alone (probing rest-frame $\sim200\,\mu$m emission at $z\sim3.5$) is insufficient to constrain spatial variations in dust temperature (see, e.g., \citealt{dacunha2021}). Future high-resolution ALMA observations at higher frequencies (Bands 8 to 10, probing rest-frame $\sim100\,\um$) will be essential for resolving dust temperature gradients, testing optically thick dust models, and refining our understanding of the stellar and dust structure of SMGs.

\begin{acknowledgments}
JL, EdC, and AJB acknowledge
support from the Australian Research Council (project DP240100589).
JH, LAB, CLL, and BW acknowledge support from
the ERC Consolidator Grant 101088676 (``VOYAJ'').
LAB further acknowledges support from the Dutch Research Council (NWO) under grant VI.Veni.242.055 (\url{https://doi.org/10.61686/LAJVP77714}).
IRS acknowledges support from STFC (ST/X001075/1).
KK acknowledges support from the Knut and Alice Wallenberg Foundation (KAW 2020.0081 and 2023.0339).
C-CC acknowledges support from the National Science and Technology Council of Taiwan (114-2628-M-001-006-MY4), as well as Academia Sinica through the Career Development Award (AS-CDA-112-M02).
MR is supported by the NWO Veni project ``Under the lens'' (VI.Veni.202.225).

This work is based in part on observations made with the
NASA/ESA/CSA James Webb Space Telescope. The data
were obtained from the Mikulski Archive for Space Telescopes
at the Space Telescope Science Institute, which is operated by
the Association of Universities for Research in Astronomy,
Inc., under NASA contract NAS 5-03127 for JWST. These
observations are associated with program No. 2516. The
specific observations analyzed can be accessed via
doi:10.17909/e33v-ga73. Support for program No. 2516 was
provided by NASA through a grant from the Space Telescope
Science Institute, which is operated by the Association of
Universities for Research in Astronomy, Inc., under NASA
contract NAS 5-03127.

This paper uses the following ALMA data: ADS/JAO.ALMA\#2016.1.00048.S, \#2015.1.00948.S, \#2012.1.00307.S, and \#2011.1.00294.S. ALMA is a partnership
of ESO (representing its member states), NSF (USA), and NINS
(Japan), together with NRC (Canada) and NSC and ASIAA
(Taiwan) and KASI (Republic of Korea), in cooperation with the
Republic of Chile. The Joint ALMA Observatory is operated by
ESO, AUI/NRAO, and NAOJ.

We acknowledge the use of data products from observations made with ESO Telescopes at the La
Silla Paranal Observatory under ESO programme ID 179.A-2006 and on
data products produced by the Cambridge Astronomy Survey Unit on
behalf of the VIDEO consortium.

ICRAR and the University of Western Australia are located on Whadjuk Noongar land, and the authors acknowledge the Whadjuk Noongar people as the cultural custodians of that land.

\end{acknowledgments}

\bibliography{bib}{}

@ARTICLE{Smail_1997,
       author = {{Smail}, Ian and {Ivison}, R.~J. and {Blain}, A.~W.},
        title = "{A Deep Sub-millimeter Survey of Lensing Clusters: A New Window on Galaxy Formation and Evolution}",
      journal = {\apjl},
         year = "1997",
        month = "Nov",
       volume = {490},
       number = {1},
        pages = {L5-L8},
          doi = {10.1086/311017},
archivePrefix = {arXiv},
       eprint = {astro-ph/9708135},
 primaryClass = {astro-ph},
       adsurl = {https://ui.adsabs.harvard.edu/abs/1997ApJ...490L...5S}
}

@ARTICLE{Hughes1998,
       author = {{Hughes}, David H. and {Serjeant}, Stephen and {Dunlop}, James and {Rowan-Robinson}, Michael and {Blain}, Andrew and {Mann}, Robert G. and {Ivison}, Rob and {Peacock}, John and {Efstathiou}, Andreas and {Gear}, Walter and {Oliver}, Seb and {Lawrence}, Andy and {Longair}, Malcolm and {Goldschmidt}, Pippa and {Jenness}, Tim},
        title = "{High-redshift star formation in the Hubble Deep Field revealed by a submillimetre-wavelength survey}",
      journal = {\nat},
         year = 1998,
        month = jul,
       volume = {394},
       number = {6690},
        pages = {241-247},
          doi = {10.1038/28328},
archivePrefix = {arXiv},
       eprint = {astro-ph/9806297},
 primaryClass = {astro-ph},
       adsurl = {https://ui.adsabs.harvard.edu/abs/1998Natur.394..241H}
}

@ARTICLE{Barger1998,
       author = {{Barger}, A.~J. and {Cowie}, L.~L. and {Sanders}, D.~B. and {Fulton}, E. and {Taniguchi}, Y. and {Sato}, Y. and {Kawara}, K. and {Okuda}, H.},
        title = "{Submillimetre-wavelength detection of dusty star-forming galaxies at high redshift}",
      journal = {\nat},
         year = 1998,
        month = jul,
       volume = {394},
       number = {6690},
        pages = {248-251},
          doi = {10.1038/28338},
archivePrefix = {arXiv},
       eprint = {astro-ph/9806317},
 primaryClass = {astro-ph},
       adsurl = {https://ui.adsabs.harvard.edu/abs/1998Natur.394..248B}
}

@ARTICLE{Eales1999,
       author = {{Eales}, Stephen and {Lilly}, Simon and {Gear}, Walter and {Dunne}, Loretta and {Bond}, J. Richard and {Hammer}, Francois and {Le F{\`e}vre}, Olivier and {Crampton}, David},
        title = "{The Canada-UK Deep Submillimeter Survey: First Submillimeter Images, the Source Counts, and Resolution of the Background}",
      journal = {\apj},
         year = 1999,
        month = apr,
       volume = {515},
       number = {2},
        pages = {518-524},
          doi = {10.1086/307069},
archivePrefix = {arXiv},
       eprint = {astro-ph/9808040},
 primaryClass = {astro-ph},
       adsurl = {https://ui.adsabs.harvard.edu/abs/1999ApJ...515..518E}
}

@ARTICLE{Blain_2002,
       author = {{Blain}, Andrew W. and {Smail}, Ian and {Ivison}, R.~J. and {Kneib}, J. -P. and {Frayer}, David T.},
        title = "{Submillimeter galaxies}",
      journal = {\physrep},
         year = 2002,
        month = oct,
       volume = {369},
       number = {2},
        pages = {111-176},
          doi = {10.1016/S0370-1573(02)00134-5},
archivePrefix = {arXiv},
       eprint = {astro-ph/0202228},
 primaryClass = {astro-ph},
       adsurl = {https://ui.adsabs.harvard.edu/abs/2002PhR...369..111B}
}

@ARTICLE{Casey_2014,
       author = {{Casey}, Caitlin M. and {Narayanan}, Desika and {Cooray}, Asantha},
        title = "{Dusty star-forming galaxies at high redshift}",
      journal = {\physrep},
         year = "2014",
        month = "Aug",
       volume = {541},
       number = {2},
        pages = {45-161},
          doi = {10.1016/j.physrep.2014.02.009},
archivePrefix = {arXiv},
       eprint = {1402.1456},
 primaryClass = {astro-ph.CO},
       adsurl = {https://ui.adsabs.harvard.edu/abs/2014PhR...541...45C}
}

@article{swinbank2014,
  title = {An {{ALMA}} Survey of Sub-Millimetre {{Galaxies}} in the {{Extended Chandra Deep Field South}}: The Far-Infrared Properties of {{SMGs}}},
  shorttitle = {An {{ALMA}} Survey of Sub-Millimetre {{Galaxies}} in the {{Extended Chandra Deep Field South}}},
  author = {Swinbank, A. M. and Simpson, J. M. and Smail, Ian and Harrison, C. M. and Hodge, J. A. and Karim, A. and Walter, F. and Alexander, D. M. and Brandt, W. N. and De Breuck, C. and Da Cunha, E. and Chapman, S. C. and K. Coppin, K. E. and Danielson, A. L. R. and Dannerbauer, H. and Decarli, R. and Greve, T. R. and Ivison, R. J. and Knudsen, K. K. and Lagos, C. D. P. and Schinnerer, E. and Thomson, A. P. and Wardlow, J. L. and Wei{\ss}, A. and Werf, P. Van Der},
  year = {2014},
  month = feb,
  journal = {Monthly Notices of the Royal Astronomical Society},
  volume = {438},
  number = {2},
  pages = {1267--1287},
  issn = {0035-8711, 1365-2966},
  doi = {10.1093/mnras/stt2273},
  urldate = {2025-09-24},
  langid = {english}
}

@ARTICLE{dacunha2021,
       author = {{da Cunha}, E. and {Hodge}, J.~A. and {Casey}, C.~M. and {Algera}, H.~S.~B. and {Kaasinen}, M. and {Smail}, I. and {Walter}, F. and {Brandt}, W.~N. and {Dannerbauer}, H. and {Decarli}, R. and {Groves}, B.~A. and {Knudsen}, K.~K. and {Swinbank}, A.~M. and {Weiss}, A. and {van der Werf}, P. and {Zavala}, J.~A.},
        title = "{Measurements of the Dust Properties in z ≃ 1-3 Submillimeter Galaxies with ALMA}",
      journal = {\apj},
         year = 2021,
        month = sep,
       volume = {919},
       number = {1},
          eid = {30},
        pages = {30},
          doi = {10.3847/1538-4357/ac0ae0},
archivePrefix = {arXiv},
       eprint = {2106.08566},
 primaryClass = {astro-ph.GA},
       adsurl = {https://ui.adsabs.harvard.edu/abs/2021ApJ...919...30D}
}

@ARTICLE{magphys2008,
       author = {{da Cunha}, Elisabete and {Charlot}, St{\'e}phane and {Elbaz}, David},
        title = "{A simple model to interpret the ultraviolet, optical and infrared emission from galaxies}",
      journal = {\mnras},
         year = 2008,
        month = aug,
       volume = {388},
       number = {4},
        pages = {1595-1617},
          doi = {10.1111/j.1365-2966.2008.13535.x},
archivePrefix = {arXiv},
       eprint = {0806.1020},
 primaryClass = {astro-ph},
       adsurl = {https://ui.adsabs.harvard.edu/abs/2008MNRAS.388.1595D}
}

@ARTICLE{magphys2019,
       author = {{Battisti}, A.~J. and {da Cunha}, E. and {Grasha}, K. and {Salvato}, M. and {Daddi}, E. and {Davies}, L. and {Jin}, S. and {Liu}, D. and {Schinnerer}, E. and {Vaccari}, M. and {COSMOS Collaboration}},
        title = "{MAGPHYS+photo-z: Constraining the Physical Properties of Galaxies with Unknown Redshifts}",
      journal = {\apj},
         year = 2019,
        month = sep,
       volume = {882},
       number = {1},
          eid = {61},
        pages = {61},
          doi = {10.3847/1538-4357/ab345d},
archivePrefix = {arXiv},
       eprint = {1908.00771},
 primaryClass = {astro-ph.GA},
       adsurl = {https://ui.adsabs.harvard.edu/abs/2019ApJ...882...61B}
}

@ARTICLE{magphys2020,
       author = {{Battisti}, A.~J. and {Cunha}, E. da and {Shivaei}, I. and {Calzetti}, D.},
        title = "{The Strength of the 2175 {\r{A}} Feature in the Attenuation Curves of Galaxies at 0.1 < z {\ensuremath{\lesssim}} 3}",
      journal = {\apj},
         year = 2020,
        month = jan,
       volume = {888},
       number = {2},
          eid = {108},
        pages = {108},
          doi = {10.3847/1538-4357/ab5fdd},
archivePrefix = {arXiv},
       eprint = {1912.05206},
 primaryClass = {astro-ph.GA},
       adsurl = {https://ui.adsabs.harvard.edu/abs/2020ApJ...888..108B}
}

@article{miller2013,
  title = {{{THE VERY LARGE ARRAY}} 1.4 {{GHz SURVEY OF THE EXTENDED CHANDRA DEEP FIELD SOUTH}}: {{SECOND DATA RELEASE}}},
  shorttitle = {{{THE VERY LARGE ARRAY}} 1.4 {{GHz SURVEY OF THE EXTENDED CHANDRA DEEP FIELD SOUTH}}},
  author = {Miller, Neal A. and Bonzini, Margherita and Fomalont, Edward B. and Kellermann, Kenneth I. and Mainieri, Vincenzo and Padovani, Paolo and Rosati, Piero and Tozzi, Paolo and Vattakunnel, Shaji},
  year = {2013},
  month = mar,
  journal = {The Astrophysical Journal Supplement Series},
  volume = {205},
  number = {2},
  pages = {13},
  issn = {0067-0049, 1538-4365},
  doi = {10.1088/0067-0049/205/2/13},
  urldate = {2025-10-16},
  copyright = {http://iopscience.iop.org/info/page/text-and-data-mining},
  langid = {english}
}

@article{simpson2014,
  title = {{{AN ALMA SURVEY OF SUBMILLIMETER GALAXIES IN THE EXTENDED CHANDRA DEEP FIELD SOUTH}}: {{THE REDSHIFT DISTRIBUTION AND EVOLUTION OF SUBMILLIMETER GALAXIES}}},
  shorttitle = {{{AN ALMA SURVEY OF SUBMILLIMETER GALAXIES IN THE EXTENDED CHANDRA DEEP FIELD SOUTH}}},
  author = {Simpson, J. M. and Swinbank, A. M. and Smail, Ian and Alexander, D. M. and Brandt, W. N. and Bertoldi, F. and De Breuck, C. and Chapman, S. C. and Coppin, K. E. K. and Da Cunha, E. and Danielson, A. L. R. and Dannerbauer, H. and Greve, T. R. and Hodge, J. A. and Ivison, R. J. and Karim, A. and Knudsen, K. K. and Poggianti, B. M. and Schinnerer, E. and Thomson, A. P. and Walter, F. and Wardlow, J. L. and Wei{\ss}, A. and Van Der Werf, P. P.},
  year = {2014},
  month = may,
  journal = {The Astrophysical Journal},
  volume = {788},
  number = {2},
  pages = {125},
  issn = {0004-637X, 1538-4357},
  doi = {10.1088/0004-637X/788/2/125},
  urldate = {2024-03-07},
  langid = {english}
}

@ARTICLE{simpson2015,
       author = {{Simpson}, J.~M. and {Smail}, Ian and {Swinbank}, A.~M. and {Chapman}, S.~C. and {Geach}, J.~E. and {Ivison}, R.~J. and {Thomson}, A.~P. and {Aretxaga}, I. and {Blain}, A.~W. and {Cowley}, W.~I. and {Chen}, Chian-Chou and {Coppin}, K.~E.~K. and {Dunlop}, J.~S. and {Edge}, A.~C. and {Farrah}, D. and {Ibar}, E. and {Karim}, A. and {Knudsen}, K.~K. and {Meijerink}, R. and {Micha{\l}owski}, M.~J. and {Scott}, D. and {Spaans}, M. and {van der Werf}, P.~P.},
        title = "{The SCUBA-2 Cosmology Legacy Survey: ALMA Resolves the Bright-end of the Sub-millimeter Number Counts}",
      journal = {\apj},
         year = 2015,
        month = jul,
       volume = {807},
       number = {2},
          eid = {128},
        pages = {128},
          doi = {10.1088/0004-637X/807/2/128},
archivePrefix = {arXiv},
       eprint = {1505.05152},
 primaryClass = {astro-ph.GA},
       adsurl = {https://ui.adsabs.harvard.edu/abs/2015ApJ...807..128S}
}

@ARTICLE{Karim2013,
       author = {{Karim}, A. and {Swinbank}, A.~M. and {Hodge}, J.~A. and {Smail}, I.~R. and {Walter}, F. and {Biggs}, A.~D. and {Simpson}, J.~M. and {Danielson}, A.~L.~R. and {Alexander}, D.~M. and {Bertoldi}, F. and {de Breuck}, C. and {Chapman}, S.~C. and {Coppin}, K.~E.~K. and {Dannerbauer}, H. and {Edge}, A.~C. and {Greve}, T.~R. and {Ivison}, R.~J. and {Knudsen}, K.~K. and {Menten}, K.~M. and {Schinnerer}, E. and {Wardlow}, J.~L. and {Wei{\ss}}, A. and {van der Werf}, P.},
        title = "{An ALMA survey of submillimetre galaxies in the Extended Chandra Deep Field South: high-resolution 870 {\ensuremath{\mu}}m source counts}",
      journal = {\mnras},
         year = 2013,
        month = jun,
       volume = {432},
       number = {1},
        pages = {2-9},
          doi = {10.1093/mnras/stt196},
archivePrefix = {arXiv},
       eprint = {1210.0249},
 primaryClass = {astro-ph.CO},
       adsurl = {https://ui.adsabs.harvard.edu/abs/2013MNRAS.432....2K}
}

@ARTICLE{statmorph2019,
       author = {{Rodriguez-Gomez}, Vicente and {Snyder}, Gregory F. and {Lotz}, Jennifer M. and {Nelson}, Dylan and {Pillepich}, Annalisa and {Springel}, Volker and {Genel}, Shy and {Weinberger}, Rainer and {Tacchella}, Sandro and {Pakmor}, R{\"u}diger and {Torrey}, Paul and {Marinacci}, Federico and {Vogelsberger}, Mark and {Hernquist}, Lars and {Thilker}, David A.},
        title = "{The optical morphologies of galaxies in the IllustrisTNG simulation: a comparison to Pan-STARRS observations}",
      journal = {\mnras},
         year = 2019,
        month = mar,
       volume = {483},
       number = {3},
        pages = {4140-4159},
          doi = {10.1093/mnras/sty3345},
archivePrefix = {arXiv},
       eprint = {1809.08239},
 primaryClass = {astro-ph.GA},
       adsurl = {https://ui.adsabs.harvard.edu/abs/2019MNRAS.483.4140R}
}

@ARTICLE{hodge2013,
       author = {{Hodge}, J.~A. and {Karim}, A. and {Smail}, I. and {Swinbank}, A.~M. and {Walter}, F. and {Biggs}, A.~D. and {Ivison}, R.~J. and {Weiss}, A. and {Alexander}, D.~M. and {Bertoldi}, F. and {Brandt}, W.~N. and {Chapman}, S.~C. and {Coppin}, K.~E.~K. and {Cox}, P. and {Danielson}, A.~L.~R. and {Dannerbauer}, H. and {De Breuck}, C. and {Decarli}, R. and {Edge}, A.~C. and {Greve}, T.~R. and {Knudsen}, K.~K. and {Menten}, K.~M. and {Rix}, H. -W. and {Schinnerer}, E. and {Simpson}, J.~M. and {Wardlow}, J.~L. and {van der Werf}, P.},
        title = "{An ALMA Survey of Submillimeter Galaxies in the Extended Chandra Deep Field South: Source Catalog and Multiplicity}",
      journal = {\apj},
         year = 2013,
        month = may,
       volume = {768},
       number = {1},
          eid = {91},
        pages = {91},
          doi = {10.1088/0004-637X/768/1/91},
archivePrefix = {arXiv},
       eprint = {1304.4266},
 primaryClass = {astro-ph.CO},
       adsurl = {https://ui.adsabs.harvard.edu/abs/2013ApJ...768...91H}
}

@article{hodge2016,
  title = {{{KILOPARSEC-SCALE DUST DISKS IN HIGH-REDSHIFT LUMINOUS SUBMILLIMETER GALAXIES}}},
  author = {Hodge, J. A. and Swinbank, A. M. and Simpson, J. M. and Smail, I. and Walter, F. and Alexander, D. M. and Bertoldi, F. and Biggs, A. D. and Brandt, W. N. and Chapman, S. C. and Chen, C. C. and Coppin, K. E. K. and Cox, P. and Dannerbauer, H. and Edge, A. C. and Greve, T. R. and Ivison, R. J. and Karim, A. and Knudsen, K. K. and Menten, K. M. and Rix, H.-W. and Schinnerer, E. and Wardlow, J. L. and Weiss, A. and van der Werf, P.},
  year = {2016},
  month = dec,
  journal = {The Astrophysical Journal},
  volume = {833},
  number = {1},
  pages = {103},
  publisher = {The American Astronomical Society},
  issn = {0004-637X},
  doi = {10.3847/1538-4357/833/1/103},
  urldate = {2023-07-26},
  langid = {english}
}

@article{hodge2019,
  title = {{{ALMA Reveals Potential Evidence}} for {{Spiral Arms}}, {{Bars}}, and {{Rings}} in {{High-redshift Submillimeter Galaxies}}},
  author = {Hodge, J. A. and Smail, I. and Walter, F. and da Cunha, E. and Swinbank, A. M. and Rybak, M. and Venemans, B. and Brandt, W. N. and Rivera, G. Calistro and Chapman, S. C. and Chen, Chian-Chou and Cox, P. and Dannerbauer, H. and Decarli, R. and Greve, T. R. and Knudsen, K. K. and Menten, K. M. and Schinnerer, E. and Simpson, J. M. and van der Werf, P. and Wardlow, J. L. and Weiss, A.},
  year = {2019},
  month = may,
  journal = {The Astrophysical Journal},
  volume = {876},
  number = {2},
  pages = {130},
  publisher = {The American Astronomical Society},
  issn = {0004-637X},
  doi = {10.3847/1538-4357/ab1846},
  urldate = {2023-07-26},
  langid = {english}
}

@ARTICLE{Hodge2020,
       author = {{Hodge}, J.~A. and {da Cunha}, E.},
        title = "{High-redshift star formation in the Atacama large millimetre/submillimetre array era}",
      journal = {Royal Society Open Science},
         year = 2020,
        month = dec,
       volume = {7},
       number = {12},
          eid = {200556},
        pages = {200556},
          doi = {10.1098/rsos.200556},
archivePrefix = {arXiv},
       eprint = {2004.00934},
 primaryClass = {astro-ph.GA},
       adsurl = {https://ui.adsabs.harvard.edu/abs/2020RSOS....700556H}
}

@article{chen2015,
  title = {{{AN ALMA SURVEY OF SUBMILLIMETER GALAXIES IN THE EXTENDED CHANDRA DEEP FIELD SOUTH}}: {{NEAR-INFRARED MORPHOLOGIES AND STELLAR SIZES}}},
  shorttitle = {{{AN ALMA SURVEY OF SUBMILLIMETER GALAXIES IN THE EXTENDED CHANDRA DEEP FIELD SOUTH}}},
  author = {Chen, Chian-Chou and Smail, Ian and Swinbank, A. M. and Simpson, J. M. and Ma, Cheng-Jiun and Alexander, D. M. and Biggs, A. D. and Brandt, W. N. and Chapman, S. C. and Coppin, K. E. K. and Danielson, A. L. R. and Dannerbauer, H. and Edge, A. C. and Greve, T. R. and Ivison, R. J. and Karim, A. and Menten, Karl M. and Schinnerer, E. and Walter, F. and Wardlow, J. L. and Wei{\ss}, A. and Van Der Werf, P. P.},
  year = {2015},
  month = jan,
  journal = {The Astrophysical Journal},
  volume = {799},
  number = {2},
  pages = {194},
  issn = {1538-4357},
  doi = {10.1088/0004-637X/799/2/194},
  urldate = {2024-04-03},
  langid = {english}
}

@ARTICLE{Chen2017,
       author = {{Chen}, Chian-Chou and {Hodge}, J.~A. and {Smail}, Ian and {Swinbank}, A.~M. and {Walter}, Fabian and {Simpson}, J.~M. and {Calistro Rivera}, Gabriela and {Bertoldi}, F. and {Brandt}, W.~N. and {Chapman}, S.~C. and {da Cunha}, Elisabete and {Dannerbauer}, H. and {De Breuck}, C. and {Harrison}, C.~M. and {Ivison}, R.~J. and {Karim}, A. and {Knudsen}, K.~K. and {Wardlow}, J.~L. and {Wei{\ss}}, A. and {van der Werf}, P.~P.},
        title = "{A Spatially Resolved Study of Cold Dust, Molecular Gas, H II Regions, and Stars in the z = 2.12 Submillimeter Galaxy ALESS67.1}",
      journal = {\apj},
         year = 2017,
        month = sep,
       volume = {846},
       number = {2},
          eid = {108},
        pages = {108},
          doi = {10.3847/1538-4357/aa863a},
archivePrefix = {arXiv},
       eprint = {1708.08937},
 primaryClass = {astro-ph.GA},
       adsurl = {https://ui.adsabs.harvard.edu/abs/2017ApJ...846..108C}
}

@article{gullberg2018,
  title = {The {{Dust}} and [{{C}} Ii] {{Morphologies}} of {{Redshift}} {$\sim$}4.5 {{Sub-millimeter Galaxies}} at {$\sim$}200 Pc {{Resolution}}: {{The Absence}} of {{Large Clumps}} in the {{Interstellar Medium}} at {{High-redshift}}},
  shorttitle = {The {{Dust}} and [{{C}} Ii] {{Morphologies}} of {{Redshift}} {$\sim$}4.5 {{Sub-millimeter Galaxies}} at {$\sim$}200 Pc {{Resolution}}},
  author = {Gullberg, B. and Swinbank, A. M. and Smail, I. and Biggs, A. D. and Bertoldi, F. and Breuck, C. De and Chapman, S. C. and Chen, C.-C. and Cooke, E. A. and Coppin, K. E. K. and Cox, P. and Dannerbauer, H. and Dunlop, J. S. and Edge, A. C. and Farrah, D. and Geach, J. E. and Greve, T. R. and Hodge, J. and Ibar, E. and Ivison, R. J. and Karim, A. and Schinnerer, E. and Scott, D. and Simpson, J. M. and Stach, S. M. and Thomson, A. P. and Van Der Werf, P. and Walter, F. and Wardlow, J. L. and Weiss, A.},
  year = {2018},
  month = may,
  journal = {The Astrophysical Journal},
  volume = {859},
  number = {1},
  pages = {12},
  issn = {0004-637X, 1538-4357},
  doi = {10.3847/1538-4357/aabe8c},
  urldate = {2025-10-16},
  langid = {english}
}

@article{gullberg2019,
  title = {An {{ALMA}} Survey of the {{SCUBA-2 Cosmology Legacy Survey UKIDSS}}/{{UDS}} Field: High-Resolution Dust Continuum Morphologies and the Link between Sub-Millimetre Galaxies and Spheroid Formation},
  shorttitle = {An {{ALMA}} Survey of the {{SCUBA-2 Cosmology Legacy Survey UKIDSS}}/{{UDS}} Field},
  author = {Gullberg, B and Smail, Ian and Swinbank, A M and Dudzevi{\v c}i{\=u}t{\.e}, U and Stach, S M and Thomson, A P and Almaini, O and Chen, C C and Conselice, C and Cooke, E A and Farrah, D and Ivison, R J and Maltby, D and Micha{\l}owski, M J and Simpson, J M and Scott, D and Wardlow, J L and Weiss, A},
  year = {2019},
  month = dec,
  journal = {Monthly Notices of the Royal Astronomical Society},
  volume = {490},
  number = {4},
  pages = {4956--4974},
  issn = {0035-8711, 1365-2966},
  doi = {10.1093/mnras/stz2835},
  urldate = {2025-07-01},
  copyright = {https://academic.oup.com/journals/pages/open\_access/funder\_policies/chorus/standard\_publication\_model},
  langid = {english}
}

@ARTICLE{chen2022b,
       author = {{Chen}, Chian-Chou and {Gao}, Zhen-Kai and {Hsu}, Qi-Ning and {Liao}, Cheng-Lin and {Ling}, Yu-Han and {Lo}, Ching-Min and {Smail}, Ian and {Wang}, Wei-Hao and {Wang}, Yu-Jan},
        title = "{JWST Sneaks a Peek at the Stellar Morphology of z 2 Submillimeter Galaxies: Bulge Formation at Cosmic Noon}",
      journal = {\apjl},
         year = 2022,
        month = nov,
       volume = {939},
       number = {1},
          eid = {L7},
        pages = {L7},
          doi = {10.3847/2041-8213/ac98c6},
archivePrefix = {arXiv},
       eprint = {2208.05296},
 primaryClass = {astro-ph.GA},
       adsurl = {https://ui.adsabs.harvard.edu/abs/2022ApJ...939L...7C}
}

@article{cheng2022,
  title = {Properties of {{Host Galaxies}} of {{Submillimeter Sources}} as {{Revealed}} by {{JWST Early Release Observations}} in {{SMACS J0723}}.3--7327},
  author = {Cheng, Cheng and Yan, Haojing and Huang, Jia-Sheng and Willmer, Christopher N. A. and Ma, Zhiyuan and {Orellana-Gonz{\'a}lez}, Gustavo},
  year = {2022},
  month = sep,
  journal = {The Astrophysical Journal Letters},
  volume = {936},
  number = {2},
  pages = {L19},
  publisher = {The American Astronomical Society},
  issn = {2041-8205},
  doi = {10.3847/2041-8213/ac8d08},
  urldate = {2024-02-27},
  langid = {english}
}

@article{gillman2023,
  title = {Sub-Millimetre Galaxies with {{{\emph{Webb}}}}: {{Near-infrared}} Counterparts and Multi-Wavelength Morphology},
  shorttitle = {Sub-Millimetre Galaxies with {{{\emph{Webb}}}}},
  author = {Gillman, Steven and Gullberg, Bitten and Brammer, Gabe and Vijayan, Aswin P. and Lee, Minju and Bl{\'a}nquez, David and Brinch, Malte and Greve, Thomas R. and Jermann, Iris and Jin, Shuowen and Kokorev, Vasily and Liu, Lijie and Magdis, Georgios and Rizzo, Francesca and Valentino, Francesco},
  year = {2023},
  month = aug,
  journal = {Astronomy \& Astrophysics},
  volume = {676},
  pages = {A26},
  issn = {0004-6361, 1432-0746},
  doi = {10.1051/0004-6361/202346531},
  urldate = {2024-09-13},
  copyright = {https://creativecommons.org/licenses/by/4.0},
  langid = {english}
}

@article{rujopakarn2023,
  title = {{{JWST}} and {{ALMA Imaging}} of {{Dust-obscured}}, {{Massive Substructures}} in a {{Typical}} z {$\sim$} 3 {{Star-forming Disk Galaxy}}},
  author = {Rujopakarn, Wiphu and Williams, Christina C. and Daddi, Emanuele and Schramm, Malte and Sun, Fengwu and Alberts, Stacey and Rieke, George H. and Tan 谈, Qing-Hua 清华 and Tacchella, Sandro and Giavalisco, Mauro and Silverman, John D.},
  year = {2023},
  month = may,
  journal = {The Astrophysical Journal Letters},
  volume = {948},
  number = {1},
  pages = {L8},
  issn = {2041-8205, 2041-8213},
  doi = {10.3847/2041-8213/accc82},
  urldate = {2025-09-05},
  langid = {english}
}

@article{wu2023,
  title = {The {{Identification}} of a {{Dusty Multiarm Spiral Galaxy}} at z = 3.06 with {{JWST}} and {{ALMA}}},
  author = {Wu, Yunjing and Cai, Zheng and Sun, Fengwu and Bian, Fuyan and Lin, Xiaojing and Li, Zihao and Li, Mingyu and Bauer, Franz E. and Egami, Eiichi and Fan, Xiaohui and {Gonz{\'a}lez-L{\'o}pez}, Jorge and Li, Jianan and Wang, Feige and Yang, Jinyi and Zhang, Shiwu and Zou, Siwei},
  year = {2023},
  month = jan,
  journal = {The Astrophysical Journal Letters},
  volume = {942},
  number = {1},
  pages = {L1},
  issn = {2041-8205, 2041-8213},
  doi = {10.3847/2041-8213/aca652},
  urldate = {2025-09-05},
  langid = {english}
}

@article{lebail2024,
  title = {{{JWST}}/{{CEERS}} Sheds Light on Dusty Star-Forming Galaxies: {{Forming}} Bulges, Lopsidedness, and Outside-in Quenching at Cosmic Noon},
  shorttitle = {{{JWST}}/{{CEERS}} Sheds Light on Dusty Star-Forming Galaxies},
  author = {Le Bail, Aur{\'e}lien and Daddi, Emanuele and Elbaz, David and Dickinson, Mark and Giavalisco, Mauro and Magnelli, Benjamin and {G{\'o}mez-Guijarro}, Carlos and Kalita, Boris S. and Koekemoer, Anton M. and Holwerda, Benne W. and Bournaud, Fr{\'e}d{\'e}ric and De La Vega, Alexander and Calabr{\`o}, Antonello and Dekel, Avishai and Cheng, Yingjie and Bisigello, Laura and Franco, Maximilien and Costantin, Luca and Lucas, Ray A. and {P{\'e}rez-Gonz{\'a}lez}, Pablo G. and Lu, Shiying and Wilkins, Stephen M. and Arrabal Haro, Pablo and Bagley, Micaela B. and Finkelstein, Steven L. and Kartaltepe, Jeyhan S. and Papovich, Casey and Pirzkal, Nor and Yung, L. Y. Aaron},
  year = {2024},
  month = aug,
  journal = {Astronomy \& Astrophysics},
  volume = {688},
  pages = {A53},
  issn = {0004-6361, 1432-0746},
  doi = {10.1051/0004-6361/202347465},
  urldate = {2025-09-05},
  copyright = {https://creativecommons.org/licenses/by/4.0},
  langid = {english}
}

@article{mckinney2023,
  title = {A {{Near-infrared-faint}}, {{Far-infrared-luminous Dusty Galaxy}} at z {$\sim$} 5 in {{COSMOS-Web}}},
  author = {McKinney, Jed and Manning, Sinclaire M. and Cooper, Olivia R. and Long, Arianna S. and Akins, Hollis and Casey, Caitlin M. and Faisst, Andreas L. and Franco, Maximilien and Hayward, Christopher C. and Lambrides, Erini and Magdis, Georgios and Whitaker, Katherine E. and Yun, Min and Champagne, Jaclyn B. and Drakos, Nicole E. and Gentile, Fabrizio and Gillman, Steven and Gozaliasl, Ghassem and Ilbert, Olivier and Jin, Shuowen and Koekemoer, Anton M. and Kokorev, Vasily and Liu, Daizhong and Rich, R. Michael and Robertson, Brant E. and Valentino, Francesco and Weaver, John R. and Zavala, Jorge A. and Allen, Natalie and Kartaltepe, Jeyhan S. and McCracken, Henry Joy and Paquereau, Louise and Rhodes, Jason and Shuntov, Marko and Toft, Sune},
  year = {2023},
  month = oct,
  journal = {The Astrophysical Journal},
  volume = {956},
  number = {2},
  pages = {72},
  issn = {0004-637X, 1538-4357},
  doi = {10.3847/1538-4357/acf614},
  urldate = {2025-10-16},
  langid = {english}
}

@ARTICLE{smail2023,
       author = {{Smail}, Ian and {Dudzevi{\v{c}}i{\={u}}t{\.{e}}}, Ugn{\.{e}} and {Gurwell}, Mark and {Fazio}, Giovanni G. and {Willner}, S.~P. and {Swinbank}, A.~M. and {Arumugam}, Vinodiran and {Summers}, Jake and {Cohen}, Seth H. and {Jansen}, Rolf A. and {Windhorst}, Rogier A. and {Meena}, Ashish and {Zitrin}, Adi and {Keel}, William C. and {Cheng}, Cheng and {Coe}, Dan and {Conselice}, Christopher J. and {D'Silva}, Jordan C.~J. and {Driver}, Simon P. and {Frye}, Brenda and {Grogin}, Norman A. and {Koekemoer}, Anton M. and {Marshall}, Madeline A. and {Nonino}, Mario and {Pirzkal}, Nor and {Robotham}, Aaron and {Rutkowski}, Michael J. and {Ryan}, Russell E., Jr. and {Tompkins}, Scott and {Willmer}, Christopher N.~A. and {Yan}, Haojing and {Broadhurst}, Thomas J. and {Diego}, Jos{\'e} M. and {Kamieneski}, Patrick and {Yun}, Min},
        title = "{Hidden Giants in JWST's PEARLS: An Ultramassive z = 4.26 Submillimeter Galaxy that Is Invisible to HST}",
      journal = {\apj},
         year = 2023,
        month = nov,
       volume = {958},
       number = {1},
          eid = {36},
        pages = {36},
          doi = {10.3847/1538-4357/acf931},
archivePrefix = {arXiv},
       eprint = {2306.16039},
 primaryClass = {astro-ph.GA},
       adsurl = {https://ui.adsabs.harvard.edu/abs/2023ApJ...958...36S}
}

@ARTICLE{hodge2025,
       author = {{Hodge}, J.~A. and {da Cunha}, E. and {Kendrew}, S. and {Li}, J. and {Smail}, I. and {Westoby}, B.~A. and {Nayak}, O. and {Swinbank}, A.~M. and {Chen}, C.-C. and {Walter}, F. and {van der Werf}, P. and {Cracraft}, M. and {Battisti}, A. and {Brandt}, W.~N. and {Calistro Rivera}, G. and {Chapman}, S.~C. and {Cox}, P. and {Dannerbauer}, H. and {Decarli}, R. and {Frias Castillo}, M. and {Greve}, T.~R. and {Knudsen}, K.~K. and {Leslie}, S. and {Menten}, K.~M. and {Rybak}, M. and {Schinnerer}, E. and {Wardlow}, J.~L. and {Weiss}, A.},
        title = "{ALESS-JWST: Joint (Sub)kiloparsec JWST and ALMA Imaging of z \raisebox{-0.5ex}\textasciitilde 3 Submillimeter Galaxies Reveals Heavily Obscured Bulge Formation Events}",
      journal = {\apj},
         year = 2025,
        month = jan,
       volume = {978},
       number = {2},
          eid = {165},
        pages = {165},
          doi = {10.3847/1538-4357/ad9a52},
archivePrefix = {arXiv},
       eprint = {2407.15846},
 primaryClass = {astro-ph.GA},
       adsurl = {https://ui.adsabs.harvard.edu/abs/2025ApJ...978..165H}
}

@ARTICLE{weiss2009,
       author = {{Wei{\ss}}, A. and {Kov{\'a}cs}, A. and {Coppin}, K. and {Greve}, T.~R. and {Walter}, F. and {Smail}, Ian and {Dunlop}, J.~S. and {Knudsen}, K.~K. and {Alexander}, D.~M. and {Bertoldi}, F. and {Brandt}, W.~N. and {Chapman}, S.~C. and {Cox}, P. and {Dannerbauer}, H. and {De Breuck}, C. and {Gawiser}, E. and {Ivison}, R.~J. and {Lutz}, D. and {Menten}, K.~M. and {Koekemoer}, A.~M. and {Kreysa}, E. and {Kurczynski}, P. and {Rix}, H. -W. and {Schinnerer}, E. and {van der Werf}, P.~P.},
        title = "{The Large Apex Bolometer Camera Survey of the Extended Chandra Deep Field South}",
      journal = {\apj},
         year = 2009,
        month = dec,
       volume = {707},
       number = {2},
        pages = {1201-1216},
          doi = {10.1088/0004-637X/707/2/1201},
archivePrefix = {arXiv},
       eprint = {0910.2821},
 primaryClass = {astro-ph.CO},
       adsurl = {https://ui.adsabs.harvard.edu/abs/2009ApJ...707.1201W}
}

@article{hodge2012,
  title = {{{EVIDENCE FOR A CLUMPY}}, {{ROTATING GAS DISK IN A SUBMILLIMETER GALAXY AT}} {\emph{z}} = 4},
  author = {Hodge, J. A. and Carilli, C. L. and Walter, F. and De Blok, W. J. G. and Riechers, D. and Daddi, E. and Lentati, L.},
  year = {2012},
  month = nov,
  journal = {The Astrophysical Journal},
  volume = {760},
  number = {1},
  pages = {11},
  issn = {0004-637X, 1538-4357},
  doi = {10.1088/0004-637X/760/1/11},
  urldate = {2025-06-13},
  langid = {english}
}

@software{jwst_pipeline_1.11.3,
       author = {{Bushouse}, Howard and {Eisenhamer}, Jonathan and {Dencheva}, Nadia and {Davies}, James and {Greenfield}, Perry and {Morrison}, Jane and {Hodge}, Phil and {Simon}, Bernie and {Grumm}, David and {Droettboom}, Michael and {Slavich}, Edward and {Sosey}, Megan and {Pauly}, Tyler and {Miller}, Todd and {Jedrzejewski}, Robert and {Hack}, Warren and {Davis}, David and {Crawford}, Steven and {Law}, David and {Gordon}, Karl and {Regan}, Michael and {Cara}, Mihai and {MacDonald}, Ken and {Bradley}, Larry and {Shanahan}, Clare and {Jamieson}, William and {Teodoro}, Mairan and {Williams}, Thomas},
        title = "{JWST Calibration Pipeline}",
         year = 2023,
        month = jul,
          eid = {10.5281/zenodo.8157276},
          doi = {10.5281/zenodo.8157276},
      version = {1.11.3},
    publisher = {Zenodo},
       adsurl = {https://ui.adsabs.harvard.edu/abs/2023zndo...8157276B}

}

@software{jwst_pipeline_1.13.4,
       author = {{Bushouse}, Howard and {Eisenhamer}, Jonathan and {Dencheva}, Nadia and {Davies}, James and {Greenfield}, Perry and {Morrison}, Jane and {Hodge}, Phil and {Simon}, Bernie and {Grumm}, David and {Droettboom}, Michael and {Slavich}, Edward and {Sosey}, Megan and {Pauly}, Tyler and {Miller}, Todd and {Jedrzejewski}, Robert and {Hack}, Warren and {Davis}, David and {Crawford}, Steven and {Law}, David and {Gordon}, Karl and {Regan}, Michael and {Cara}, Mihai and {MacDonald}, Ken and {Bradley}, Larry and {Shanahan}, Clare and {Jamieson}, William and {Teodoro}, Mairan and {Williams}, Thomas and {Pena-Guerrero}, Maria},
        title = "{JWST Calibration Pipeline}",
         year = 2024,
        month = jan,
          eid = {10.5281/zenodo.10569856},
          doi = {10.5281/zenodo.10569856},
      version = {1.13.4},
    publisher = {Zenodo},
       adsurl = {https://ui.adsabs.harvard.edu/abs/2024zndo..10569856B}
}

@ARTICLE{ceers_survey,
       author = {{Bagley}, Micaela B. and {Finkelstein}, Steven L. and {Koekemoer}, Anton M. and {Ferguson}, Henry C. and {Arrabal Haro}, Pablo and {Dickinson}, Mark and {Kartaltepe}, Jeyhan S. and {Papovich}, Casey and {P{\'e}rez-Gonz{\'a}lez}, Pablo G. and {Pirzkal}, Nor and {Somerville}, Rachel S. and {Willmer}, Christopher N.~A. and {Yang}, Guang and {Yung}, L.~Y. Aaron and {Fontana}, Adriano and {Grazian}, Andrea and {Grogin}, Norman A. and {Hirschmann}, Michaela and {Kewley}, Lisa J. and {Kirkpatrick}, Allison and {Kocevski}, Dale D. and {Lotz}, Jennifer M. and {Medrano}, Aubrey and {Morales}, Alexa M. and {Pentericci}, Laura and {Ravindranath}, Swara and {Trump}, Jonathan R. and {Wilkins}, Stephen M. and {Calabr{\`o}}, Antonello and {Cooper}, M.~C. and {Costantin}, Luca and {de la Vega}, Alexander and {Hilbert}, Bryan and {Hutchison}, Taylor A. and {Larson}, Rebecca L. and {Lucas}, Ray A. and {McGrath}, Elizabeth J. and {Ryan}, Russell and {Wang}, Xin and {Wuyts}, Stijn},
        title = "{CEERS Epoch 1 NIRCam Imaging: Reduction Methods and Simulations Enabling Early JWST Science Results}",
      journal = {\apjl},
         year = 2023,
        month = mar,
       volume = {946},
       number = {1},
          eid = {L12},
        pages = {L12},
          doi = {10.3847/2041-8213/acbb08},
archivePrefix = {arXiv},
       eprint = {2211.02495},
 primaryClass = {astro-ph.IM},
       adsurl = {https://ui.adsabs.harvard.edu/abs/2023ApJ...946L..12B}
}

@article{danielson2017,
  title = {An {{ALMA Survey}} of {{Submillimeter Galaxies}} in the {{Extended Chandra Deep Field South}}: {{Spectroscopic Redshifts}}},
  shorttitle = {An {{ALMA Survey}} of {{Submillimeter Galaxies}} in the {{Extended Chandra Deep Field South}}},
  author = {Danielson, A. L. R. and Swinbank, A. M. and Smail, Ian and Simpson, J. M. and Casey, C. M. and Chapman, S. C. and Cunha, E. Da and Hodge, J. A. and Walter, F. and Wardlow, J. L. and Alexander, D. M. and Brandt, W. N. and Breuck, C. De and Coppin, K. E. K. and Dannerbauer, H. and Dickinson, M. and Edge, A. C. and Gawiser, E. and Ivison, R. J. and Karim, A. and Kovacs, A. and Lutz, D. and Menten, K. and Schinnerer, E. and Wei{\ss}, A. and Werf, P. Van Der},
  year = {2017},
  month = may,
  journal = {The Astrophysical Journal},
  volume = {840},
  number = {2},
  pages = {78},
  issn = {0004-637X, 1538-4357},
  doi = {10.3847/1538-4357/aa6caf},
  urldate = {2024-03-07},
  langid = {english}
}

@ARTICLE{birkin2021,
       author = {{Birkin}, Jack E. and {Weiss}, Axel and {Wardlow}, J.~L. and {Smail}, Ian and {Swinbank}, A.~M. and {Dudzevi{\v{c}}i{\={u}}t{\.{e}}}, U. and {An}, Fang Xia and {Ao}, Y. and {Chapman}, S.~C. and {Chen}, Chian-Chou and {da Cunha}, E. and {Dannerbauer}, H. and {Gullberg}, B. and {Hodge}, J.~A. and {Ikarashi}, S. and {Ivison}, R.~J. and {Matsuda}, Y. and {Stach}, S.~M. and {Walter}, F. and {Wang}, W. -H. and {van der Werf}, P.},
        title = "{An ALMA/NOEMA survey of the molecular gas properties of high-redshift star-forming galaxies}",
      journal = {\mnras},
         year = 2021,
        month = mar,
       volume = {501},
       number = {3},
        pages = {3926-3950},
          doi = {10.1093/mnras/staa3862},
archivePrefix = {arXiv},
       eprint = {2009.03341},
 primaryClass = {astro-ph.GA},
       adsurl = {https://ui.adsabs.harvard.edu/abs/2021MNRAS.501.3926B}
}

@article{li2024,
  title = {The {{ALMA-CRISTAL Survey}}: {{Spatially Resolved Star Formation Activity}} and {{Dust Content}} in 4 {$<$} z {$<$} 6 {{Star-forming Galaxies}}},
  shorttitle = {The {{ALMA-CRISTAL Survey}}},
  author = {Li, Juno and Da Cunha, Elisabete and {Gonz{\'a}lez-L{\'o}pez}, Jorge and Aravena, Manuel and De Looze, Ilse and F{\"o}rster Schreiber, N. M. and {Herrera-Camus}, Rodrigo and Spilker, Justin and Tadaki, Ken-ichi and {Barcos-Munoz}, Loreto and Battisti, Andrew J. and Birkin, Jack E. and Bowler, Rebecca A. A. and Davies, Rebecca and {D{\'i}az-Santos}, Tanio and Ferrara, Andrea and Fisher, Deanne B. and Hodge, Jacqueline and Ikeda, Ryota and Killi, Meghana and Lee, Lilian and Liu, Daizhong and Lutz, Dieter and Mitsuhashi, Ikki and Naab, Thorsten and Posses, Ana and Rela{\~n}o, Monica and Solimano, Manuel and {\"U}bler, Hannah and Van Der Giessen, Stefan Anthony and Villanueva, Vicente},
  year = {2024},
  month = nov,
  journal = {The Astrophysical Journal},
  volume = {976},
  number = {1},
  pages = {70},
  issn = {0004-637X, 1538-4357},
  doi = {10.3847/1538-4357/ad7fee},
  urldate = {2025-04-29},
  langid = {english}
}

@ARTICLE{cunha2015,
       author = {{da Cunha}, E. and {Walter}, F. and {Smail}, I.~R. and {Swinbank}, A.~M. and {Simpson}, J.~M. and {Decarli}, R. and {Hodge}, J.~A. and {Weiss}, A. and {van der Werf}, P.~P. and {Bertoldi}, F. and {Chapman}, S.~C. and {Cox}, P. and {Danielson}, A.~L.~R. and {Dannerbauer}, H. and {Greve}, T.~R. and {Ivison}, R.~J. and {Karim}, A. and {Thomson}, A.},
        title = "{An ALMA Survey of Sub-millimeter Galaxies in the Extended Chandra Deep Field South: Physical Properties Derived from Ultraviolet-to-radio Modeling}",
      journal = {\apj},
         year = 2015,
        month = jun,
       volume = {806},
       number = {1},
          eid = {110},
        pages = {110},
          doi = {10.1088/0004-637X/806/1/110},
archivePrefix = {arXiv},
       eprint = {1504.04376},
 primaryClass = {astro-ph.GA},
       adsurl = {https://ui.adsabs.harvard.edu/abs/2015ApJ...806..110D}
}

@ARTICLE{Speagle2014,
       author = {{Speagle}, J.~S. and {Steinhardt}, C.~L. and {Capak}, P.~L. and {Silverman}, J.~D.},
        title = "{A Highly Consistent Framework for the Evolution of the Star-Forming ``Main Sequence'' from z \raisebox{-0.5ex}\textasciitilde 0-6}",
      journal = {\apjs},
         year = 2014,
        month = oct,
       volume = {214},
       number = {2},
          eid = {15},
        pages = {15},
          doi = {10.1088/0067-0049/214/2/15},
archivePrefix = {arXiv},
       eprint = {1405.2041},
 primaryClass = {astro-ph.GA},
       adsurl = {https://ui.adsabs.harvard.edu/abs/2014ApJS..214...15S}
}

@ARTICLE{Cardamone2010,
       author = {{Cardamone}, Carolin N. and {van Dokkum}, Pieter G. and {Urry}, C. Megan and {Taniguchi}, Yoshi and {Gawiser}, Eric and {Brammer}, Gabriel and {Taylor}, Edward and {Damen}, Maaike and {Treister}, Ezequiel and {Cobb}, Bethany E. and {Bond}, Nicholas and {Schawinski}, Kevin and {Lira}, Paulina and {Murayama}, Takashi and {Saito}, Tomoki and {Sumikawa}, Kentaro},
        title = "{The Multiwavelength Survey by Yale-Chile (MUSYC): Deep Medium-band Optical Imaging and High-quality 32-band Photometric Redshifts in the ECDF-S}",
      journal = {\apjs},
         year = 2010,
        month = aug,
       volume = {189},
       number = {2},
        pages = {270-285},
          doi = {10.1088/0067-0049/189/2/270},
archivePrefix = {arXiv},
       eprint = {1008.2974},
 primaryClass = {astro-ph.CO},
       adsurl = {https://ui.adsabs.harvard.edu/abs/2010ApJS..189..270C}
}

@ARTICLE{Boogaard2024,
       author = {{Boogaard}, Leindert A. and {Gillman}, Steven and {Melinder}, Jens and {Walter}, Fabian and {Colina}, Luis and {{\"O}stlin}, G{\"o}ran and {Caputi}, Karina I. and {Iani}, Edoardo and {P{\'e}rez-Gonz{\'a}lez}, Pablo and {van der Werf}, Paul and {Greve}, Thomas R. and {Wright}, Gillian and {Alonso-Herrero}, Almudena and {{\'A}lvarez-M{\'a}rquez}, Javier and {Annunziatella}, Marianna and {Bik}, Arjan and {Bosman}, Sarah and {Costantin}, Luca and {Crespo G{\'o}mez}, Alejandro and {Dicken}, Dan and {Eckart}, Andreas and {Hjorth}, Jens and {Jermann}, Iris and {Labiano}, Alvaro and {Langeroodi}, Danial and {Meyer}, Romain A. and {Moutard}, Thibaud and {Pei{\ss}ker}, Florian and {Pye}, John P. and {Rinaldi}, Pierluigi and {Tikkanen}, Tuomo V. and {Topinka}, Martin and {Henning}, Thomas},
        title = "{MIDIS: JWST/MIRI Reveals the Stellar Structure of ALMA-selected Galaxies in the Hubble Ultra Deep Field at Cosmic Noon}",
      journal = {\apj},
         year = 2024,
        month = jul,
       volume = {969},
       number = {1},
          eid = {27},
        pages = {27},
          doi = {10.3847/1538-4357/ad43e5},
archivePrefix = {arXiv},
       eprint = {2308.16895},
 primaryClass = {astro-ph.GA},
       adsurl = {https://ui.adsabs.harvard.edu/abs/2024ApJ...969...27B}
}

@ARTICLE{Hodge2015,
       author = {{Hodge}, J.~A. and {Riechers}, D. and {Decarli}, R. and {Walter}, F. and {Carilli}, C.~L. and {Daddi}, E. and {Dannerbauer}, H.},
        title = "{The Kiloparsec-scale Star Formation Law at Redshift 4: Widespread, Highly Efficient Star Formation in the Dust-obscured Starburst Galaxy GN20}",
      journal = {\apjl},
         year = 2015,
        month = jan,
       volume = {798},
       number = {1},
          eid = {L18},
        pages = {L18},
          doi = {10.1088/2041-8205/798/1/L18},
archivePrefix = {arXiv},
       eprint = {1412.2132},
 primaryClass = {astro-ph.GA},
       adsurl = {https://ui.adsabs.harvard.edu/abs/2015ApJ...798L..18H}
}

@ARTICLE{Cano-Diaz2016,
       author = {{Cano-D{\'\i}az}, M. and {S{\'a}nchez}, S.~F. and {Zibetti}, S. and {Ascasibar}, Y. and {Bland-Hawthorn}, J. and {Ziegler}, B. and {Gonz{\'a}lez Delgado}, R.~M. and {Walcher}, C.~J. and {Garc{\'\i}a-Benito}, R. and {Mast}, D. and {Mendoza-P{\'e}rez}, M.~A. and {Falc{\'o}n-Barroso}, J. and {Galbany}, L. and {Husemann}, B. and {Kehrig}, C. and {Marino}, R.~A. and {S{\'a}nchez-Bl{\'a}zquez}, P. and {L{\'o}pez-Cob{\'a}}, C. and {L{\'o}pez-S{\'a}nchez}, {\'A}. R. and {Vilchez}, J.~M.},
        title = "{Spatially Resolved Star Formation Main Sequence of Galaxies in the CALIFA Survey}",
      journal = {\apjl},
         year = 2016,
        month = apr,
       volume = {821},
       number = {2},
          eid = {L26},
        pages = {L26},
          doi = {10.3847/2041-8205/821/2/L26},
archivePrefix = {arXiv},
       eprint = {1602.02770},
 primaryClass = {astro-ph.GA},
       adsurl = {https://ui.adsabs.harvard.edu/abs/2016ApJ...821L..26C}
}

@ARTICLE{Abdurrouf2017,
       author = {{Abdurro'uf} and {Akiyama}, Masayuki},
        title = "{Understanding the scatter in the spatially resolved star formation main sequence of local massive spiral galaxies}",
      journal = {\mnras},
         year = 2017,
        month = aug,
       volume = {469},
       number = {3},
        pages = {2806-2820},
          doi = {10.1093/mnras/stx936},
archivePrefix = {arXiv},
       eprint = {1704.04571},
 primaryClass = {astro-ph.GA},
       adsurl = {https://ui.adsabs.harvard.edu/abs/2017MNRAS.469.2806A}
}

@ARTICLE{pessa2021,
       author = {{Pessa}, I. and {Schinnerer}, E. and {Belfiore}, F. and {Emsellem}, E. and {Leroy}, A.~K. and {Schruba}, A. and {Kruijssen}, J.~M.~D. and {Pan}, H. -A. and {Blanc}, G.~A. and {Sanchez-Blazquez}, P. and {Bigiel}, F. and {Chevance}, M. and {Congiu}, E. and {Dale}, D. and {Faesi}, C.~M. and {Glover}, S.~C.~O. and {Grasha}, K. and {Groves}, B. and {Ho}, I. and {Jim{\'e}nez-Donaire}, M. and {Klessen}, R. and {Kreckel}, K. and {Koch}, E.~W. and {Liu}, D. and {Meidt}, S. and {Pety}, J. and {Querejeta}, M. and {Rosolowsky}, E. and {Saito}, T. and {Santoro}, F. and {Sun}, J. and {Usero}, A. and {Watkins}, E.~J. and {Williams}, T.~G.},
        title = "{Star formation scaling relations at {\ensuremath{\sim}}100 pc from PHANGS: Impact of completeness and spatial scale}",
      journal = {\aap},
         year = 2021,
        month = jun,
       volume = {650},
          eid = {A134},
        pages = {A134},
          doi = {10.1051/0004-6361/202140733},
archivePrefix = {arXiv},
       eprint = {2104.09536},
 primaryClass = {astro-ph.GA},
       adsurl = {https://ui.adsabs.harvard.edu/abs/2021A&A...650A.134P}
}

@ARTICLE{Sanchez2021,
       author = {{S{\'a}nchez}, S.~F. and {Barrera-Ballesteros}, J.~K. and {Colombo}, D. and {Wong}, T. and {Bolatto}, A. and {Rosolowsky}, E. and {Vogel}, S. and {Levy}, R. and {Kalinova}, V. and {Alvarez-Hurtado}, P. and {Luo}, Y. and {Cao}, Y.},
        title = "{The EDGE-CALIFA survey: the local and global relations between {\ensuremath{\Sigma}}$_{*}$, {\ensuremath{\Sigma}}$_{SFR}$, and {\ensuremath{\Sigma}}$_{mol}$ that regulate star formation}",
      journal = {\mnras},
         year = 2021,
        month = may,
       volume = {503},
       number = {2},
        pages = {1615-1635},
          doi = {10.1093/mnras/stab442},
archivePrefix = {arXiv},
       eprint = {2102.06226},
 primaryClass = {astro-ph.GA},
       adsurl = {https://ui.adsabs.harvard.edu/abs/2021MNRAS.503.1615S}
}

@ARTICLE{Mun2024,
       author = {{Mun}, Marcie and {Wisnioski}, Emily and {Battisti}, Andrew J. and {Mendel}, J. Trevor and {Ellison}, Sara L. and {Taylor}, Edward N. and {Lagos}, Claudia D.~P. and {Harborne}, Katherine E. and {Foster}, Caroline and {Croom}, Scott M. and {Bellstedt}, Sabine and {Barsanti}, Stefania and {Gupta}, Anshu and {Valenzuela}, Lucas M. and {Chen}, Qian-Hui and {Grasha}, Kathryn and {Mukherjee}, Tamal and {Park}, Hye-Jin and {Sharda}, Piyush and {Sweet}, Sarah M. and {Remus}, Rhea-Silvia and {Zafar}, Tayyaba},
        title = "{The MAGPI survey: evolution of radial trends in star formation activity across cosmic time}",
      journal = {\mnras},
         year = 2024,
        month = jun,
       volume = {530},
       number = {4},
        pages = {5072-5090},
          doi = {10.1093/mnras/stae1132},
archivePrefix = {arXiv},
       eprint = {2404.16319},
 primaryClass = {astro-ph.GA},
       adsurl = {https://ui.adsabs.harvard.edu/abs/2024MNRAS.530.5072M}
}

@ARTICLE{Koller2024,
       author = {{Koller}, M. and {Ziegler}, B. and {Ciocan}, B.~I. and {Thater}, S. and {Mendel}, J.~T. and {Wisnioski}, E. and {Battisti}, A.~J. and {Harborne}, K.~E. and {Foster}, C. and {Lagos}, C. and {Croom}, S.~M. and {Grasha}, K. and {Papaderos}, P. and {Remus}, R.~S. and {Sharma}, G. and {Sweet}, S.~M. and {Valenzuela}, L.~M. and {van de Ven}, G. and {Zafar}, T.},
        title = "{The MAGPI survey: The interdependence of the mass, star formation rate, and metallicity in galaxies at z\raisebox{-0.5ex}\textasciitilde0.3}",
      journal = {arXiv e-prints},
         year = 2024,
        month = jun,
          eid = {arXiv:2406.20017},
        pages = {arXiv:2406.20017},
          doi = {10.48550/arXiv.2406.20017},
archivePrefix = {arXiv},
       eprint = {2406.20017},
 primaryClass = {astro-ph.GA},
       adsurl = {https://ui.adsabs.harvard.edu/abs/2024arXiv240620017K}
}

@ARTICLE{Wuyts2013,
       author = {{Wuyts}, Stijn and {F{\"o}rster Schreiber}, Natascha M. and {Nelson}, Erica J. and {van Dokkum}, Pieter G. and {Brammer}, Gabe and {Chang}, Yu-Yen and {Faber}, Sandra M. and {Ferguson}, Henry C. and {Franx}, Marijn and {Fumagalli}, Mattia and {Genzel}, Reinhard and {Grogin}, Norman A. and {Kocevski}, Dale D. and {Koekemoer}, Anton M. and {Lundgren}, Britt and {Lutz}, Dieter and {McGrath}, Elizabeth J. and {Momcheva}, Ivelina and {Rosario}, David and {Skelton}, Rosalind E. and {Tacconi}, Linda J. and {van der Wel}, Arjen and {Whitaker}, Katherine E.},
        title = "{A CANDELS-3D-HST synergy: Resolved Star Formation Patterns at 0.7 < z < 1.5}",
      journal = {\apj},
         year = 2013,
        month = dec,
       volume = {779},
       number = {2},
          eid = {135},
        pages = {135},
          doi = {10.1088/0004-637X/779/2/135},
archivePrefix = {arXiv},
       eprint = {1310.5702},
 primaryClass = {astro-ph.CO},
       adsurl = {https://ui.adsabs.harvard.edu/abs/2013ApJ...779..135W}
}

@ARTICLE{Baker2022,
       author = {{Baker}, William M. and {Maiolino}, Roberto and {Bluck}, Asa F.~L. and {Lin}, Lihwai and {Ellison}, Sara L. and {Belfiore}, Francesco and {Pan}, Hsi-An and {Thorp}, Mallory},
        title = "{The ALMaQUEST survey IX: the nature of the resolved star forming main sequence}",
      journal = {\mnras},
         year = 2022,
        month = mar,
       volume = {510},
       number = {3},
        pages = {3622-3628},
          doi = {10.1093/mnras/stab3672},
archivePrefix = {arXiv},
       eprint = {2201.03592},
 primaryClass = {astro-ph.GA},
       adsurl = {https://ui.adsabs.harvard.edu/abs/2022MNRAS.510.3622B}
}

@ARTICLE{Sorba2015,
       author = {{Sorba}, R. and {Sawicki}, M.},
        title = "{Missing stellar mass in SED fitting: spatially unresolved photometry can underestimate galaxy masses}",
      journal = {\mnras},
         year = 2015,
        month = sep,
       volume = {452},
       number = {1},
        pages = {235-245},
          doi = {10.1093/mnras/stv1235},
archivePrefix = {arXiv},
       eprint = {1506.01653},
 primaryClass = {astro-ph.GA},
       adsurl = {https://ui.adsabs.harvard.edu/abs/2015MNRAS.452..235S}
}

@ARTICLE{Sorba2018,
       author = {{Sorba}, Robert and {Sawicki}, Marcin},
        title = "{Spatially unresolved SED fitting can underestimate galaxy masses: a solution to the missing mass problem}",
      journal = {\mnras},
         year = 2018,
        month = may,
       volume = {476},
       number = {2},
        pages = {1532-1547},
          doi = {10.1093/mnras/sty186},
archivePrefix = {arXiv},
       eprint = {1801.07368},
 primaryClass = {astro-ph.GA},
       adsurl = {https://ui.adsabs.harvard.edu/abs/2018MNRAS.476.1532S}
}

@ARTICLE{gimenez-arteaga2024,
       author = {{Gim{\'e}nez-Arteaga}, C. and {Fujimoto}, S. and {Valentino}, F. and {Brammer}, G.~B. and {Mason}, C.~A. and {Rizzo}, F. and {Rusakov}, V. and {Colina}, L. and {Prieto-Lyon}, G. and {Oesch}, P.~A. and {Espada}, D. and {Heintz}, K.~E. and {Knudsen}, K.~K. and {Dessauges-Zavadsky}, M. and {Laporte}, N. and {Lee}, M. and {Magdis}, G.~E. and {Ono}, Y. and {Ao}, Y. and {Ouchi}, M. and {Kohno}, K. and {Koekemoer}, A.~M.},
        title = "{Outshining in the spatially resolved analysis of a strongly lensed galaxy at z = 6.072 with JWST NIRCam}",
      journal = {\aap},
         year = 2024,
        month = jun,
       volume = {686},
          eid = {A63},
        pages = {A63},
          doi = {10.1051/0004-6361/202349135},
archivePrefix = {arXiv},
       eprint = {2402.17875},
 primaryClass = {astro-ph.GA},
       adsurl = {https://ui.adsabs.harvard.edu/abs/2024A&A...686A..63G}
}

@ARTICLE{Lines2025,
       author = {{Lines}, N.~E.~P. and {Bowler}, R.~A.~A. and {Adams}, N.~J. and {Fisher}, R. and {Varadaraj}, R.~G. and {Nakazato}, Y. and {Aravena}, M. and {Assef}, R.~J. and {Birkin}, J.~E. and {Ceverino}, D. and {da Cunha}, E. and {Cullen}, F. and {De Looze}, I. and {Donnan}, C.~T. and {Dunlop}, J.~S. and {Ferrara}, A. and {Grogin}, N.~A. and {Herrera-Camus}, R. and {Ikeda}, R. and {Koekemoer}, A.~M. and {Killi}, M. and {Li}, J. and {McLeod}, D.~J. and {McLure}, R.~J. and {Mitsuhashi}, I. and {P{\'e}rez-Gonz{\'a}lez}, P.~G. and {Relano}, M. and {Solimano}, M. and {Spilker}, J.~S. and {Villanueva}, V. and {Yoshida}, N.},
        title = "{JWST PRIMER: a lack of outshining in four normal z = 4 ‑ 6 galaxies from the ALMA-CRISTAL Survey}",
      journal = {\mnras},
         year = 2025,
        month = may,
       volume = {539},
       number = {3},
        pages = {2685-2706},
          doi = {10.1093/mnras/staf627},
archivePrefix = {arXiv},
       eprint = {2409.10963},
 primaryClass = {astro-ph.GA},
       adsurl = {https://ui.adsabs.harvard.edu/abs/2025MNRAS.539.2685L}
}

@article{harvey2025,
  title = {Behind the {{Spotlight}}: {{A}} Systematic Assessment of Outshining Using {{NIRCam}} Medium-Bands in the {{JADES Origins Field}}},
  shorttitle = {Behind the {{Spotlight}}},
  author = {Harvey, Thomas and Conselice, Christopher J. and Adams, Nathan J. and Austin, Duncan and Li, Qiong and Rusakov, Vadim and Westcott, Lewi and Goolsby, Caio M. and Lovell, Christopher C. and Cochrane, Rachel K. and Vijayan, Aswin P. and Trussler, James},
  year = {2025},
  month = sep,
  journal = {Monthly Notices of the Royal Astronomical Society},
  volume = {542},
  number = {4},
  eprint = {2504.05244},
  primaryclass = {astro-ph},
  pages = {2998--3027},
  issn = {0035-8711, 1365-2966},
  doi = {10.1093/mnras/staf1396},
  urldate = {2025-09-23},
  archiveprefix = {arXiv}
}

@ARTICLE{Cochrane_2021,
       author = {{Cochrane}, R.~K. and {Best}, P.~N. and {Smail}, I. and {Ibar}, E. and {Cheng}, C. and {Swinbank}, A.~M. and {Molina}, J. and {Sobral}, D. and {Dudzevi{\v{c}}i{\={u}}t{\.{e}}}, U.},
        title = "{Resolving a dusty, star-forming SHiZELS galaxy at z = 2.2 with HST, ALMA, and SINFONI on kiloparsec scales}",
      journal = {\mnras},
         year = 2021,
        month = may,
       volume = {503},
       number = {2},
        pages = {2622-2638},
          doi = {10.1093/mnras/stab467},
archivePrefix = {arXiv},
       eprint = {2102.07791},
 primaryClass = {astro-ph.GA},
       adsurl = {https://ui.adsabs.harvard.edu/abs/2021MNRAS.503.2622C}
}

@ARTICLE{inami2022,
       author = {{Inami}, Hanae and {Algera}, Hiddo S.~B. and {Schouws}, Sander and {Sommovigo}, Laura and {Bouwens}, Rychard and {Smit}, Renske and {Stefanon}, Mauro and {Bowler}, Rebecca A.~A. and {Endsley}, Ryan and {Ferrara}, Andrea and {Oesch}, Pascal and {Stark}, Daniel and {Aravena}, Manuel and {Barrufet}, Laia and {da Cunha}, Elisabete and {Dayal}, Pratika and {De Looze}, Ilse and {Fudamoto}, Yoshinobu and {Gonzalez}, Valentino and {Graziani}, Luca and {Hodge}, Jacqueline A. and {Hygate}, Alexander P.~S. and {Nanayakkara}, Themiya and {Pallottini}, Andrea and {Riechers}, Dominik A. and {Schneider}, Raffaella and {Topping}, Michael and {van der Werf}, Paul},
        title = "{The ALMA REBELS Survey: dust continuum detections at z > 6.5}",
      journal = {\mnras},
         year = 2022,
        month = sep,
       volume = {515},
       number = {3},
        pages = {3126-3143},
          doi = {10.1093/mnras/stac1779},
archivePrefix = {arXiv},
       eprint = {2203.15136},
 primaryClass = {astro-ph.GA},
       adsurl = {https://ui.adsabs.harvard.edu/abs/2022MNRAS.515.3126I}
}

@ARTICLE{Killi2024,
       author = {{Killi}, Meghana and {Ginolfi}, Michele and {Popping}, Gerg{\"o} and {Watson}, Darach and {Zamorani}, Giovanni and {Lemaux}, Brian C. and {Fujimoto}, Seiji and {Faisst}, Andreas and {Bethermin}, Matthieu and {Romano}, Michael and {Fudamoto}, Yoshinobu and {Bardelli}, Sandro and {Boquien}, M{\'e}d{\'e}ric and {Carniani}, Stefano and {Dessauges-Zavadsky}, Miroslava and {Gruppioni}, Carlotta and {Hathi}, Nimish and {Ibar}, Eduardo and {Jones}, Gareth C. and {Koekemoer}, Anton M. and {Langan}, Ivanna and {M{\'e}ndez-Hern{\'a}ndez}, Hugo and {Sugahara}, Yuma and {Vallini}, Livia and {Vergani}, Daniela},
        title = "{The ALPINE-ALMA [C II] survey: Characterisation of Spatial Offsets in Main-Sequence Galaxies at $z \sim$ 4-6}",
      journal = {arXiv e-prints},
         year = 2024,
        month = feb,
          eid = {arXiv:2402.07982},
        pages = {arXiv:2402.07982},
          doi = {10.48550/arXiv.2402.07982},
archivePrefix = {arXiv},
       eprint = {2402.07982},
 primaryClass = {astro-ph.GA},
       adsurl = {https://ui.adsabs.harvard.edu/abs/2024arXiv240207982K}
}

@ARTICLE{CalistroRivera_2018,
       author = {{Calistro Rivera}, Gabriela and {Hodge}, J.~A. and {Smail}, Ian and {Swinbank}, A.~M. and {Weiss}, A. and {Wardlow}, J.~L. and {Walter}, F. and {Rybak}, M. and {Chen}, Chian-Chou and {Brandt}, W.~N. and {Coppin}, K. and {da Cunha}, E. and {Dannerbauer}, H. and {Greve}, T.~R. and {Karim}, A. and {Knudsen}, K.~K. and {Schinnerer}, E. and {Simpson}, J.~M. and {Venemans}, B. and {van der Werf}, P.~P.},
        title = "{Resolving the ISM at the Peak of Cosmic Star Formation with ALMA: The Distribution of CO and Dust Continuum in z {\ensuremath{\sim}} 2.5 Submillimeter Galaxies}",
      journal = {\apj},
         year = 2018,
        month = aug,
       volume = {863},
       number = {1},
          eid = {56},
        pages = {56},
          doi = {10.3847/1538-4357/aacffa},
archivePrefix = {arXiv},
       eprint = {1804.06852},
 primaryClass = {astro-ph.GA},
       adsurl = {https://ui.adsabs.harvard.edu/abs/2018ApJ...863...56C}
}

@ARTICLE{Cheng_2020,
       author = {{Cheng}, Cheng and {Ibar}, Edo and {Smail}, Ian and {Molina}, Juan and {Sobral}, David and {Escala}, Andr{\'e}s and {Best}, Philip and {Cochrane}, Rachel and {Gillman}, Steven and {Swinbank}, Mark and {Ivison}, R.~J. and {Huang}, Jia-Sheng and {Hughes}, Thomas M. and {Villard}, Eric and {Cirasuolo}, Michele},
        title = "{A kpc-scale-resolved study of unobscured and obscured star formation activity in normal galaxies at z = 1.5 and 2.2 from ALMA and HiZELS}",
      journal = {\mnras},
         year = 2020,
        month = dec,
       volume = {499},
       number = {4},
        pages = {5241-5256},
          doi = {10.1093/mnras/staa3036},
archivePrefix = {arXiv},
       eprint = {2011.00686},
 primaryClass = {astro-ph.GA},
       adsurl = {https://ui.adsabs.harvard.edu/abs/2020MNRAS.499.5241C}
}

@article{cochrane2019,
	title = {Predictions for the spatial distribution of the dust continuum emission in \${\textbackslash}boldsymbol \{1{\textbackslash},{\textbackslash}lt{\textbackslash}, z{\textbackslash},{\textbackslash}lt{\textbackslash}, 5\}\$ star-forming galaxies},
	volume = {488},
	issn = {0035-8711, 1365-2966},
	url = {https://academic.oup.com/mnras/article/488/2/1779/5523144},
	doi = {10.1093/mnras/stz1736},
	language = {en},
	number = {2},
	urldate = {2022-11-29},
	journal = {Monthly Notices of the Royal Astronomical Society},
	author = {Cochrane, R K and Hayward, C C and Anglés-Alcázar, D and Lotz, J and Parsotan, T and Ma, X and Kereš, D and Feldmann, R and Faucher-Giguère, C A and Hopkins, P F},
	month = sep,
	year = {2019},
	pages = {1779--1789},
}

@ARTICLE{popping2022,
       author = {{Popping}, Gerg{\"o} and {Pillepich}, Annalisa and {Calistro Rivera}, Gabriela and {Schulz}, Sebastian and {Hernquist}, Lars and {Kaasinen}, Melanie and {Marinacci}, Federico and {Nelson}, Dylan and {Vogelsberger}, Mark},
        title = "{The dust-continuum size of TNG50 galaxies at z = 1-5: a comparison with the distribution of stellar light, stars, dust, and H$_{2}$}",
      journal = {\mnras},
         year = 2022,
        month = mar,
       volume = {510},
       number = {3},
        pages = {3321-3334},
          doi = {10.1093/mnras/stab3312},
archivePrefix = {arXiv},
       eprint = {2101.12218},
 primaryClass = {astro-ph.GA},
       adsurl = {https://ui.adsabs.harvard.edu/abs/2022MNRAS.510.3321P}
}

@ARTICLE{Bershady2000,
       author = {{Bershady}, Matthew A. and {Jangren}, Anna and {Conselice}, Christopher J.},
        title = "{Structural and Photometric Classification of Galaxies. I. Calibration Based on a Nearby Galaxy Sample}",
      journal = {\aj},
         year = 2000,
        month = jun,
       volume = {119},
       number = {6},
        pages = {2645-2663},
          doi = {10.1086/301386},
archivePrefix = {arXiv},
       eprint = {astro-ph/0002262},
 primaryClass = {astro-ph},
       adsurl = {https://ui.adsabs.harvard.edu/abs/2000AJ....119.2645B}
}

@ARTICLE{Conselice2003,
       author = {{Conselice}, Christopher J. and {Chapman}, Scott C. and {Windhorst}, Rogier A.},
        title = "{Evidence for a Major Merger Origin of High-Redshift Submillimeter Galaxies}",
      journal = {\apjl},
         year = 2003,
        month = oct,
       volume = {596},
       number = {1},
        pages = {L5-L8},
          doi = {10.1086/379109},
archivePrefix = {arXiv},
       eprint = {astro-ph/0308198},
 primaryClass = {astro-ph},
       adsurl = {https://ui.adsabs.harvard.edu/abs/2003ApJ...596L...5C}
}

@ARTICLE{Conselice2014,
       author = {{Conselice}, Christopher J},
        title = "{The Evolution of Galaxy Structure Over Cosmic Time}",
      journal = {\araa},
         year = 2014,
        month = aug,
       volume = {52},
        pages = {291-337},
          doi = {10.1146/annurev-astro-081913-040037},
archivePrefix = {arXiv},
       eprint = {1403.2783},
 primaryClass = {astro-ph.GA},
       adsurl = {https://ui.adsabs.harvard.edu/abs/2014ARA&A..52..291C}
}

@ARTICLE{Lotz2004,
       author = {{Lotz}, Jennifer M. and {Primack}, Joel and {Madau}, Piero},
        title = "{A New Nonparametric Approach to Galaxy Morphological Classification}",
      journal = {\aj},
         year = 2004,
        month = jul,
       volume = {128},
       number = {1},
        pages = {163-182},
          doi = {10.1086/421849},
archivePrefix = {arXiv},
       eprint = {astro-ph/0311352},
 primaryClass = {astro-ph},
       adsurl = {https://ui.adsabs.harvard.edu/abs/2004AJ....128..163L}
}

@ARTICLE{Lotz2008,
       author = {{Lotz}, Jennifer M. and {Davis}, M. and {Faber}, S.~M. and {Guhathakurta}, P. and {Gwyn}, S. and {Huang}, J. and {Koo}, D.~C. and {Le Floc'h}, E. and {Lin}, Lihwai and {Newman}, J. and {Noeske}, K. and {Papovich}, C. and {Willmer}, C.~N.~A. and {Coil}, A. and {Conselice}, C.~J. and {Cooper}, M. and {Hopkins}, A.~M. and {Metevier}, A. and {Primack}, J. and {Rieke}, G. and {Weiner}, B.~J.},
        title = "{The Evolution of Galaxy Mergers and Morphology at z < 1.2 in the Extended Groth Strip}",
      journal = {\apj},
         year = 2008,
        month = jan,
       volume = {672},
       number = {1},
        pages = {177-197},
          doi = {10.1086/523659},
archivePrefix = {arXiv},
       eprint = {astro-ph/0602088},
 primaryClass = {astro-ph},
       adsurl = {https://ui.adsabs.harvard.edu/abs/2008ApJ...672..177L}
}

@ARTICLE{erwin2015,
       author = {{Erwin}, Peter},
        title = "{IMFIT: A Fast, Flexible New Program for Astronomical Image Fitting}",
      journal = {\apj},
         year = 2015,
        month = feb,
       volume = {799},
       number = {2},
          eid = {226},
        pages = {226},
          doi = {10.1088/0004-637X/799/2/226},
archivePrefix = {arXiv},
       eprint = {1408.1097},
 primaryClass = {astro-ph.IM},
       adsurl = {https://ui.adsabs.harvard.edu/abs/2015ApJ...799..226E}
}

@ARTICLE{Baes2024,
       author = {{Baes}, Maarten and {Mosenkov}, Aleksandr and {Kelly}, Raymond and {Abdurro'uf} and {Andreadis}, Nick and {Bokona Tulu}, Sena and {Camps}, Peter and {Tassama Emana}, Abdissa and {Fritz}, Jacopo and {Gebek}, Andrea and {Kova{\v{c}}i{\'c}}, Inja and {La Marca}, Antonio and {Martorano}, Marco and {Nersesian}, Angelos and {Rodriguez-Gomez}, Vicente and {Tortora}, Crescenzo and {Tr{\v{c}}ka}, Ana and {Vander Meulen}, Bert and {van der Wel}, Arjen and {Wang}, Lingyu},
        title = "{The TNG50-SKIRT Atlas: Wavelength dependence of the effective radius}",
      journal = {\aap},
         year = 2024,
        month = mar,
       volume = {683},
          eid = {A182},
        pages = {A182},
          doi = {10.1051/0004-6361/202348419},
archivePrefix = {arXiv},
       eprint = {2401.04225},
 primaryClass = {astro-ph.GA},
       adsurl = {https://ui.adsabs.harvard.edu/abs/2024A&A...683A.182B}
}

@article{kokorev2023a,
  title = {\${{JWST}}\$ {{Insight Into}} a {{Lensed}} \${{HST}}\$-Dark {{Galaxy}} and Its {{Quiescent Companion}} at \$z=2.58\$},
  author = {Kokorev, Vasily and Jin, Shuowen and Magdis, Georgios E. and Caputi, Karina I. and Valentino, Francesco and Dayal, Pratika and Trebitsch, Maxime and Brammer, Gabriel and Fujimoto, Seiji and Bauer, Franz and Iani, Edoardo and Kohno, Kotaro and Sese, David Blanquez and {G{\'o}mez-Guijarro}, Carlos and Rinaldi, Pierluigi and {Navarro-Carrera}, Rafael},
  year = {2023},
  month = mar,
  journal = {The Astrophysical Journal Letters},
  volume = {945},
  number = {2},
  eprint = {2301.04158},
  primaryclass = {astro-ph},
  pages = {L25},
  issn = {2041-8205, 2041-8213},
  doi = {10.3847/2041-8213/acbd9d},
  urldate = {2024-08-30},
  archiveprefix = {arXiv},
  langid = {english}
}

@article{sun2024,
  title = {{{JADES}}: {{Resolving}} the {{Stellar Component}} and {{Filamentary Overdense Environment}} of {{Hubble Space Telescope}} ({{HST}})-Dark {{Submillimeter Galaxy HDF850}}.1 at z = 5.18},
  shorttitle = {{{JADES}}},
  author = {Sun, Fengwu and Helton, Jakob M. and Egami, Eiichi and Hainline, Kevin N. and Rieke, George H. and Willmer, Christopher N. A. and Eisenstein, Daniel J. and Johnson, Benjamin D. and Rieke, Marcia J. and Robertson, Brant and Tacchella, Sandro and Alberts, Stacey and Baker, William M. and Bhatawdekar, Rachana and Boyett, Kristan and Bunker, Andrew J. and Charlot, Stephane and Chen, Zuyi and Chevallard, Jacopo and {Curtis-Lake}, Emma and Danhaive, A. Lola and DeCoursey, Christa and Ji, Zhiyuan and Lyu, Jianwei and Maiolino, Roberto and Rujopakarn, Wiphu and Sandles, Lester and Shivaei, Irene and {\"U}bler, Hannah and Willott, Chris and Witstok, Joris},
  year = {2024},
  month = jan,
  journal = {The Astrophysical Journal},
  volume = {961},
  number = {1},
  pages = {69},
  issn = {0004-637X, 1538-4357},
  doi = {10.3847/1538-4357/ad07e3},
  urldate = {2025-09-05},
  langid = {english}
}

@ARTICLE{bodansky2025,
       author = {{Bodansky}, Sarah and {Whitaker}, Katherine E. and {Abdullah}, Ayesha and {Lin}, Jamie and {Oesch}, Pascal A. and {Pope}, Alexandra and {Xiao}, Mengyuan and {Covelo-Paz}, Alba and {Cutler}, Sam and {Garcia Diaz}, Carlos and {Lee}, Minju M. and {Manning}, Sinclaire M. and {Meyer}, Romain A. and {Narayanan}, Desika and {Nelson}, Erica and {Shivaei}, Irene and {van Dokkum}, Pieter},
        title = "{JWST+ALMA reveal the build up of stellar mass in the cores of dusty star-forming galaxies at Cosmic Noon}",
      journal = {arXiv e-prints},
         year = 2025,
        month = jul,
          eid = {arXiv:2507.19472},
        pages = {arXiv:2507.19472},
          doi = {10.48550/arXiv.2507.19472},
archivePrefix = {arXiv},
       eprint = {2507.19472},
 primaryClass = {astro-ph.GA},
       adsurl = {https://ui.adsabs.harvard.edu/abs/2025arXiv250719472B}
}

@ARTICLE{price2025,
       author = {{Price}, Sedona H. and {Suess}, Katherine A. and {Williams}, Christina C. and {Bezanson}, Rachel and {Khullar}, Gourav and {Nelson}, Erica J. and {Wang}, Bingjie and {Weaver}, John R. and {Fujimoto}, Seiji and {Kokorev}, Vasily and {Greene}, Jenny E. and {Brammer}, Gabriel and {Cutler}, Sam E. and {Dayal}, Pratika and {Furtak}, Lukas J. and {Labbe}, Ivo and {Leja}, Joel and {Miller}, Tim B. and {Nanayakkara}, Themiya and {Pan}, Richard and {Whitaker}, Katherine E.},
        title = "{UNCOVER: The Rest-ultraviolet to Near-infrared Multiwavelength Structures and Dust Distributions of Submillimeter-detected Galaxies in A2744}",
      journal = {\apj},
         year = 2025,
        month = feb,
       volume = {980},
       number = {1},
          eid = {11},
        pages = {11},
          doi = {10.3847/1538-4357/ada0b1},
archivePrefix = {arXiv},
       eprint = {2310.02500},
 primaryClass = {astro-ph.GA},
       adsurl = {https://ui.adsabs.harvard.edu/abs/2025ApJ...980...11P}
}

@ARTICLE{chan2025,
       author = {{Chan}, Siu-Wang and {Ao}, Yiping and {Tan}, Qinghua},
        title = "{Disclosing Submillimeter Galaxy Formation: Mergers or Secular Evolution?}",
      journal = {arXiv e-prints},
         year = 2025,
        month = sep,
          eid = {arXiv:2509.07913},
        pages = {arXiv:2509.07913},
          doi = {10.48550/arXiv.2509.07913},
archivePrefix = {arXiv},
       eprint = {2509.07913},
 primaryClass = {astro-ph.GA},
       adsurl = {https://ui.adsabs.harvard.edu/abs/2025arXiv250907913C}
}

@article{cheng2023,
  title = {{{JWST}}'s {{PEARLS}}: {{A JWST}}/{{NIRCam View}} of {{ALMA Sources}}},
  shorttitle = {{{JWST}}'s {{PEARLS}}},
  author = {Cheng, Cheng and Huang, Jia-Sheng and Smail, Ian and Yan, Haojing and Cohen, Seth H. and Jansen, Rolf A. and Windhorst, Rogier A. and Ma, Zhiyuan and Koekemoer, Anton and Willmer, Christopher N. A. and Willner, S. P. and Diego, Jose M. and Frye, Brenda and Conselice, Christopher J. and Ferreira, Leonardo and Petric, Andreea and Yun, Min and Gim, Hansung B. and Polletta, Maria del Carmen and Duncan, Kenneth J. and Holwerda, Benne W. and R{\"o}ttgering, Huub J. A. and Honor, Rachel and Hathi, Nimish P. and Kamieneski, Patrick S. and Adams, Nathan J. and Coe, Dan and Broadhurst, Tom and Summers, Jake and Tompkins, Scott and Driver, Simon P. and Grogin, Norman A. and Marshall, Madeline A. and Pirzkal, Nor and Robotham, Aaron and Ryan, Russell E.},
  year = {2023},
  month = jan,
  journal = {The Astrophysical Journal Letters},
  volume = {942},
  number = {1},
  pages = {L19},
  publisher = {The American Astronomical Society},
  issn = {2041-8205},
  doi = {10.3847/2041-8213/aca9d0},
  urldate = {2024-02-27},
  langid = {english}
}

@article{lang2019,
  title = {Revealing the {{Stellar Mass}} and {{Dust Distributions}} of {{Submillimeter Galaxies}} at {{Redshift}} 2},
  author = {Lang, Philipp and Schinnerer, E. and Smail, Ian and Dudzevi{\v c}i{\=u}t{\.e}, U. and Swinbank, A. M. and Liu, Daizhong and Leslie, S. K. and Almaini, O. and An, Fang Xia and Bertoldi, F. and Blain, A. W. and Chapman, S. C. and Chen, Chian-Chou and Conselice, C. and Cooke, E. A. and Coppin, K. E. K. and Dunlop, J. S. and Farrah, D. and Fudamoto, Y. and Geach, J. E. and Gullberg, B. and Harrington, K. C. and Hodge, J. A. and Ivison, R. J. and {Jim{\'e}nez-Andrade}, E. F. and Magnelli, B. and Micha{\l}owski, M. J. and Oesch, P. and Scott, D. and Simpson, J. M. and Smol{\v c}i{\'c}, V. and Stach, S. M. and Thomson, A. P. and Toft, S. and Vardoulaki, E. and Wardlow, J. L. and Weiss, A. and Werf, P. Van Der},
  year = {2019},
  month = jul,
  journal = {The Astrophysical Journal},
  volume = {879},
  number = {1},
  pages = {54},
  issn = {0004-637X, 1538-4357},
  doi = {10.3847/1538-4357/ab1f77},
  urldate = {2024-05-23},
  langid = {english}
}

@ARTICLE{sun2021,
       author = {{Sun}, Fengwu and {Egami}, Eiichi and {Rawle}, Timothy D. and {Walth}, Gregory L. and {Smail}, Ian and {Dessauges-Zavadsky}, Miroslava and {P{\'e}rez-Gonz{\'a}lez}, Pablo G. and {Richard}, Johan and {Combes}, Francoise and {Ebeling}, Harald and {Pell{\'o}}, Roser and {Van der Werf}, Paul and {Altieri}, Bruno and {Boone}, Fr{\'e}d{\'e}ric and {Cava}, Antonio and {Chapman}, Scott C. and {Cl{\'e}ment}, Benjamin and {Finoguenov}, Alexis and {Nakajima}, Kimihiko and {Rujopakarn}, Wiphu and {Schaerer}, Daniel and {Valtchanov}, Ivan},
        title = "{ALMA 1.3 mm Survey of Lensed Submillimeter Galaxies Selected by Herschel: Discovery of Spatially Extended SMGs and Implications}",
      journal = {\apj},
         year = 2021,
        month = feb,
       volume = {908},
       number = {2},
          eid = {192},
        pages = {192},
          doi = {10.3847/1538-4357/abd6e4},
archivePrefix = {arXiv},
       eprint = {2101.03677},
 primaryClass = {astro-ph.GA},
       adsurl = {https://ui.adsabs.harvard.edu/abs/2021ApJ...908..192S}
}

@ARTICLE{Tadaki2020ApJ,
       author = {{Tadaki}, Ken-ichi and {Belli}, Sirio and {Burkert}, Andreas and {Dekel}, Avishai and {F{\"o}rster Schreiber}, Natascha M. and {Genzel}, Reinhard and {Hayashi}, Masao and {Herrera-Camus}, Rodrigo and {Kodama}, Tadayuki and {Kohno}, Kotaro and {Koyama}, Yusei and {Lee}, Minju M. and {Lutz}, Dieter and {Mowla}, Lamiya and {Nelson}, Erica J. and {Renzini}, Alvio and {Suzuki}, Tomoko L. and {Tacconi}, Linda J. and {{\"U}bler}, Hannah and {Wisnioski}, Emily and {Wuyts}, Stijn},
        title = "{Structural Evolution in Massive Galaxies at z {\ensuremath{\sim}} 2}",
      journal = {\apj},
         year = 2020,
        month = sep,
       volume = {901},
       number = {1},
          eid = {74},
        pages = {74},
          doi = {10.3847/1538-4357/abaf4a},
archivePrefix = {arXiv},
       eprint = {2009.01976},
 primaryClass = {astro-ph.GA},
       adsurl = {https://ui.adsabs.harvard.edu/abs/2020ApJ...901...74T}
}

@ARTICLE{Barro2017,
       author = {{Barro}, Guillermo and {Faber}, S.~M. and {Koo}, David C. and {Dekel}, Avishai and {Fang}, Jerome J. and {Trump}, Jonathan R. and {P{\'e}rez-Gonz{\'a}lez}, Pablo G. and {Pacifici}, Camilla and {Primack}, Joel R. and {Somerville}, Rachel S. and {Yan}, Haojing and {Guo}, Yicheng and {Liu}, Fengshan and {Ceverino}, Daniel and {Kocevski}, Dale D. and {McGrath}, Elizabeth},
        title = "{Structural and Star-forming Relations since z {\ensuremath{\sim}} 3: Connecting Compact Star-forming and Quiescent Galaxies}",
      journal = {\apj},
         year = 2017,
        month = may,
       volume = {840},
       number = {1},
          eid = {47},
        pages = {47},
          doi = {10.3847/1538-4357/aa6b05},
archivePrefix = {arXiv},
       eprint = {1509.00469},
 primaryClass = {astro-ph.GA},
       adsurl = {https://ui.adsabs.harvard.edu/abs/2017ApJ...840...47B}
}

@ARTICLE{Hopkins2009,
       author = {{Hopkins}, Philip F. and {Bundy}, Kevin and {Murray}, Norman and {Quataert}, Eliot and {Lauer}, Tod R. and {Ma}, Chung-Pei},
        title = "{Compact high-redshift galaxies are the cores of the most massive present-day spheroids}",
      journal = {\mnras},
         year = 2009,
        month = sep,
       volume = {398},
       number = {2},
        pages = {898-910},
          doi = {10.1111/j.1365-2966.2009.15062.x},
archivePrefix = {arXiv},
       eprint = {0903.2479},
 primaryClass = {astro-ph.CO},
       adsurl = {https://ui.adsabs.harvard.edu/abs/2009MNRAS.398..898H}
}

@ARTICLE{Ito2024,
       author = {{Ito}, Kei and {Valentino}, Francesco and {Brammer}, Gabriel and {Faisst}, Andreas L. and {Gillman}, Steven and {G{\'o}mez-Guijarro}, Carlos and {Gould}, Katriona M.~L. and {Heintz}, Kasper E. and {Ilbert}, Olivier and {Jespersen}, Christian Kragh and {Kokorev}, Vasily and {Kubo}, Mariko and {Magdis}, Georgios E. and {McPartland}, Conor J.~R. and {Onodera}, Masato and {Rizzo}, Francesca and {Tanaka}, Masayuki and {Toft}, Sune and {Vijayan}, Aswin P. and {Weaver}, John R. and {Whitaker}, Katherine E. and {Wright}, Lillian},
        title = "{Size{\textendash}Stellar Mass Relation and Morphology of Quiescent Galaxies at z {\ensuremath{\geq}} 3 in Public JWST Fields}",
      journal = {\apj},
         year = 2024,
        month = apr,
       volume = {964},
       number = {2},
          eid = {192},
        pages = {192},
          doi = {10.3847/1538-4357/ad2512},
archivePrefix = {arXiv},
       eprint = {2307.06994},
 primaryClass = {astro-ph.GA},
       adsurl = {https://ui.adsabs.harvard.edu/abs/2024ApJ...964..192I}
}

@ARTICLE{Lustig2021,
       author = {{Lustig}, Peter and {Strazzullo}, Veronica and {D'Eugenio}, Chiara and {Daddi}, Emanuele and {Pannella}, Maurilio and {Renzini}, Alvio and {Cimatti}, Andrea and {Gobat}, Raphael and {Jin}, Shuowen and {Mohr}, Joseph J. and {Onodera}, Masato},
        title = "{Compact, bulge-dominated structures of spectroscopically confirmed quiescent galaxies at z {\ensuremath{\approx}} 3}",
      journal = {\mnras},
         year = 2021,
        month = feb,
       volume = {501},
       number = {2},
        pages = {2659-2676},
          doi = {10.1093/mnras/staa3766},
archivePrefix = {arXiv},
       eprint = {2012.02766},
 primaryClass = {astro-ph.GA},
       adsurl = {https://ui.adsabs.harvard.edu/abs/2021MNRAS.501.2659L}
}

@ARTICLE{Naab2009,
       author = {{Naab}, Thorsten and {Johansson}, Peter H. and {Ostriker}, Jeremiah P.},
        title = "{Minor Mergers and the Size Evolution of Elliptical Galaxies}",
      journal = {\apjl},
         year = 2009,
        month = jul,
       volume = {699},
       number = {2},
        pages = {L178-L182},
          doi = {10.1088/0004-637X/699/2/L178},
archivePrefix = {arXiv},
       eprint = {0903.1636},
 primaryClass = {astro-ph.CO},
       adsurl = {https://ui.adsabs.harvard.edu/abs/2009ApJ...699L.178N}
}

@ARTICLE{Trujillo2006,
       author = {{Trujillo}, I. and {Feulner}, G. and {Goranova}, Y. and {Hopp}, U. and {Longhetti}, M. and {Saracco}, P. and {Bender}, R. and {Braito}, V. and {Della Ceca}, R. and {Drory}, N. and {Mannucci}, F. and {Severgnini}, P.},
        title = "{Extremely compact massive galaxies at z \raisebox{-0.5ex}\textasciitilde 1.4}",
      journal = {\mnras},
         year = 2006,
        month = nov,
       volume = {373},
       number = {1},
        pages = {L36-L40},
          doi = {10.1111/j.1745-3933.2006.00238.x},
archivePrefix = {arXiv},
       eprint = {astro-ph/0608657},
 primaryClass = {astro-ph},
       adsurl = {https://ui.adsabs.harvard.edu/abs/2006MNRAS.373L..36T}
}

@ARTICLE{Trujillo2007,
       author = {{Trujillo}, Ignacio and {Conselice}, C.~J. and {Bundy}, Kevin and {Cooper}, M.~C. and {Eisenhardt}, P. and {Ellis}, Richard S.},
        title = "{Strong size evolution of the most massive galaxies since z {\ensuremath{\sim}} 2}",
      journal = {\mnras},
         year = 2007,
        month = nov,
       volume = {382},
       number = {1},
        pages = {109-120},
          doi = {10.1111/j.1365-2966.2007.12388.x},
archivePrefix = {arXiv},
       eprint = {0709.0621},
 primaryClass = {astro-ph},
       adsurl = {https://ui.adsabs.harvard.edu/abs/2007MNRAS.382..109T}
}

@ARTICLE{vanDokkum2014,
       author = {{van Dokkum}, Pieter G. and {Bezanson}, Rachel and {van der Wel}, Arjen and {Nelson}, Erica June and {Momcheva}, Ivelina and {Skelton}, Rosalind E. and {Whitaker}, Katherine E. and {Brammer}, Gabriel and {Conroy}, Charlie and {F{\"o}rster Schreiber}, Natascha M. and {Fumagalli}, Mattia and {Kriek}, Mariska and {Labb{\'e}}, Ivo and {Leja}, Joel and {Marchesini}, Danilo and {Muzzin}, Adam and {Oesch}, Pascal and {Wuyts}, Stijn},
        title = "{Dense Cores in Galaxies Out to z = 2.5 in SDSS, UltraVISTA, and the Five 3D-HST/CANDELS Fields}",
      journal = {\apj},
         year = 2014,
        month = aug,
       volume = {791},
       number = {1},
          eid = {45},
        pages = {45},
          doi = {10.1088/0004-637X/791/1/45},
archivePrefix = {arXiv},
       eprint = {1404.4874},
 primaryClass = {astro-ph.GA},
       adsurl = {https://ui.adsabs.harvard.edu/abs/2014ApJ...791...45V}
}

@ARTICLE{Gomez-Guijarro_2018,
       author = {{G{\'o}mez-Guijarro}, C. and {Toft}, S. and {Karim}, A. and {Magnelli}, B. and {Magdis}, G.~E. and {Jim{\'e}nez-Andrade}, E.~F. and {Capak}, P.~L. and {Fraternali}, F. and {Fujimoto}, S. and {Riechers}, D.~A. and {Schinnerer}, E. and {Smol{\v{c}}i{\'c}}, V. and {Aravena}, M. and {Bertoldi}, F. and {Cortzen}, I. and {Hasinger}, G. and {Hu}, E.~M. and {Jones}, G.~C. and {Koekemoer}, A.~M. and {Lee}, N. and {McCracken}, H.~J. and {Micha{\l}owski}, M.~J. and {Navarrete}, F. and {Povi{\'c}}, M. and {Puglisi}, A. and {Romano-D{\'\i}az}, E. and {Sheth}, K. and {Silverman}, J.~D. and {Staguhn}, J. and {Steinhardt}, C.~L. and {Stockmann}, M. and {Tanaka}, M. and {Valentino}, F. and {van Kampen}, E. and {Zirm}, A.},
        title = "{Starburst to Quiescent from HST/ALMA: Stars and Dust Unveil Minor Mergers in Submillimeter Galaxies at z {\ensuremath{\sim}} 4.5}",
      journal = {\apj},
         year = 2018,
        month = apr,
       volume = {856},
       number = {2},
          eid = {121},
        pages = {121},
          doi = {10.3847/1538-4357/aab206},
archivePrefix = {arXiv},
       eprint = {1802.07751},
 primaryClass = {astro-ph.GA},
       adsurl = {https://ui.adsabs.harvard.edu/abs/2018ApJ...856..121G}
}

@ARTICLE{Newman2012,
       author = {{Newman}, Andrew B. and {Ellis}, Richard S. and {Bundy}, Kevin and {Treu}, Tommaso},
        title = "{Can Minor Merging Account for the Size Growth of Quiescent Galaxies? New Results from the CANDELS Survey}",
      journal = {\apj},
         year = 2012,
        month = feb,
       volume = {746},
       number = {2},
          eid = {162},
        pages = {162},
          doi = {10.1088/0004-637X/746/2/162},
archivePrefix = {arXiv},
       eprint = {1110.1637},
 primaryClass = {astro-ph.CO},
       adsurl = {https://ui.adsabs.harvard.edu/abs/2012ApJ...746..162N}
}

@ARTICLE{vanderWel2014,
       author = {{van der Wel}, A. and {Franx}, M. and {van Dokkum}, P.~G. and {Skelton}, R.~E. and {Momcheva}, I.~G. and {Whitaker}, K.~E. and {Brammer}, G.~B. and {Bell}, E.~F. and {Rix}, H. -W. and {Wuyts}, S. and {Ferguson}, H.~C. and {Holden}, B.~P. and {Barro}, G. and {Koekemoer}, A.~M. and {Chang}, Yu-Yen and {McGrath}, E.~J. and {H{\"a}ussler}, B. and {Dekel}, A. and {Behroozi}, P. and {Fumagalli}, M. and {Leja}, J. and {Lundgren}, B.~F. and {Maseda}, M.~V. and {Nelson}, E.~J. and {Wake}, D.~A. and {Patel}, S.~G. and {Labb{\'e}}, I. and {Faber}, S.~M. and {Grogin}, N.~A. and {Kocevski}, D.~D.},
        title = "{3D-HST+CANDELS: The Evolution of the Galaxy Size-Mass Distribution since z = 3}",
      journal = {\apj},
         year = 2014,
        month = jun,
       volume = {788},
       number = {1},
          eid = {28},
        pages = {28},
          doi = {10.1088/0004-637X/788/1/28},
archivePrefix = {arXiv},
       eprint = {1404.2844},
 primaryClass = {astro-ph.GA},
       adsurl = {https://ui.adsabs.harvard.edu/abs/2014ApJ...788...28V}
}

@ARTICLE{Longhetti2007,
       author = {{Longhetti}, M. and {Saracco}, P. and {Severgnini}, P. and {Della Ceca}, R. and {Mannucci}, F. and {Bender}, R. and {Drory}, N. and {Feulner}, G. and {Hopp}, U.},
        title = "{The Kormendy relation of massive elliptical galaxies at z \raisebox{-0.5ex}\textasciitilde 1.5: evidence for size evolution}",
      journal = {\mnras},
         year = 2007,
        month = jan,
       volume = {374},
       number = {2},
        pages = {614-626},
          doi = {10.1111/j.1365-2966.2006.11171.x},
archivePrefix = {arXiv},
       eprint = {astro-ph/0610241},
 primaryClass = {astro-ph},
       adsurl = {https://ui.adsabs.harvard.edu/abs/2007MNRAS.374..614L}
}

@ARTICLE{Martorano2024,
       author = {{Martorano}, Marco and {van der Wel}, Arjen and {Baes}, Maarten and {Bell}, Eric F. and {Brammer}, Gabriel and {Franx}, Marijn and {Nersesian}, Angelos},
        title = "{The Size{\textendash}Mass Relation at Rest-frame 1.5 {\ensuremath{\mu}}m from JWST/NIRCam in the COSMOS-WEB and PRIMER-COSMOS Fields}",
      journal = {\apj},
         year = 2024,
        month = sep,
       volume = {972},
       number = {2},
          eid = {134},
        pages = {134},
          doi = {10.3847/1538-4357/ad5c6a},
archivePrefix = {arXiv},
       eprint = {2406.17756},
 primaryClass = {astro-ph.GA},
       adsurl = {https://ui.adsabs.harvard.edu/abs/2024ApJ...972..134M}
}

@ARTICLE{Cappellari2011,
       author = {{Cappellari}, Michele and {Emsellem}, Eric and {Krajnovi{\'c}}, Davor and {McDermid}, Richard M. and {Scott}, Nicholas and {Verdoes Kleijn}, G.~A. and {Young}, Lisa M. and {Alatalo}, Katherine and {Bacon}, R. and {Blitz}, Leo and {Bois}, Maxime and {Bournaud}, Fr{\'e}d{\'e}ric and {Bureau}, M. and {Davies}, Roger L. and {Davis}, Timothy A. and {de Zeeuw}, P.~T. and {Duc}, Pierre-Alain and {Khochfar}, Sadegh and {Kuntschner}, Harald and {Lablanche}, Pierre-Yves and {Morganti}, Raffaella and {Naab}, Thorsten and {Oosterloo}, Tom and {Sarzi}, Marc and {Serra}, Paolo and {Weijmans}, Anne-Marie},
        title = "{The ATLAS$^{3D}$ project - I. A volume-limited sample of 260 nearby early-type galaxies: science goals and selection criteria}",
      journal = {\mnras},
         year = 2011,
        month = may,
       volume = {413},
       number = {2},
        pages = {813-836},
          doi = {10.1111/j.1365-2966.2010.18174.x},
archivePrefix = {arXiv},
       eprint = {1012.1551},
 primaryClass = {astro-ph.CO},
       adsurl = {https://ui.adsabs.harvard.edu/abs/2011MNRAS.413..813C}
}

@ARTICLE{Cappellari2013,
       author = {{Cappellari}, Michele and {McDermid}, Richard M. and {Alatalo}, Katherine and {Blitz}, Leo and {Bois}, Maxime and {Bournaud}, Fr{\'e}d{\'e}ric and {Bureau}, M. and {Crocker}, Alison F. and {Davies}, Roger L. and {Davis}, Timothy A. and {de Zeeuw}, P.~T. and {Duc}, Pierre-Alain and {Emsellem}, Eric and {Khochfar}, Sadegh and {Krajnovi{\'c}}, Davor and {Kuntschner}, Harald and {Morganti}, Raffaella and {Naab}, Thorsten and {Oosterloo}, Tom and {Sarzi}, Marc and {Scott}, Nicholas and {Serra}, Paolo and {Weijmans}, Anne-Marie and {Young}, Lisa M.},
        title = "{The ATLAS$^{3D}$ project - XX. Mass-size and mass-{\ensuremath{\sigma}} distributions of early-type galaxies: bulge fraction drives kinematics, mass-to-light ratio, molecular gas fraction and stellar initial mass function}",
      journal = {\mnras},
         year = 2013,
        month = jul,
       volume = {432},
       number = {3},
        pages = {1862-1893},
          doi = {10.1093/mnras/stt644},
archivePrefix = {arXiv},
       eprint = {1208.3523},
 primaryClass = {astro-ph.CO},
       adsurl = {https://ui.adsabs.harvard.edu/abs/2013MNRAS.432.1862C}
}

@ARTICLE{Koekemoer2011,
       author = {{Koekemoer}, Anton M. and {Faber}, S.~M. and {Ferguson}, Henry C. and {Grogin}, Norman A. and {Kocevski}, Dale D. and {Koo}, David C. and {Lai}, Kamson and {Lotz}, Jennifer M. and {Lucas}, Ray A. and {McGrath}, Elizabeth J. and {Ogaz}, Sara and {Rajan}, Abhijith and {Riess}, Adam G. and {Rodney}, Steve A. and {Strolger}, Louis and {Casertano}, Stefano and {Castellano}, Marco and {Dahlen}, Tomas and {Dickinson}, Mark and {Dolch}, Timothy and {Fontana}, Adriano and {Giavalisco}, Mauro and {Grazian}, Andrea and {Guo}, Yicheng and {Hathi}, Nimish P. and {Huang}, Kuang-Han and {van der Wel}, Arjen and {Yan}, Hao-Jing and {Acquaviva}, Viviana and {Alexander}, David M. and {Almaini}, Omar and {Ashby}, Matthew L.~N. and {Barden}, Marco and {Bell}, Eric F. and {Bournaud}, Fr{\'e}d{\'e}ric and {Brown}, Thomas M. and {Caputi}, Karina I. and {Cassata}, Paolo and {Challis}, Peter J. and {Chary}, Ranga-Ram and {Cheung}, Edmond and {Cirasuolo}, Michele and {Conselice}, Christopher J. and {Roshan Cooray}, Asantha and {Croton}, Darren J. and {Daddi}, Emanuele and {Dav{\'e}}, Romeel and {de Mello}, Duilia F. and {de Ravel}, Loic and {Dekel}, Avishai and {Donley}, Jennifer L. and {Dunlop}, James S. and {Dutton}, Aaron A. and {Elbaz}, David and {Fazio}, Giovanni G. and {Filippenko}, Alexei V. and {Finkelstein}, Steven L. and {Frazer}, Chris and {Gardner}, Jonathan P. and {Garnavich}, Peter M. and {Gawiser}, Eric and {Gruetzbauch}, Ruth and {Hartley}, Will G. and {H{\"a}ussler}, Boris and {Herrington}, Jessica and {Hopkins}, Philip F. and {Huang}, Jia-Sheng and {Jha}, Saurabh W. and {Johnson}, Andrew and {Kartaltepe}, Jeyhan S. and {Khostovan}, Ali A. and {Kirshner}, Robert P. and {Lani}, Caterina and {Lee}, Kyoung-Soo and {Li}, Weidong and {Madau}, Piero and {McCarthy}, Patrick J. and {McIntosh}, Daniel H. and {McLure}, Ross J. and {McPartland}, Conor and {Mobasher}, Bahram and {Moreira}, Heidi and {Mortlock}, Alice and {Moustakas}, Leonidas A. and {Mozena}, Mark and {Nandra}, Kirpal and {Newman}, Jeffrey A. and {Nielsen}, Jennifer L. and {Niemi}, Sami and {Noeske}, Kai G. and {Papovich}, Casey J. and {Pentericci}, Laura and {Pope}, Alexandra and {Primack}, Joel R. and {Ravindranath}, Swara and {Reddy}, Naveen A. and {Renzini}, Alvio and {Rix}, Hans-Walter and {Robaina}, Aday R. and {Rosario}, David J. and {Rosati}, Piero and {Salimbeni}, Sara and {Scarlata}, Claudia and {Siana}, Brian and {Simard}, Luc and {Smidt}, Joseph and {Snyder}, Diana and {Somerville}, Rachel S. and {Spinrad}, Hyron and {Straughn}, Amber N. and {Telford}, Olivia and {Teplitz}, Harry I. and {Trump}, Jonathan R. and {Vargas}, Carlos and {Villforth}, Carolin and {Wagner}, Cory R. and {Wandro}, Pat and {Wechsler}, Risa H. and {Weiner}, Benjamin J. and {Wiklind}, Tommy and {Wild}, Vivienne and {Wilson}, Grant and {Wuyts}, Stijn and {Yun}, Min S.},
        title = "{CANDELS: The Cosmic Assembly Near-infrared Deep Extragalactic Legacy Survey{\textemdash}The Hubble Space Telescope Observations, Imaging Data Products, and Mosaics}",
      journal = {\apjs},
         year = 2011,
        month = dec,
       volume = {197},
       number = {2},
          eid = {36},
        pages = {36},
          doi = {10.1088/0067-0049/197/2/36},
archivePrefix = {arXiv},
       eprint = {1105.3754},
 primaryClass = {astro-ph.CO},
       adsurl = {https://ui.adsabs.harvard.edu/abs/2011ApJS..197...36K}
}

@ARTICLE{Dudze2020,
       author = {{Dudzevi{\v{c}}i{\={u}}t{\.{e}}}, U. and {Smail}, Ian and {Swinbank}, A.~M. and {Stach}, S.~M. and {Almaini}, O. and {da Cunha}, E. and {An}, Fang Xia and {Arumugam}, V. and {Birkin}, J. and {Blain}, A.~W. and {Chapman}, S.~C. and {Chen}, C.-C. and {Conselice}, C.~J. and {Coppin}, K.~E.~K. and {Dunlop}, J.~S. and {Farrah}, D. and {Geach}, J.~E. and {Gullberg}, B. and {Hartley}, W.~G. and {Hodge}, J.~A. and {Ivison}, R.~J. and {Maltby}, D.~T. and {Scott}, D. and {Simpson}, C.~J. and {Simpson}, J.~M. and {Thomson}, A.~P. and {Walter}, F. and {Wardlow}, J.~L. and {Weiss}, A. and {van der Werf}, P.},
        title = "{An ALMA survey of the SCUBA-2 CLS UDS field: physical properties of 707 sub-millimetre galaxies}",
      journal = {\mnras},
         year = 2020,
        month = may,
       volume = {494},
       number = {3},
        pages = {3828-3860},
          doi = {10.1093/mnras/staa769},
archivePrefix = {arXiv},
       eprint = {1910.07524},
 primaryClass = {astro-ph.GA},
       adsurl = {https://ui.adsabs.harvard.edu/abs/2020MNRAS.494.3828D}
}

@ARTICLE{Chabrier2003,
       author = {{Chabrier}, Gilles},
        title = "{Galactic Stellar and Substellar Initial Mass Function}",
      journal = {\pasp},
         year = 2003,
        month = jul,
       volume = {115},
       number = {809},
        pages = {763-795},
          doi = {10.1086/376392},
archivePrefix = {arXiv},
       eprint = {astro-ph/0304382},
 primaryClass = {astro-ph},
       adsurl = {https://ui.adsabs.harvard.edu/abs/2003PASP..115..763C}
}

@article{wang2013,
  title = {{{AN ALMA SURVEY OF SUBMILLIMETER GALAXIES IN THE EXTENDED CHANDRA DEEP FIELD-SOUTH}}: {{THE AGN FRACTION AND X-RAY PROPERTIES OF SUBMILLIMETER GALAXIES}}},
  shorttitle = {{{AN ALMA SURVEY OF SUBMILLIMETER GALAXIES IN THE EXTENDED CHANDRA DEEP FIELD-SOUTH}}},
  author = {Wang, S. X. and Brandt, W. N. and Luo, B. and Smail, I. and Alexander, D. M. and Danielson, A. L. R. and Hodge, J. A. and Karim, A. and Lehmer, B. D. and Simpson, J. M. and Swinbank, A. M. and Walter, F. and Wardlow, J. L. and Xue, Y. Q. and Chapman, S. C. and Coppin, K. E. K. and Dannerbauer, H. and De Breuck, C. and Menten, K. M. and Van Der Werf, P.},
  year = 2013,
  month = nov,
  journal = {The Astrophysical Journal},
  volume = {778},
  number = {2},
  pages = {179},
  issn = {0004-637X, 1538-4357},
  doi = {10.1088/0004-637X/778/2/179},
  urldate = {2024-04-03},
  copyright = {http://iopscience.iop.org/info/page/text-and-data-mining},
  langid = {english}
}

@ARTICLE{Popesso2023,
       author = {{Popesso}, P. and {Concas}, A. and {Cresci}, G. and {Belli}, S. and {Rodighiero}, G. and {Inami}, H. and {Dickinson}, M. and {Ilbert}, O. and {Pannella}, M. and {Elbaz}, D.},
        title = "{The main sequence of star-forming galaxies across cosmic times}",
      journal = {\mnras},
         year = 2023,
        month = feb,
       volume = {519},
       number = {1},
        pages = {1526-1544},
          doi = {10.1093/mnras/stac3214},
archivePrefix = {arXiv},
       eprint = {2203.10487},
 primaryClass = {astro-ph.GA},
       adsurl = {https://ui.adsabs.harvard.edu/abs/2023MNRAS.519.1526P}
}

@ARTICLE{Gomez2024,
       author = {{Crespo G{\'o}mez}, A. and {Colina}, L. and {{\'A}lvarez-M{\'a}rquez}, J. and {Bik}, A. and {Boogaard}, L. and {{\"O}stlin}, G. and {Pei{\ss}ker}, F. and {Walter}, F. and {Labiano}, A. and {P{\'e}rez-Gonz{\'a}lez}, P.~G. and {Greve}, T.~R. and {Wright}, G. and {Alonso-Herrero}, A. and {Caputi}, K.~I. and {Costantin}, L. and {Eckart}, A. and {Garc{\'\i}a-Mar{\'\i}n}, M. and {Gillman}, S. and {Hjorth}, J. and {Iani}, E. and {Langeroodi}, D. and {Pye}, J.~P. and {Rinaldi}, P. and {Tikkanen}, T. and {van der Werf}, P. and {Lagage}, P.~O. and {van Dishoeck}, E.~F.},
        title = "{JWST/MIRI unveils the stellar component of the GN20 dusty galaxy overdensity at z = 4.05}",
      journal = {\aap},
         year = 2024,
        month = nov,
       volume = {691},
          eid = {A325},
        pages = {A325},
          doi = {10.1051/0004-6361/202449750},
archivePrefix = {arXiv},
       eprint = {2402.18672},
 primaryClass = {astro-ph.GA},
       adsurl = {https://ui.adsabs.harvard.edu/abs/2024A&A...691A.325C}
}

@ARTICLE{McKinney2025,
       author = {{McKinney}, Jed and {Casey}, Caitlin M. and {Long}, Arianna S. and {Cooper}, Olivia R. and {Manning}, Sinclaire M. and {Franco}, Maximilien and {Akins}, Hollis and {Lambrides}, Erini and {Gammon}, Elaine and {Silva}, Camila and {Gentile}, Fabrizio and {Zavala}, Jorge A. and {Amvrosiadis}, Aristeidis and {Andika}, Irham and {Brinch}, Malte and {Champagne}, Jaclyn B. and {Chartab}, Nima and {Drakos}, Nicole E. and {Faisst}, Andreas L. and {Fujimoto}, Seiji and {Gillman}, Steven and {Gozaliasl}, Ghassem and {Greve}, Thomas R. and {Harish}, Santosh and {Hayward}, Christopher C. and {Hirschmann}, Michaela and {Ilbert}, Olivier and {Kalita}, Boris S. and {Kartaltepe}, Jeyhan S. and {Koekemoer}, Anton M. and {Kokorev}, Vasily and {Liu}, Daizhong and {Magdis}, Georgios and {McCracken}, Henry Joy and {Rhodes}, Jason and {Robertson}, Brant E. and {Talia}, Margherita and {Valentino}, Francesco and {Vijayan}, Aswin P.},
        title = "{SCUBADive. I. JWST+ALMA Analysis of 289 Submillimeter Galaxies in COSMOS-web}",
      journal = {\apj},
         year = 2025,
        month = feb,
       volume = {979},
       number = {2},
          eid = {229},
        pages = {229},
          doi = {10.3847/1538-4357/ada357},
archivePrefix = {arXiv},
       eprint = {2408.08346},
 primaryClass = {astro-ph.GA},
       adsurl = {https://ui.adsabs.harvard.edu/abs/2025ApJ...979..229M}
}

@ARTICLE{Umehata2026,
       author = {{Umehata}, Hideki and {Kubo}, Mariko and {Smail}, Ian and {Lehmer}, Bret D. and {Monson}, Erik B. and {Nakanishi}, Kouichiro and {Matsuda}, Yuichi},
        title = "{ADF22-WEB: ALMA and JWST (Sub)kiloparsec-scale Views of Dusty Star-forming Galaxies in a z ≍ 3 Protocluster}",
      journal = {\apj},
         year = 2026,
        month = jan,
       volume = {997},
       number = {1},
          eid = {79},
        pages = {79},
          doi = {10.3847/1538-4357/ae1a84},
archivePrefix = {arXiv},
       eprint = {2502.01868},
 primaryClass = {astro-ph.GA},
       adsurl = {https://ui.adsabs.harvard.edu/abs/2026ApJ...997...79U}
}

@ARTICLE{Ikeda2026,
       author = {{Ikeda}, Ryota and {Iono}, Daisuke and {Tadaki}, Ken-ichi and {Franco}, Maximilien and {Yun}, Min S. and {Zavala}, Jorge A. and {Tamura}, Yoichi and {Tsukui}, Takafumi and {Williams}, Christina C. and {Hatsukade}, Bunyo and {Lee}, Minju M. and {Michiyama}, Tomonari and {Mitsuhashi}, Ikki and {Nakanishi}, Kouichiro and {Casey}, Caitlin M. and {Ikarashi}, Soh and {Lee}, Kianhong and {Matsuda}, Yuichi and {Saito}, Toshiki and {Silva}, Andrea and {Umehata}, Hideki and {Yajima}, Hidenobu},
        title = "{Formation of Substructure in Luminous Submillimeter Galaxies (FOSSILS): Evidence of Multiple Pathways to Trigger Starbursts in Luminous Submillimeter Galaxies}",
      journal = {\apj},
         year = 2026,
        month = jan,
       volume = {996},
       number = {2},
          eid = {121},
        pages = {121},
          doi = {10.3847/1538-4357/ae157e},
archivePrefix = {arXiv},
       eprint = {2510.18006},
 primaryClass = {astro-ph.GA},
       adsurl = {https://ui.adsabs.harvard.edu/abs/2026ApJ...996..121I}
}

@ARTICLE{Gaia2023,
       author = {{Gaia Collaboration} and {Vallenari}, A. and {Brown}, A.~G.~A. and {Prusti}, T. and {de Bruijne}, J.~H.~J. and {Arenou}, F. and {Babusiaux}, C. and {Biermann}, M. and {Creevey}, O.~L. and {Evans}, D.~W. and et al.},
        title = "{Gaia Data Release 3: Summary of the content and survey properties}",
      journal = {\aap},
         year = 2023,
        month = jun,
       volume = {674},
          eid = {A1},
        pages = {A1},
          doi = {10.1051/0004-6361/202243940},
       adsurl = {https://ui.adsabs.harvard.edu/abs/2023A&A...674A...1G},
}

@INPROCEEDINGS{perrin2014,
       author = {{Perrin}, Marshall D. and {Sivaramakrishnan}, Anand and {Lajoie}, Charles-Philippe and {Elliott}, Erin and {Pueyo}, Laurent and {Ravindranath}, Swara and {Albert}, Lo{\"\i}c.},
        title = "{Updated point spread function simulations for JWST with WebbPSF}",
    booktitle = {Space Telescopes and Instrumentation 2014: Optical, Infrared, and Millimeter Wave},
         year = 2014,
       editor = {{Oschmann}, Jacobus M., Jr. and {Clampin}, Mark and {Fazio}, Giovanni G. and {MacEwen}, Howard A.},
       series = {Society of Photo-Optical Instrumentation Engineers (SPIE) Conference Series},
       volume = {9143},
        month = aug,
          eid = {91433X},
        pages = {91433X},
          doi = {10.1117/12.2056689},
       adsurl = {https://ui.adsabs.harvard.edu/abs/2014SPIE.9143E..3XP}
}

@software{photutils2022,
author       = {Larry Bradley and
                Brigitta Sipőcz and
                Thomas Robitaille and
                Erik Tollerud and
                Zé Vinícius and
                Christoph Deil and
                Kyle Barbary and
                Tom J Wilson and
                Ivo Busko and
                Axel Donath and
                Hans Moritz Günther and
                Mihai Cara and
                P. L. Lim and
                Sebastian Meßlinger and
                Simon Conseil and
                Azalee Bostroem and
                Michael Droettboom and
                E. M. Bray and
                Lars Andersen Bratholm and
                Geert Barentsen and
                Matt Craig and
                Shivangee Rathi and
                Sergio Pascual and
                Gabriel Perren and
                Iskren Y. Georgiev and
                Miguel de Val-Borro and
                Wolfgang Kerzendorf and
                Yoonsoo P. Bach and
                Bruno Quint and
                Harrison Souchereau},
title        = {astropy/photutils: 1.5.0},
month        = jul,
year         = 2022,
publisher    = {Zenodo},
version      = {1.5.0},
doi          = {10.5281/zenodo.6825092},
url          = {https://doi.org/10.5281/zenodo.6825092}
}

@ARTICLE{Simpson2020,
       author = {{Simpson}, J.~M. and {Smail}, Ian and {Dudzevi{\v{c}}i{\={u}}t{\.{e}}}, U. and {Matsuda}, Y. and {Hsieh}, B.-C. and {Wang}, W.-H. and {Swinbank}, A.~M. and {Stach}, S.~M. and {An}, Fang Xia and {Birkin}, J.~E. and {Ao}, Y. and {Bunker}, A.~J. and {Chapman}, S.~C. and {Chen}, Chian-Chou and {Coppin}, K.~E.~K. and {Ikarashi}, S. and {Ivison}, R.~J. and {Mitsuhashi}, I. and {Saito}, T. and {Umehata}, H. and {Wang}, R. and {Zhao}, Y.},
        title = "{An ALMA survey of the brightest sub-millimetre sources in the SCUBA-2-COSMOS field}",
      journal = {\mnras},
         year = 2020,
        month = jul,
       volume = {495},
       number = {3},
        pages = {3409-3430},
          doi = {10.1093/mnras/staa1345},
archivePrefix = {arXiv},
       eprint = {2003.05484},
 primaryClass = {astro-ph.GA},
       adsurl = {https://ui.adsabs.harvard.edu/abs/2020MNRAS.495.3409S}
}

@ARTICLE{Swinbank2010,
       author = {{Swinbank}, A.~M. and {Smail}, Ian and {Chapman}, S.~C. and {Borys}, C. and {Alexander}, D.~M. and {Blain}, A.~W. and {Conselice}, C.~J. and {Hainline}, L.~J. and {Ivison}, R.~J.},
        title = "{A Hubble Space Telescope NICMOS and ACS morphological study of z \raisebox{-0.5ex}\textasciitilde 2 submillimetre galaxies}",
      journal = {\mnras},
         year = 2010,
        month = jun,
       volume = {405},
       number = {1},
        pages = {234-244},
          doi = {10.1111/j.1365-2966.2010.16485.x},
archivePrefix = {arXiv},
       eprint = {1002.2518},
 primaryClass = {astro-ph.CO},
       adsurl = {https://ui.adsabs.harvard.edu/abs/2010MNRAS.405..234S}
}
\bibliographystyle{aasjournalv7}

\appendix

\section{JWST and ALMA imaging of our sources.}
\label{appendix_images}

In Fig.\,\ref{fig:montage}, we present the JWST/NIRCam and MIRI imaging of our sources (with superimposed ALMA contours).

\begin{figure*}
\centering
    \includegraphics[width=0.95\linewidth]{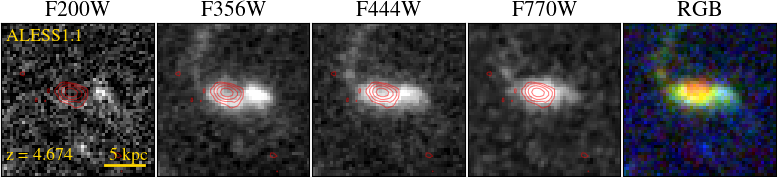}
    \includegraphics[width=0.95\linewidth]{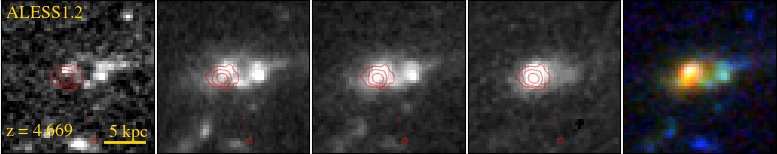}
    \includegraphics[width=0.95\linewidth]{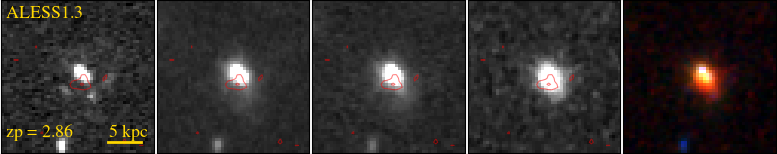}
    \includegraphics[width=0.95\linewidth]{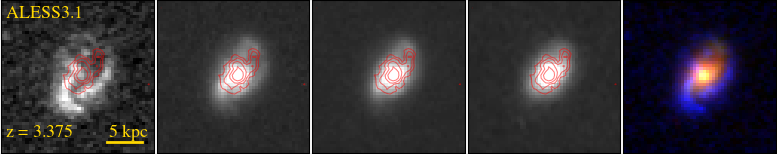}
    \includegraphics[width=0.95\linewidth]{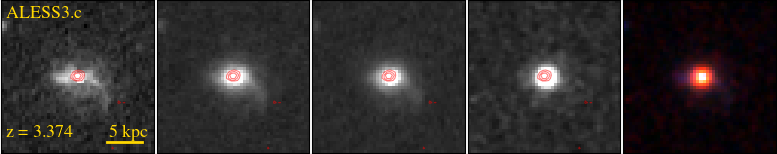}
    \includegraphics[width=0.95\linewidth]{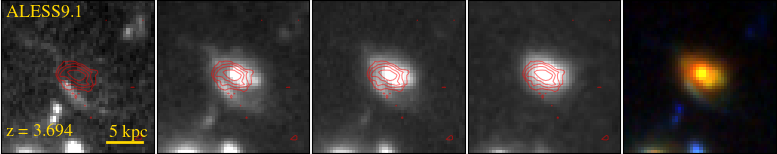}
\caption{JWST images of our SMGs (including the ALESS~3.1 companion). For each galaxy, columns show F200W, F356W, F444W, F770W, and an RGB composite (B: F200W, G: F444W, R: F770W; linear stretch). Redshifts are indicated in the RGB panel, with `zp' marking photometric values. Red contours show ALMA 870\um\ continuum emission, starting at $3\sigma$ and increasing by factors of two up to $24\sigma$. \label{fig:montage}}
\end{figure*}
\addtocounter{figure}{-1}
\begin{figure*}
\centering
    \includegraphics[width=0.95\linewidth]{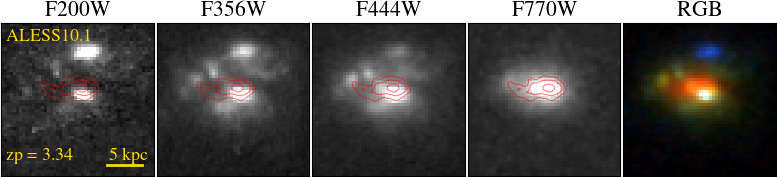}
    \includegraphics[width=0.95\linewidth]{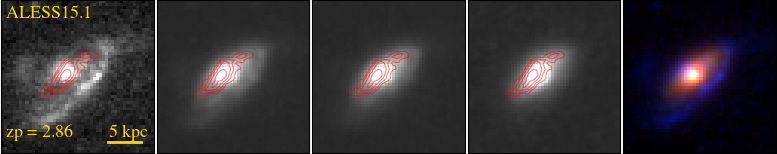}
    \includegraphics[width=0.95\linewidth]{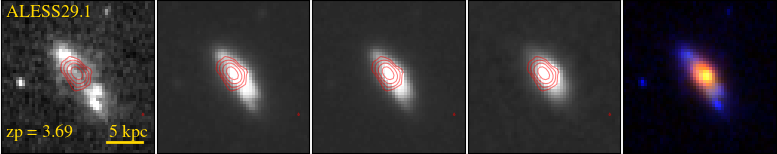}
    \includegraphics[width=0.95\linewidth]{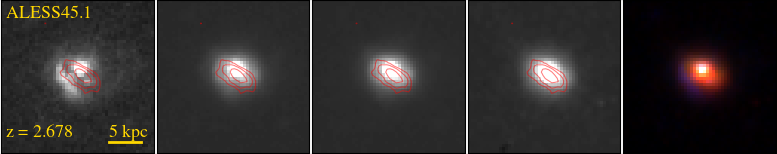}
    \includegraphics[width=0.95\linewidth]{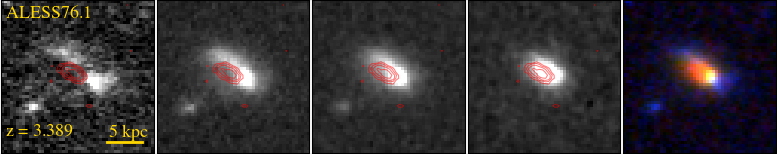}
    \includegraphics[width=0.95\linewidth]{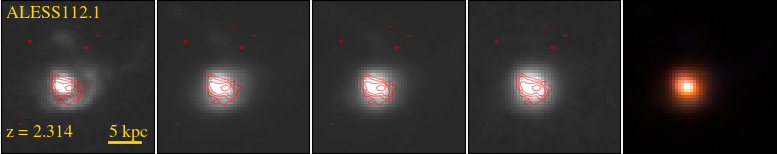}
\caption{Continued.}
\end{figure*}

\section{S\'ersic modelling before and after dust correction}\label{app:sersic}

\begin{figure*}[t]
    \centering
    \includegraphics[width=\linewidth]{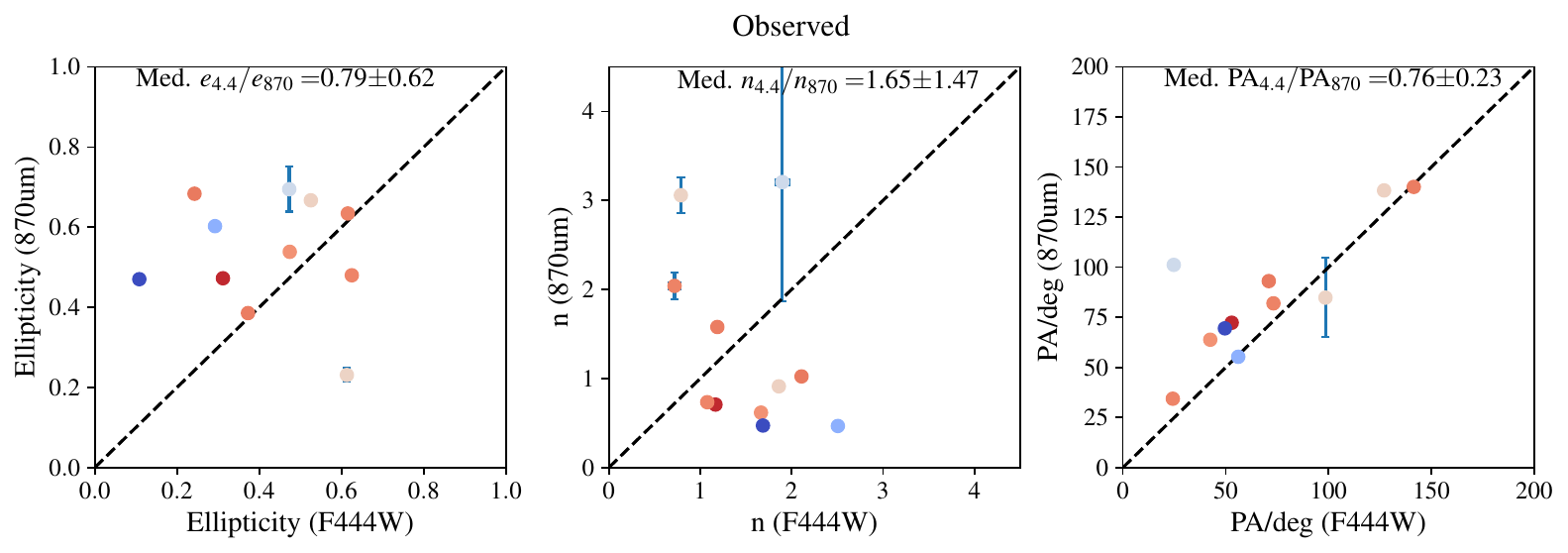}
    \includegraphics[width=\linewidth]{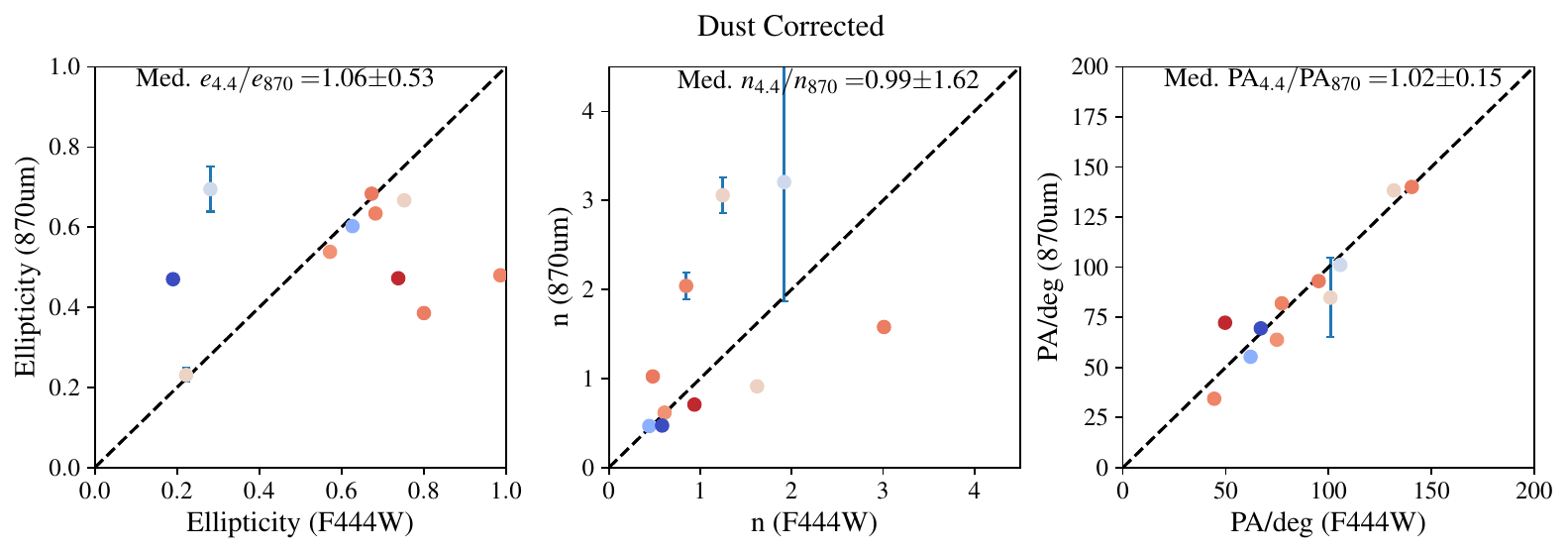}
    \caption{Comparison of S\'rsic parameters measured from the 870\um\ continuum and NIRCam/F444W imaging, before (top row) and after (bottom row) dust correction. Panels show ellipticity ($e$), S\'ersic index ($n$), and position angle (PA). Points are color-coded by central \av\ (blue to red; see Figures~\ref{fig:offset} and~\ref{fig:cog}). Median parameter ratios and their scatter are indicated in each panel.}
\label{fig:Sersic-alma-f444}
\end{figure*}

\begin{figure*}[t]
    \centering
    \includegraphics[width=\linewidth]{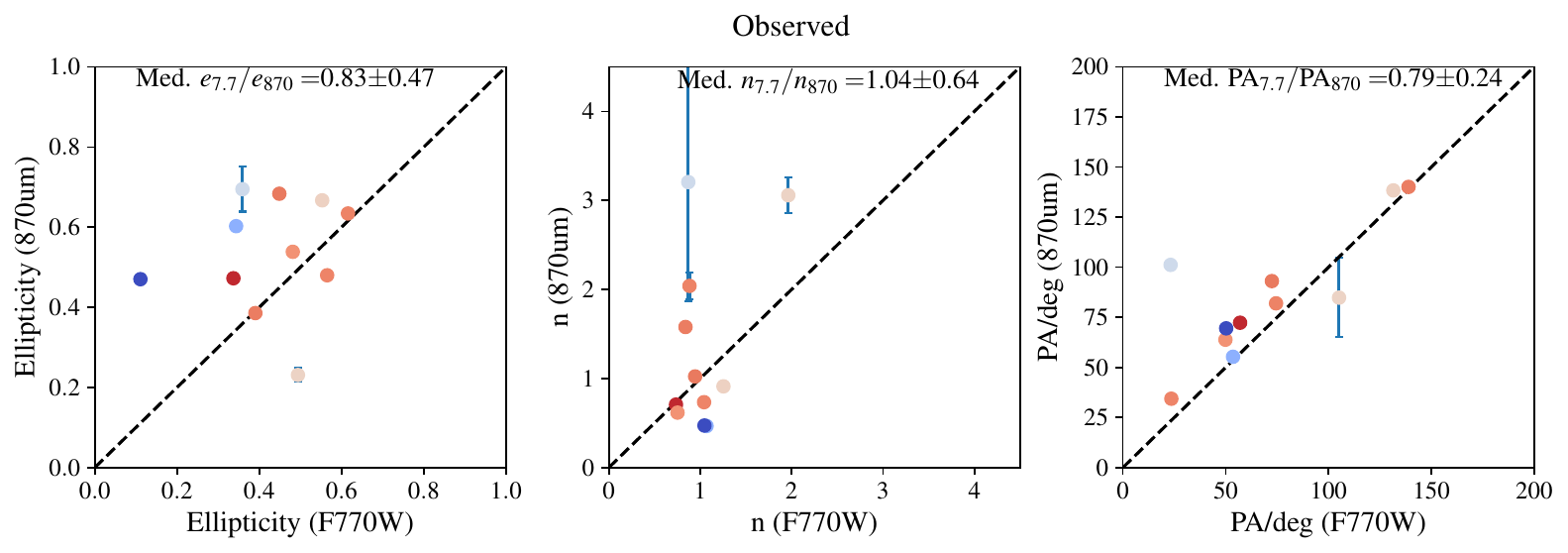}
    \includegraphics[width=\linewidth]{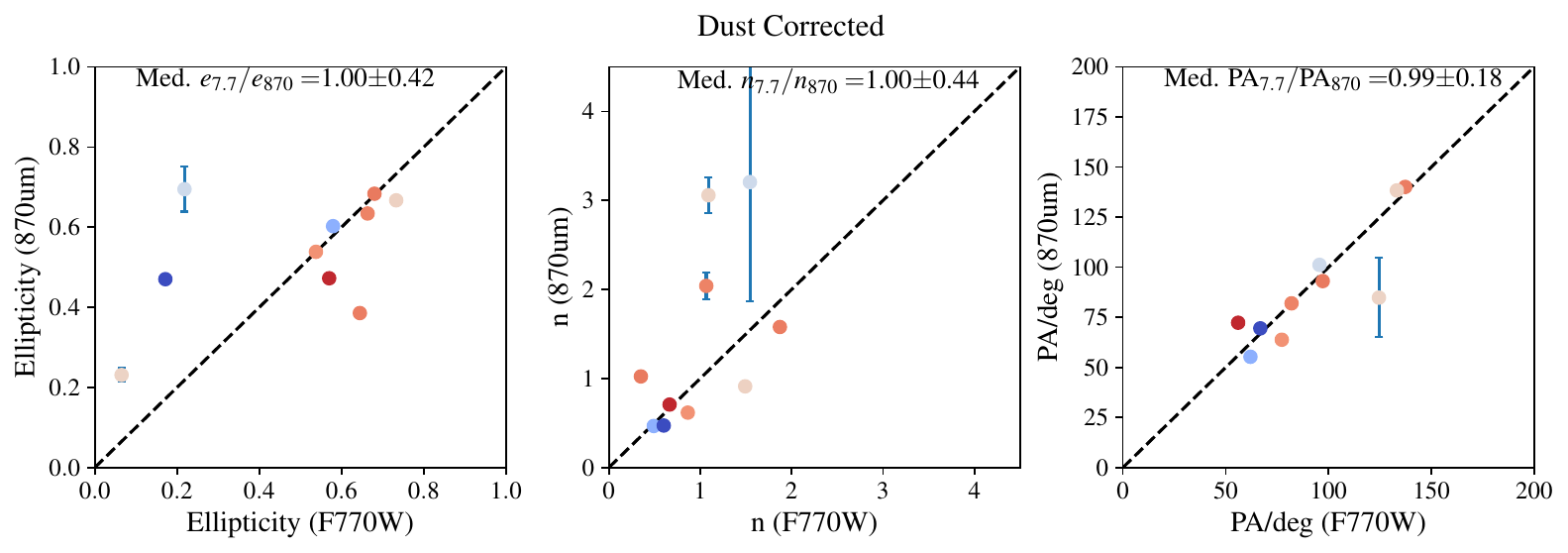}
    \caption{Same as Figure~\ref{fig:Sersic-alma-f444}, but for MIRI/F770W.
    }
    \label{fig:Sersic-alma-f770}
\end{figure*}

We performed S\'ersic modeling of the dust-corrected F444W and F770W images using \texttt{statmorph}, alongside fits to the observed images, to isolate the impact of dust attenuation on structural parameters. Figures \ref{fig:Sersic-alma-f444} and \ref{fig:Sersic-alma-f770} compare the resulting S\'ersic parameters (ellipticity, S\'ersic index, and position angle) with those from the ALMA 870\um\ continuum.

For F444W, dust correction significantly alters the inferred structure: the corrected parameters are more consistent with the ALMA measurements, although the scatter remains large. The position angle shows the clearest improvement, tightening from an already significant correlation in the observed images. This is consistent with \citet{hodge2025}, who identified position angle as the most robust parameter linking rest-frame near-IR and sub-millimeter emission.
F770W shows the same trend. Dust-corrected images yield S\'ersic parameters in closer agreement with ALMA than the observed images. Ellipticities shift toward unity relative to 870\um, while S\'ersic indices remain broadly consistent before and after correction, albeit with significant scatter. As in F444W, the position angle exhibits the strongest and most stable correlation with ALMA, largely independent of dust correction.

Overall, dust strongly perturbs structural parameters at rest-frame optical and near-IR wavelengths, particularly ellipticity and S\'ersic index, while the global position angle remains comparatively robust.



\end{document}